\documentclass[a4paper,fleqn,usenatbib]{mnras}

\usepackage{mathptmx}

\usepackage[T1]{fontenc}
\usepackage{ae,aecompl}

\usepackage{graphicx}	
\usepackage{amsmath}	
\usepackage{amssymb}	
\usepackage{natbib}
\usepackage{bm} 

\title[Tidally induced oscillations]{Non--adiabatic tidal oscillations induced by a planetary companion}

\author[A. Bunting, J. C. B. Papaloizou, C. Terquem]{
Andrew Bunting$^{1}$\thanks{andrew.bunting@physics.ox.ac.uk},
John C. B. Papaloizou$^{2}$\thanks{ J.C.B.Papaloizou@damtp.cam.ac.uk} and
Caroline Terquem$^{1,3}$\thanks{caroline.terquem@physics.ox.ac.uk}
\\
$^{1}$Department of Physics, Oxford University, Keble Road, Oxford OX1 3RH, UK\\
$^{2}$DAMPT, Cambridge University, Wilberforce Road, Cambridge, CB3 0WA, UK\\
$^{3}$ Institut d'Astrophysique de Paris, UPMC Univ Paris 06, CNRS,
  UMR7095, 98 bis bd Arago, F-75014, Paris, France \\
}

\date{Accepted XXX. Received YYY; in original form ZZZ}

\pubyear{2018}

\begin{document}
\label{firstpage}
\pagerange{\pageref{firstpage}--\pageref{lastpage}}
\maketitle

\begin{abstract}
  We calculate the dynamical tides raised by a close planetary companion on non--rotating stars of $1$~$\text{M}_{\odot}$ and $1.4$~$\text{M}_{\odot}$.  Using the Henyey method, we solve the fully non--adiabatic equations throughout the star.  The horizontal Lagrangian displacement is found to be 10 to 100 times larger than the equilibrium tide value in a thin region near the surface of the star.  {  This is because non--adiabatic effects dominate in a region that extends from below the outer edge of the convection zone up to the stellar surface, and the equilibrium tide approximation is inconsistent with non--adiabaticity.  Although this approximation generally applies in the low frequency limit, it also fails in the parts of the convection zone where the forcing frequency is small but larger than the Brunt-V\"ais\"al\"a frequency.  
 We derive  analytical estimates 
 which give a good approximation to the numerical values of the magnitude of the ratio of the horizontal and radial displacements at the surface.}  The relative surface flux perturbation is also significant, on the order of 0.1\% {for a system modelled on 51 Pegasi b}.  Observations affected by the horizontal displacement may therefore be more achievable than previously thought, and brightness perturbations may be the result of flux perturbations rather than due to the radial displacement.  We discuss the implication of this on the possibility of detecting such tidally excited oscillations, including the prospect of utilising the large horizontal motion for observations of systems such as 51 Pegasi.  \end{abstract}




\section{Introduction}

It is becoming increasingly apparent that planets are a common phenomenon, with observations suggesting that they may be about as common as stars~\citep{Borucki2011}. The first outright exoplanet discovery came in 1995: 51 Pegasi b~\citep{Mayor1995}, a gas giant orbiting a solar-type star. Since then, many more exoplanets have been discovered, with many of these being gas giants in close orbits~\citep{Winn2015}. Such planets are preferentially selected for in the two most successful exoplanet detection techniques: radial velocity and transit detection. The same characteristics which lead to this preference for hot Jupiters
(planets with a mass on the order of that of Jupiter, and with a close orbit, with semi-major axis up to $\sim 0.1$~au) also increase the possibility of detecting tidally induced oscillations.

Both detection methods depend upon variations in the light from the host star, but in both cases the star is supposed to be unchanged in its own reference frame. The fact that stars are variable is of course well known, and stellar modes of oscillation have been measured for the sun~(\citet{Ulrich1970}; \citet{DiMauro2017}) and many other stars (such as by~\citet{Brown1991} and~\citet{Kjeldsen2003}).

In these cases,  the driving mechanism for the oscillating modes is somewhat unclear. The gravitational effect of a hot Jupiter, however, provides  external forcing of the 
oscillation through the time--varying tidal potential. Such tidally excited oscillations have been studied in various regimes, from stellar binaries~\citep{Quataert1995} to planetary companions~\citep{Terquem1998}, including studies with orbital eccentricities both small~\citep{Arras2012} and large (\citet{Burkart2012}; \citet{Fuller2017}; \citet{Penoyre2018}).

Often, the equilibrium tide is used as a simple approximation. Focus has largely been upon the radial displacement induced by the tidal potential, and the horizontal displacement has been somewhat neglected, potentially because it does not contribute to the disc--integrated luminosity variations.  Conditions very near the
surface, which is where we are able to observe, are difficult to model, and the simple approximation may break down. Previous works have suggested that the horizontal displacement may be particularly poorly described by the equilibrium tide in this region (\citet{Savonije1983}; \citet{Arras2012}).

The work presented in this paper focusses upon the behaviour of the perturbation near the surface of the star.   The fully non--adiabatic oscillation equations are solved  in order to attempt to capture the surface behaviour as accurately as possible.   Both cases with frozen convection and convection { adjusting instantaneously to the tidal perturbation}  are considered.   

The paper is structured as follows: Section~\ref{sec:Methods} discusses the methods, set--up and governing equations and provides some analytical comparisons to assess how well the equations have been solved numerically; Section~\ref{sec:Results} describes the response of the star to the tidal perturbation, focussing upon the surface region, for a system similar to Pegasi 51 b; Section~\ref{sec:Discussion} discusses these results and their implications; Section~\ref{sec:Conclusion} summarises our conclusions.

\section{Methods}
\label{sec:Methods}

Section\,\ref{sec:Methods:convection} discusses the treatment of convection used throughout this work. We detail the set--up of the system and the equations which are solved in section\,\ref{sec:Methods:set-up}.  In section\,\ref{sec:Methods:analytical}, we derive analytical estimates that are used to ensure that the numerical solution is self--consistent.

\subsection{Treatment of convection}
\label{sec:Methods:convection}

For low mass stars, convection occurs close to the stellar surface and is likely to have an impact on the observable behaviour of the star. 
{ Here we adopt a description based on the local time dependent  Mixing Length Theory (MLT) that has been  discussed by many authors
\citep[see eg.][]{Unno1967, Gough1977, Salaris&Cassisi2008, Houdek2015}.
According to the MLT, under steady state conditions, or when it can be assumed to have relaxed
 to an equilibrium value, the convective flux is given by an expression of the form:}

\begin{equation}
{\bf F}_{\text{c}} = - A \bm{\nabla} s,\label{CONVFLUX}
  \end{equation}

  \noindent where $s$ is the entropy and  $A$ depends on the convective velocity, the mixing length and various other parameters (see appendix~\ref{sec:app:derivation} and eq.~[\ref{appB:Fconv}] for more details).  

In this work, we first { consider the case when} convection is frozen, i.e. the convective flux is unchanged by the perturbation.  
This leads to a value of the Lagrangian displacement at the surface of the star which differs { in magnitude  from the standard equilibrium tide
 value by more than an order of magnitude.   We then calculate the response to forcing   allowing for  the perturbation of the convective flux. }  Again, significant departure from the  equilibrium tide is obtained in that case, although the results are qualitatively different.  This enables us to assess the effect of convection on the Lagrangian displacement, within the framework of the MLT.
 
In order to calculate the perturbed convective flux in the calculations for which the convection has assumed to have relaxed to quasi-steady conditions, given the uncertainties in the MLT, we have followed two approaches.  In the first, which we label approach~A, we assume that it is dominated by the perturbation to the entropy gradient
 displayed on the right hand side of equation (\ref {CONVFLUX}).  That is to say,
 its linearised form  is approximated as: 

\begin{equation}
{\bf  F}'_{\text{c}} = - A  \bm{\nabla} s' ,  
\end{equation}

\noindent (see appendix \ref{sec:app:derivation} for more details).
We also use the fact that gradients are dominated by the radial component, so that the perturbed convective flux is radial { with that component being given by:}

\begin{equation}
 F'_{\text{c},r}  = - A \frac{\partial s'}{\partial r}.
 \label{eq:F_prime_conv_radial}
\end{equation}

\noindent This is because the variations in the radial direction turn out to be  over a scale $\lambda_{\rm r}$ which is very small compared to the radius $R$ of the star, and the ratio of radial to horizontal gradients near the surface of the star is on the order of $R/\lambda_{\rm r} \gg 1$.

{ In the second approach, which we label approach~B, we have allowed for variations in $A$ that incorporate perturbations to the state variables and also to the convective velocity.  This also has a dependence on the entropy gradient (see appendix~\ref{sec:app:derivation}).  We further remark that the procedures we have adopted are consistent with those used to calculate the stellar model we have chosen as our equilibrium background.}

\subsection{Set--up}
\label{sec:Methods:set-up}

The system studied here is that of a Jupiter--mass planet on a circular orbit around a central star.  We shall present detailed results for an orbital period of $4.23$ days, { which corresponds to that of the planet around 51~Peg. The choice of this period allows comparisons to be made with the results of Terquem et al. (1998).  We shall also scan neighbouring periods to search for resonances associated with $g$--modes that are confined in an interior radiative region.}  For the most part, this work focusses on a star of mass $M=1$~$ \text{M}_{\odot}$ with parameters close to those of the actual sun, but some results are shown for a typical main sequence star of mass $M=1.4$~$\text{M}_{\odot}$ for comparison. Some parameters describing the two stars are given in Table~\ref{tab:stellar_parameters}. 

\begin{table}
 \caption{The parameters of the stellar models used throughout this work.}
 \label{tab:stellar_parameters}
 \begin{tabular}{ccccc}
  \hline
  Mass & Luminosity & Radius & Age & $T_{\text{eff}}$ \\
  M$_{\sun}$ & L$_{\sun}$ & R$_{\sun}$ & Gyr & K \\

  \hline
  1.00 & 1.02 & 1.01 & 4.24 & 5790\\
  \hline
  1.40 & 4.66 & 1.67 & 1.58 & 6560\\
  \hline
 \end{tabular}
\end{table}

 We adopt spherical coordinates, $(r, \theta, \phi)$, centred on the non--rotating star,
 with the orbital angular momentum vector aligned with $\theta = 0$. 
 The associated unit vectors are $\bm{\hat{r}}$, $\bm{\hat{\theta}}$ and $\bm{\hat{\phi}}$, respectively.

The perturbing tidal potential introduces the only source of non--radial dependence,
 which has the form of a spherical harmonic of degree and order 2 ( i.e., in standard notation, $l=m=2$).
 More details can be found in appendix~\ref{sec:app:Tidal}.
 Because the equations are linear, the variables specifying the response that we solve for will
 therefore also have this non--radial dependence through a multiplicative factor.

The following assumptions are adopted in order to derive a set of governing equations for the forced response problem:
\begin{enumerate}
\item Time independence { of the background}: the equilibrium state of the star changes on a timescale much longer than the period
 of the oscillations.
\item Spherical symmetry: the equilibrium structure of the star is spherically symmetric, parametrised only as a function of radius.
{ This is assumed to apply after horizontal averaging in convection zones.}
\item Cowling approximation: the perturbation to the gravitational potential of the star is neglected,
 which is justified in the outer regions of the star   which has  low density  and in the inner parts of the star
 where the wavelength of the oscillatory response is small.
\item Small perturbations: the departures from equilibrium are everywhere small  such that 
the linear regime is a valid approximation.
\end{enumerate}

To calculate the response, the linear non--adiabatic stellar oscillation equations
 (derived in, e.g., \citet{Unno1989}) are solved for the case when 
 the star is perturbed by a regular tidal potential, due to the planet.
 The variables directly solved for are: $\xi_{r}$, the radial { component of the }  Lagrangian displacement; $F_{r}'$,
 the Eulerian perturbation to the radial total flux, $F_{r}$ (with contribution from both the radiative and convective fluxes); $p'$,
 the Eulerian perturbation to the pressure, $p;$  and $T'$,
 the Eulerian perturbation to the temperature, $T.$
 More details can be found in appendix~\ref{sec:app:derivation}.  These equations are:

\begin{multline} \label{eq:cont_osc}
\frac{1}{r^{2}} \frac{\partial}{\partial r} ( r^{2} \rho_{0} \xi_{r} )  
+ \left( \frac{\rho_{0}}{\chi_{\rho} p_{0}} - \frac{l (l+1)}{m^{2} \omega^{2} r^{2}} \right) p' \\
- \frac{\rho_{0}}{T_{0}} \frac{\chi_{T}}{\chi_{\rho}} T'
=
\frac{l (l+1)}{m^{2} \omega^{2} r^{2}} \rho_{0} \Phi_{P} ,
\end{multline}

\begin{multline} \label{eq:ent_osc}
\left( i \rho_{0} m \omega c_{p}  + \frac{l (l+1)}{r^{2}} K_{0} \right) T'
- \left( i m \omega c_{p} \nabla_{\text{ad}} \rho_{0} T_{0}  \right) \frac{p'}{p_{0}}\\
+ i m \omega \rho_{0} T_{0} \frac{d s_{0}}{d r} \xi_{r}
+ \frac{1}{r^{2}} \frac{\partial}{\partial r} ( r^{2} F_{r}') 
= 0 ,
\end{multline}

\begin{multline} \label{eq:flux_osc}
 \frac{F_{r}'}{K_{0}}
+\frac{\partial T'}{\partial r} - \frac{1}{T_{0}} \frac{d T_{0}}{d r}
 \left[ -3 + \frac{\kappa_{T}}{\kappa_{0}} 
 - \frac{\chi_{T}}{\chi_{\rho}} \left( 1 + \frac{\kappa_{\rho}}{\kappa_{0}}  \right) \right]T' \\
-  \frac{dr}{ds_{0}} \frac{F_{\text{c}, r,0}}{K_{0}}  \frac{d}{dr} \left(c_{p} \frac{T'}{T_{0}} \right)
- \frac{d T_{0}}{d r} \frac{1}{p_{0} \chi_{\rho}} \left( 1 + \frac{\kappa_{\rho}}{\kappa_{0}}  \right) p' \\
+ \frac{dr}{ds_{0}} \frac{F_{\text{c}, r,0}}{K_{0}}  \frac{d}{dr} \left( c_{p} \nabla_{\text{ad}} \frac{p'}{p_{0}} \right)
= 0 ,
\end{multline}

\begin{multline} \label{eq:mom_osc}
- m^{2} \omega^{2} \rho_{0} \xi_{r} 
+ \left( \frac{\partial}{\partial r} + \frac{\rho_{0}}{\chi_{\rho} p_{0}} \frac{d \Phi_{0}}{d r} \right) p' \\
-  \frac{d \Phi_{0}}{d r} \frac{\rho_{0}}{T_{0}} \frac{\chi_{T}}{\chi_{\rho}} T'
=
- \rho_{0} \frac{\partial \Phi_{P}}{\partial r} ,
\end{multline}

\noindent which correspond to the linearised continuity, entropy, and radiative diffusion equations together with the radial component of the equation of motion, respectively.
 The subscript $0$ refers to the equilibrium state. 
 Here $\rho$ is the density;
 $\omega$ is the angular frequency of the planet's orbit; $\Phi_{P} = - G m_{P} r^2/ (4 D^{3}) $ is
 the  tidal potential after removal of the spherical harmonic factor
 (see Appendix~\ref{sec:app:Tidal}), with $m_{P}$ being  the planetary mass, $G$ the gravitational constant 
and $D$ the orbital radius; $c_{p}$  is the specific heat capacity at constant pressure; $K$ is the radiative thermal conductivity;
 $\chi_{\rho} \equiv \left( \partial \ln p/\partial \ln \rho \right)_{T}$ and  $\chi_{T} \equiv \left( \partial \ln p/{\partial \ln T} \right)_{\rho}$;  
$\kappa$ is the opacity;   $\kappa_{\rho} \equiv \left( \partial \kappa/\partial \ln \rho \right)_{T}$ and $\kappa_{T} \equiv \left( \partial \kappa/\partial \ln T \right)_{\rho}$; $s$ is the specific entropy;  $F_{\text{c}, r,0}$ is the radial component of the equilibrium convective flux; 
 $\nabla_{\text{ad}} = \left( \partial \ln T_{0}/{\partial \ln p_{0}} \right)_{s}$.
 The radiative thermal  conductivity is related to the opacity, temperature and density through:
$K = 4 a c_{*} T^3 / (3 \kappa \rho)$; here
$a =4 \sigma/c_{*} = 7.5657 \times 10^{-15}$ erg cm$^{-3}$ K$^{-4}$ is the radiation density constant, with 
$\sigma$ being the Stefan--Boltzmann constant and $c_{*}$ being the speed of light
 (the subscript--asterisk is to differentiate it from the sound speed).

All the perturbed quantities are proportional to $Y_{2}^{-2}(\theta, \phi) \text{e}^{2 \text{i} \omega t}$, where $Y_{2}^{-2}$ is a spherical harmonic (see Appendix~\ref{sec:app:Tidal}).
 This factor is taken as read and, following standard practice, can be removed 
 so that  all the perturbation  quantities which appear in equations~(\ref{eq:cont_osc})--(\ref{eq:mom_osc})
 depend only upon $r$. Note that at this stage these quantities are in general complex.
Once the solution has been calculated, the spherical harmonic factors can be restored and the real part taken to obtain a physical solution.

The boundary conditions are split, two apply at the centre and two at the surface.  At the centre, $\xi_{r} = 0$ and $F_{r}' = 0$, which ensure regularity.  In our calculations, we assume that the surface is free, so that $\Delta p = 0$ there, where $\Delta$ denotes the Lagrangian perturbation.  We however also test different conditions on $\Delta p$ to check that our results do not depend on it.   Finally, assuming that the star is a blackbody emitter, and defining the surface as where the temperature is equal to the effective temperature, $\left( 4 \Delta T/T_{0} - \Delta F_{r}/F_{r_{0}} \right) = 0$ there.  

The Henyey method \citep{Henyey1964} is used to solve these equations, with a stellar model produced by MESA (\citet{Paxton2011}, \citet{Paxton2013}, \citet{Paxton2015}, \citet{Paxton2018}) as the equilibrium background star which we perturb.  For a more extended account of the numerical method see \citet{Savonije1983}, where it was used to evaluate the tidal response of a massive star.  In order to ensure that the behaviour at the surface is accurately captured, the resolution is smoothly increased using linear interpolation, leading to a grid cell width, $\delta r,$ such that $\delta r/r = 3.5 \times 10^{-6}$ there.

\begin{figure*}
        \centering
        \includegraphics[width=7cm]{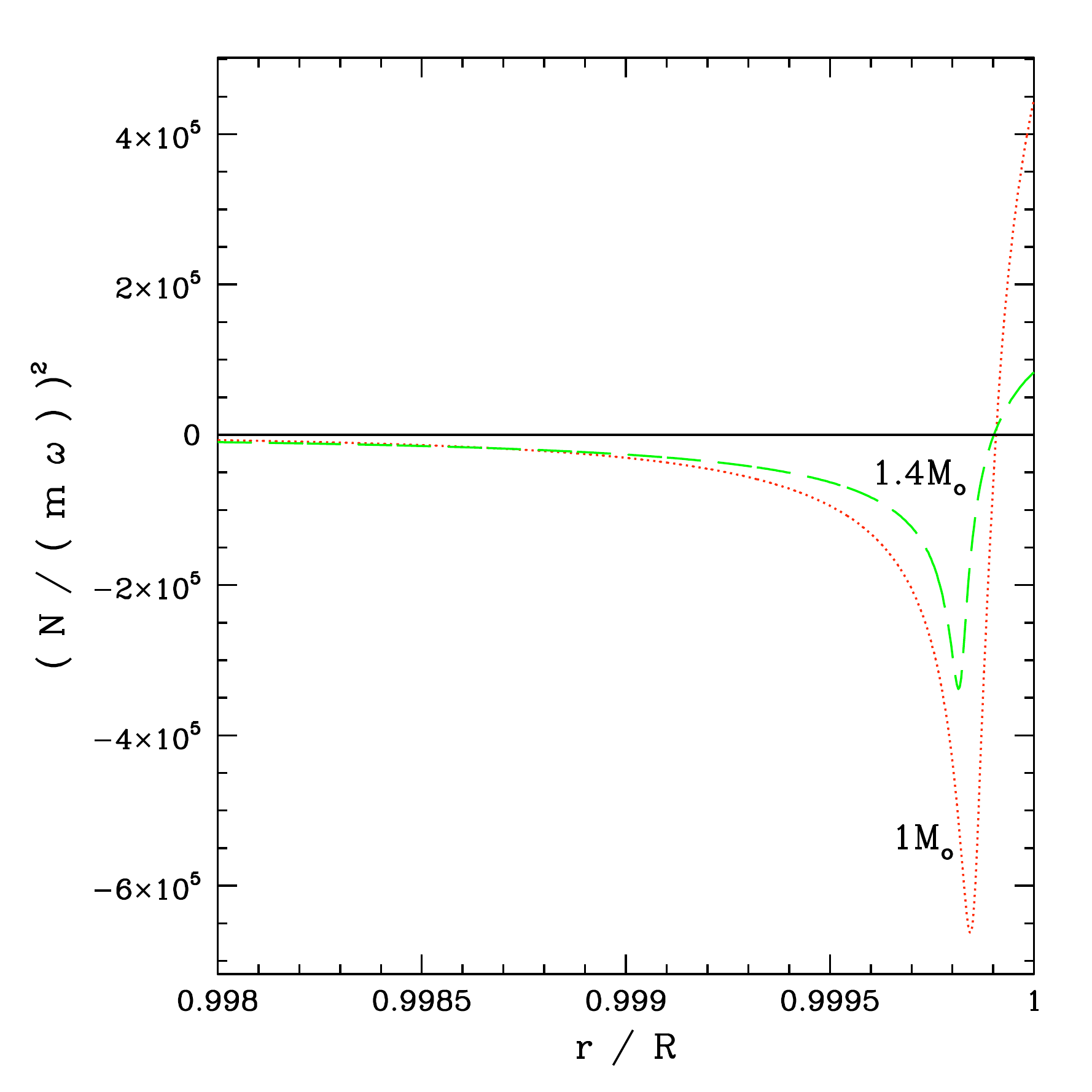}
        \includegraphics[width=7cm]{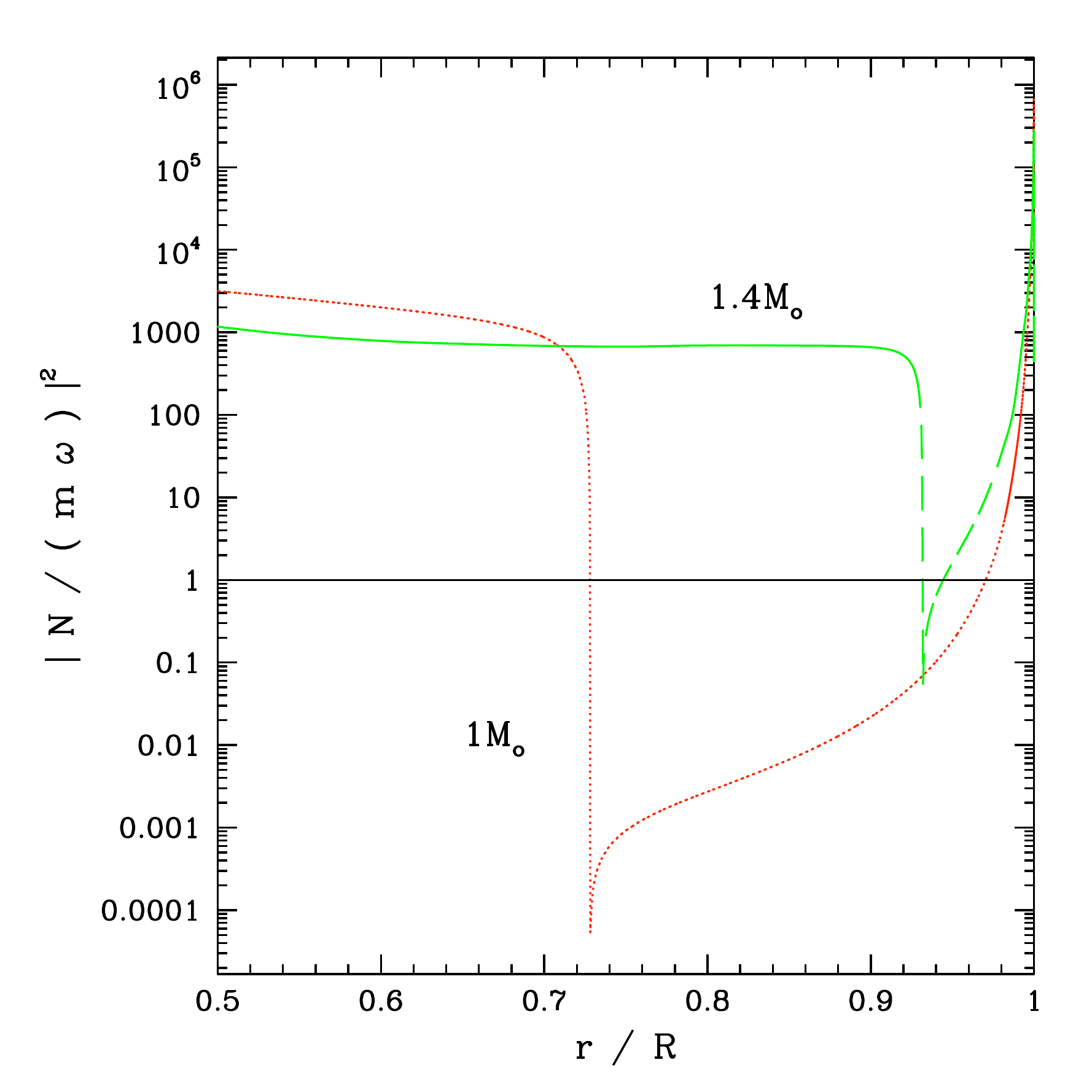}
        \caption{The left panel shows $N^{2}/(m^{2} \omega^{2})$ as a function of $r/R$, where $R$ is the stellar radius, near the surface, for an orbital period of $4.23$~days.
          The red dotted line corresponds to a $1$~$\text{M}_{\odot}$ star, and the green dashed line corresponds to a $1.4$~$\text{M}_{\odot}$ star.  The squared, normalised Brunt-V\"ais\"al\"a frequency (in units of the angular frequency of the oscillations -- twice that of the orbit) indicates the structure of the star: a negative value corresponds to an imaginary frequency and implies convection; a positive value indicates a real frequency and therefore a stratified, radiative region. { The right panel shows $|N^{2}/(m^{2} \omega^{2})|$ in logarithmic scale  for the outer 50\% of the stellar radius  with the same colour coding.  This quantity passes through zero near $ r/R = 0.73$ and $0.94$ for the lower and higher mass, respectively,  which correspond to the inner boundaries of their  convective envelopes. This behaviour is not fully resolved in the plots.  In convection zones, the quantity  $|N/(m \omega)|$ measures the ratio of the inverse forcing frequency to the convective time scale. It can be seen that this exceeds unity for the solar mass model  for $ r/R > 0.97$ and for   $r/R > 0.94$ for the  $1.4$~$\text{M}_{\odot}$ star.
          } }
    \label{fig:N2_surface}
\end{figure*}

At the surface of both the $1$~$\text{M}_{\odot}$ and $1.4$~$\text{M}_{\odot}$ stars, there is a thin, strongly stably stratified layer.  This is illustrated 
in Figure~\ref{fig:N2_surface}, which shows $N^{2}/(m^{2} \omega^{2})$ against $r / R$ in the surface region, where $R$ is the stellar radius and  $N^2$ is  the square of the Brunt--V\"ais\"al\"a frequency, defined through:
\begin{equation}
N^2 = \frac{1}{\rho_0}\frac{dp_0}{dr}\left( \frac{1}{\rho_0}\frac{d\rho_0}{dr}-  \frac{1}{\Gamma_1p_0}\frac{dp_0}{dr}\right),
\end{equation} 
Here $\Gamma_1 = d\ln p_0/d \ln \rho_0|_s$ is the adiabatic exponent,  with the derivative being taken at constant specific entropy.  { Negative values of  $N^2$  indicate a convection zone, which extends from $r/R=0.73$ to $r/R=0.9999$ for the $1$~$\text{M}_{\odot}$ star and from $r/R=0.93$ to $r/R=0.9999$ for the $1.4$~$\text{M}_{\odot}$  star.}

\noindent Near the surface, $N^2$ transitions suddenly from being large and negative to being large and positive, reflecting a
 transition from a strongly superadiabatic convection region to a strongly stably stratified region. In this thin, radiative layer, the local thermal time scale is much shorter than the period of the forced oscillations,
 enabling  heat to escape
 on a timescale much shorter than this period.  Accordingly, we expect that non--adiabatic effects will be important in this region.  {  However, the region near the surface in which non--adiabaticity is important extends below this stably stratified layer into the outer parts of the convection zone, as we show in Section~\ref{sec:Methods:analytical:r_na}.  }
 
 \subsection{ Extent of the non--adiabatic zone below the stellar surface}
 \label{sec:Methods:analytical:r_na}

{
 
 We define $r_{\text{na,f}}$ and $r_{\text{na,p}}$ as the radius of  the base of the non--adiabatic zone when convection is frozen and perturbed, respectively. To evaluate this radius, we start by writing the perturbed energy equation under the form:

 \begin{equation}
\rho T \left( \frac{\partial s'}{\partial t} + \bm{u} \bm{\cdot} \bm{\nabla}  s \right)
= - \bm{\nabla} \bm{\cdot} \bm{F}' ,
\label{secondthermo2}
\end{equation}

\noindent where $\bm{u}$ is the vector velocity.  (Note that although the subscript `0' does not appear here as we are only concerned with estimates, background state variables are used).
In this section, we are interested in the cases where either the perturbed radiative flux, which we denote $\bm{F}'_{\text{rad}}$, or the
perturbed convective flux, $\bm{F}'_{\text{c}}$, dominates.
As here we are only concerned with an estimate of the extent of the non--adiabatic zone, we approximate $\bm{F}'_{\text{rad}}$ and $\bm{F}'_{\text{c}}$ by assuming that most of the contributions to these perturbed quantities come from the radial gradients of temperature and entropy, respectively.  Therefore:
\begin{equation}
{F}'_{\text{c}} \sim -A \frac{ \partial s'}{\partial r}, \; \;  {\text{with}}  \; \;A \sim \rho v_{\text{c}} l T,
\end{equation}
to within a factor of order unity, where $v_{\text{c}}$ is the convective velocity and $l$ is the mixing length (see eq.~[\ref{appB:Fconv}], where we remark that we have neglected the cooling term at the denominator).  For the radiative flux, we have ${F}'_{\text{rad}} \sim - K \partial T'/ \partial r$.  {We remark that perturbations to the opacity may also contribute to ${F}'_{\text{rad}}$. However, this is not expected to change the estimates given below significantly when, as in our case, the non adiabatic layer extends below the very  strongly superadiabatic region into regions where  the opacity is no longer increasing very rapidly with temperature.} Using thermodynamic relations, changes in entropy can be related to changes in temperature through $T \bm{\nabla} s=\epsilon c_p \bm{\nabla} T$, where $\epsilon= 1- \nabla_{\text{ad}}/ \nabla$ with $\nabla= (\partial \ln T / \partial r)/( \partial \ln P/ \partial r)$.
(In the deep parts of the convection zone, convection is efficient and $|\epsilon| \ll 1$.)
Neglecting variations of $\epsilon$, this relation implies $ \partial T'/ \partial r = T (\partial s'/ \partial r) / (\epsilon c_p)$, and therefore:
\begin{equation}
F'_{\text{rad}} \sim -D \frac{ \partial s'}{ \partial r}, \; \; {\text{with}} \; \;  D \sim \frac{K T}{\epsilon c_p}.
\end{equation}
So both the convective and radiative fluxes can be viewed as arising from the diffusion of entropy with different diffusion coefficients.  Noting $\lambda_r $ the radial scale on which perturbations vary   (near the stellar surface,  $\lambda_r \ll r$)
 we may then set $\partial / \partial r \sim 1/\lambda_r$.   This yields $ |\bm{\nabla} \bm{\cdot} \bm{F}' | \sim  |F'|/ \lambda_r$, so that:
 \begin{equation}
 |\bm{\nabla} \bm{\cdot} \bm{F}'_{\text{rad}} | \sim \frac{Ds'}{\lambda_r^2} \; \; {\text{and}} \; \;  |\bm{\nabla} \bm{\cdot} \bm{F}'_{\text{c}} | \sim \frac{As'}{\lambda_r^2};
 \label{eq:divF_estimates}
 \end{equation} 



 Note that, in the left hand side of equation~(\ref{secondthermo2}):
 \begin{equation}
\left| \frac{\partial s'}{\partial t} \right| = m \omega |s'| = \frac{2 \pi}{P} |s'|,
 \end{equation}
  where $m=2$ and $P=2 \pi/(m \omega)$ is the period of the oscillations.


We now look at the two  cases of frozen and perturbed convection in turn, and discuss where convection is relaxed.  

\subsubsection{Frozen convection:}

When convection is frozen, $\bm{F}' = \bm{F}'_{\text{rad}}$.  In this case, equation~(\ref{secondthermo2}) yields:

\begin{multline}
\left| \frac{1}{s'} \frac{\partial s'}{ \partial t} + \bm{u} \bm{\cdot} \frac{ \bm{\nabla}  s}{s'} \right| \sim \frac{D}{\rho T \lambda_r^2}   \sim  \frac{K T }{\epsilon c_p T \rho \lambda_r^2}   \sim
\frac{F_{\text{rad}}H_p}{\epsilon c_p T \rho \lambda_r^2} \\  \sim \frac{4 \pi r^2 F_{\text{rad}}}{ c_p T 4 \pi r^2 \rho \lambda_r} \frac{H_p}{\epsilon \lambda_r}, 
\end{multline}
  

\noindent where we have used $F_{\text{rad}} \sim K T /H_p$, where $H_p$ is the pressure scale height.  In the numerator, the term $4 \pi r^2 F_{\text{rad}}$ is the radiative luminosity $L_{\text{rad}}(r)$.  In the denominator, $c_p T$ is approximately the internal energy per unit mass, $4 \pi r^2 \rho \lambda_r$ is the mass within a spherical shell of radius $r$ and width $\lambda_r$, and therefore the product is approximately the internal energy $E_{\text{int}}(r)$ contained within that shell.  This relation can therefore be written as:
\begin{equation}
  \left| \frac{1}{s'} \frac{\partial s'}{\partial t} + \bm{u} \bm{\cdot} \frac{\bm{\nabla}  s}{s'} \right|  \sim \frac{H_p}{\epsilon \lambda_r} \frac{1}{t_{\text{rad}}} ,
  \label{balancerad}
  \end{equation}
  where $t_{\text{rad}} = E_{\text{int}}(r) /L_{\text{rad}}(r) $ is the timescale on which energy is transported by radiation through a distance $\lambda_r$ at radius $r$.  In the outer parts of the convection zone where convection is not efficient, $\epsilon \sim 1$.  We also have $\lambda_r \sim H_p$ there.  The perturbation is non--adiabatic{, and approximately isothermal,} if $t_{\text{rad}} \ll P$, which means that the right hand side of equation~(\ref{balancerad}) is large compared  to $| (\partial s'/ \partial t)/s'|$.  In that case, the balance is between $\rho T \bm{u} \bm{\cdot} \bm{\nabla} s$ and $- \bm{\nabla} \cdot \bm{F}'_{\text{rad}}$.  In the opposite regime, in the adiabatic parts of the convective zone, $t_{\text{rad}} \gg P$ and the balance is between $\partial s'/ \partial t$ and $\bm{u} \bm{\cdot} \bm{\nabla} s$ (that is to say, $\bm{\nabla} \bm{\cdot} \bm{F}'_{\text{rad}} \simeq 0$).  The transition between the two regimes is therefore where $| (\partial s'/  \partial t)/s'| \sim 1/ t_{\text{rad}} $, which means that the radius $r_{\text{na,f}}$  of the base of the non--adiabatic zone 
  is determined by $L_{\text{rad}}(r_{\text{na,f}})=m \omega E_{\text{int}}(r_{\text{na,f}} ).$

\noindent  The internal energy in a spherical shell of width $\lambda_r \sim H_p$ at $r_{\text{na,f}}$ is comparable to that between $r_{\text{na,f}}$ and the stellar surface $R$ (since the mass density decreases sharply towards the surface).  Therefore  we approximate  $r_{\text{na,f}}$ by writing that the internal energy of the material in the region above that radius is equal to the energy transported by radiation through that radius during $P/(2\pi)$:

\begin{equation} \label{eq:Methods:analytical:r_na}
\frac{L_{\text{rad}}}{m \omega} = \int_{r_{\text{na,f}}}^{R} \text{d}E_{\text{int}}(r) = \int_{r_{\text{na,f}}}^{R} \frac{6 \pi k_{\text{B}} \rho T r^{2} \, \text{d}r}{\mu m_{\text{H}}} , \end{equation}

\noindent where $k_{\text{B}}$ is the Boltzmann constant, $\mu$ is the mean molecular weight per gas particle (including both ions and free electrons), and $m_{\text{H}}$ is the mass of a Hydrogen atom. For simplicity, the stellar material is taken to be a monatomic ideal gas.   This gives estimates of $r_{\text{na,f}}/R = 0.9995$ for $\text{M} = 1$~$\text{M}_{\odot}$ and $r_{\text{na,f}}/R= 0.9991$ for $\text{M} = 1.4$~$\text{M}_{\odot}$.   
Figure~\ref{fig:N2_surface} shows that $ r_{\text{na,f}}$ is roughly the inner edge of the super--adiabatic region, where $N^2$ has large variations.  

\subsubsection{Perturbed convection:}\label{Pertcon}

  In the parts of the convection zone where the perturbed convective flux dominates,
equation~(\ref{secondthermo2}) yields:

\begin{equation}
\left| \frac{1}{s'} \frac{\partial s'}{\partial t} + \bm{u} \bm{\cdot} \frac{ \bm{\nabla}  s}{s'} \right| \sim \frac{A}{\rho T  \lambda_r^2}   \sim \left( \frac{l}{ \lambda_r^2} \right) \frac{v_{\text{c}}  }{l} \sim \left( \frac{l}{ \lambda_r^2} \right) \frac{1}{t_{\text{c}}},
\end{equation}
where $t_{\text{c}} \sim l/ v_{\text{c}}$ is the convective timescale.  Within the MLT, this timescale is also given by $t_{\text{c}} \sim 1/\sqrt{|N^2|}$.  Note that  $l \sim H_p$, and $\lambda_r \sim H_p$ in the outer parts of the convection zone.  Above the radius $r_{\text{na,f}}$ calculated in the case of frozen convection, the radiative flux is not negligible so the equation above does not apply.
In the parts of the convection zone below this radius and where $ t_{\text{c}} \ll P$, which means that the perturbation is {governed by diffusion rather than advection}, the balance is between $\rho T \bm{u} \bm{\cdot} \bm{\nabla} s$ and $- \bm{\nabla} \cdot \bm{F}'_{\text{c}}$. 
In the opposite regime, in the parts of the convection zone which are {governed by advection} , $t_{\text{c}} \gg P$ and, as in the case of frozen convection, the balance is between $\partial s'/ \partial t$ and $\bm{u} \bm{\cdot} \bm{\nabla} s$, which again implies $\bm{\nabla} \bm{\cdot} \bm{F}'_{\text{c}} \simeq 0$. 
Therefore, the transition between the two regimes is where $| (\partial s'/  \partial t)/s'| \sim 1/ ( t_{\text{c}}) $, the {corresponding} radius $r_{\text{na,p}}$  is approximately determined by $\sqrt{|N^2|} = m \omega $. {In some circumstances (see below) this can correspond to the base of the non--adiabatic zone.}

 {The above criterion is for the transition between the advective and diffusive transport of entropy. If the background is nearly isentropic on large scales
 this transition may not correspond to a transition to non adiabatic behaviour. To consider this further we consider a transition radius determined by
 the condition that the unperturbed convective luminosity can replenish the internal energy of the upper layers in a time  $\sim 1/(m\omega).$ 
 This is determined by the condition  \\
 $m \omega \sim F_{\text{c}} / ( p H_{p} ) \sim \rho v_{c}^{3} / ( p H_{p}) \sim \rho v_{c}^{2} \sqrt{|N|^{2}} / p  \sim (v_{c}^{2}/c_s^2)/t_c.$ 
 This  takes a similar form to  the criterion given by equation~(\ref{eq:Methods:analytical:r_na}), but with the radiative luminosity replaced by the convective luminosity.
It may be written in the form
$\sqrt{|N^2|} = m \omega f_s $ where
$ f_s = ( \rho g^{2}) / ( |N^{2}| p ) \sim c_s^2 / v_{c}^{2}$ greatly exceeds unity in the  deepest regions of the  convection zone.
The magnitude of  $f_s$ relates to the efficiency  of  the convection. In particular the parameter $1/f_{s}$ measures how superadiabatic
  the convection  is. The associated thermal time scale is then $f_{s} t_c$,  as only a small amount of effective heating can occur in one turn over time 
  of the almost adiabatic convection when $f_s$ is large. 
The radius specified by this condition is  defined to be $r_{\text{na,p1}}.$ 

If this is to determine the location of the transition from adiabatic to non adiabatic behaviour of the perturbations, the small degree of superadiabaticity must be maintained for the perturbations as well as the background.
This condition may be approached when the background is approximately isentropic over large scales as might be expected 
in the inner regions of the convection zone. However, we remark that the scale over which entropy perturbations  diffuse increases from $H_p$ over 
the time  $t_c$  for a transition at $r=r_{\text{na,p}}$ to $H_p f_s$  over the time  $t_c f_s$ for a transition at $r=r_{\text{na,p1}}$ 
increasing the scale of the required almost isentropic background. If this is untenable transition to non adiabatic behaviour may occur 
below this radius.}


\noindent Figure~\ref{fig:N2_surface} shows that, {for $\sqrt{|N^2|} = m \omega $,} $r_{\text{na,p}}/R =0.97$ for $M=1$~M$_{\odot}$ and $r_{\text{na,p}}/R =0.94$ for $M=1.4$~M$_{\odot}$, which is significantly deeper than in the case of frozen convection. {In addition we find that $r_{\text{na,p1}}/R =0.998$ for $M=1$~M$_{\odot}$ and $r_{\text{na,p1}}/R =0.995$ for $M=1.4$~M$_{\odot}$.  } Therefore non--adiabatic effects are potentially significant in a layer of the convection zone larger than could have been expected.  As will be shown below, this has important consequences on the response of the star to the tidal forcing.


\subsubsection{Relaxed convection}

In both the frozen and perturbed convection cases, the convective timescale is smaller than $P/(2 \pi)$ in the non--adiabatic surface region (in the case of frozen convection it is even very small compared to $P/(2 \pi)$).  Accordingly, we assume that convection relaxes to an equilibrium state in the non--adiabatic region, whereas below $ r_{\text{na,f}}$ for frozen convection and $ r_{\text{na,p}}$ for perturbed convection,  the response is nearly adiabatic.

}

\subsection{Illustrative estimates of $| \xi_{h}|/| \xi_{r}|$}
\label{sec:Methods:analytical}

Here, we focus upon the behaviour of the Lagrangian displacement from the equilibrium position, $\bm{\xi}$. 
In particular, we highlight the  characteristic dominance of the horizontal displacement in the non--adiabatic surface layers,
which is a significant departure from the behaviour expected  from the standard equilibrium tide under adiabatic conditions.
{ We argue that such a departure is expected because the strong constraint of hydrostatic equilibrium together
with zero variation of the Lagrangian pressure and density perturbations, that is required to obtain the standard equilibrium tide, cannot be satisfied when the perturbation is non--adiabatic.  
 The equilibrium tide approximation only applies for adiabatic perturbations in regions with $N^2 \ne 0$ (see appendix  \ref{sec:app:eq}).}

The radial component, $\xi_{r}$, is simply output in the solution of the oscillation equations (\ref{eq:cont_osc}) - (\ref{eq:mom_osc}).  Given this complete solution, the horizontal displacement, $\bm{\xi}_{h} = \bm{\xi} - \xi_{r} \bm{\hat{r}}$, can be derived either from the linearised continuity equation or the linearised horizontal component of the equation of motion.

\subsubsection{The radial and horizontal components of the Lagrangian displacement}

The  linearised continuity equation, with the angular dependence retained:

\begin{equation} \label{eq:Methods:analytical:cont_lin}
\frac{\partial}{\partial t} \left( \rho' + \bm{\nabla} \cdot \left( \rho_{0} \bm{\xi} \right) \right) = 0,
\end{equation}
with 
\begin{equation}
 \rho' = \frac{\rho_{0}}{\chi_{\rho} p_{0}} p'- \frac{\rho_{0}}{T_{0}} \frac{\chi_{T}}{\chi_{\rho}} T' ,
 \label{rhop}
\end{equation}
\noindent can be rearranged to give an expression for the divergence of the horizontal displacement:

\begin{equation} \label{eq:Methods:analytical:rearranged_cont}
\bm{\nabla \cdot \xi}_{h} = - \frac{\rho'}{\rho_{0}} - \frac{1}{\rho_{0} r^{2}} \frac{\partial}{\partial r} \left(  r^{2} \rho_{0} \xi_{r}  \right).
\end{equation}

\noindent The horizontal components of the equation of motion, with the angular dependence retained, give:

\begin{equation} \label{eq:Methods:analytical:rearranged_eom}
\bm{\xi}_{h} = \frac{1}{m^{2} \omega^{2}} \bm{\nabla}_{\perp} \left(  \frac{p'}{\rho_{0}}  +  \Phi_{\text{P}}  \right) ,
\end{equation}

\noindent where $\bm{\nabla}_{\perp}$ is the horizontal component of the gradient. 
 As we have mentioned above, $\xi_r$, $p'$, $\rho'$ and $\Phi_{\text{P}}$ are proportional to $Y_{2}^{-2}(\theta, \phi)$.  However, equation
 ~(\ref{eq:Methods:analytical:rearranged_eom}) shows that this is not the case for $\bm{\xi}_{h}$.  By introducing $V$ such that $\bm{\xi}_{h} \equiv r \bm{\nabla}_{\perp} V$, equation~(\ref{eq:Methods:analytical:rearranged_cont}) becomes:

\begin{equation} \label{eq:Methods:analytical:rearranged_cont_V}
  V = \frac{r}{\rho_{0} l(l+1)} \left[ \rho' + \frac{1}{r^{2}} \frac{\partial}{\partial r} \left( r^{2} \rho_{0} \xi_{r} \right) \right] ,
\end{equation}

\noindent and equation~(\ref{eq:Methods:analytical:rearranged_eom}) becomes:

\begin{equation} \label{eq:Methods:analytical:rearranged_eom_V}
  V = \frac{1}{r m^{2} \omega^{2}} \left( \frac{p'}{\rho_{0}} + \Phi_{\text{P}} \right).
\end{equation}

\noindent As $V$ has the same angular dependence as the other quantities with which it is being compared, and is a scalar instead of a vector, it is much easier to use in the analysis than $\bm{\xi}_{h}$.  We can calculate $\bm{\xi}_{h}$ from $V$ by using $V(r, \theta, \phi, t)= V(r) Y_{2}^{-2}(\theta, \phi) \text{e}^{2 \text{i} \omega t }$, with $ Y_{2}^{-2}(\theta, \phi)=3 \sin^2 \theta \text{e}^{-2 \text{i} \phi}$.  Then  $\bm{\xi}_{h} = r \bm{\nabla}_{\perp} V$ yields:

\begin{equation} \label{eq:V_spatial}
\bm{\xi}_{h} (r, \theta, \phi, t) =  6 V(r) \sin \theta \text{e}^{2 \text{i} (\omega t - \phi)} \left( \cos \theta \bm{\hat{\theta}} - \text{i} \bm{\hat{\phi}} \right).
\end{equation}

\noindent This can be combined with the radial displacement to give the full (complex) vector displacement as:

\begin{multline}
\label{eq:xi_vec}
\bm{\xi}  (r, \theta, \phi, t) = \\ 3 \sin(\theta) \text{e}^{2 \text{i} (\omega t - \phi)} \left[  \xi_{r}(r) \sin \theta \bm{\hat{r}}  +  2 V(r) \left( \cos \theta \bm{\hat{\theta}}  -   \text{i} \bm{\hat{\phi}} \right) \right],
\end{multline}

\noindent which shows that, when estimating contributions to the displacement vector, $V$ can be compared with $\xi_{r}$ up to a trigonometric factor and a factor of $2$.

\subsubsection{Numerical calculation of the  horizontal displacement }

\begin{figure*}
        \includegraphics[width=\columnwidth]{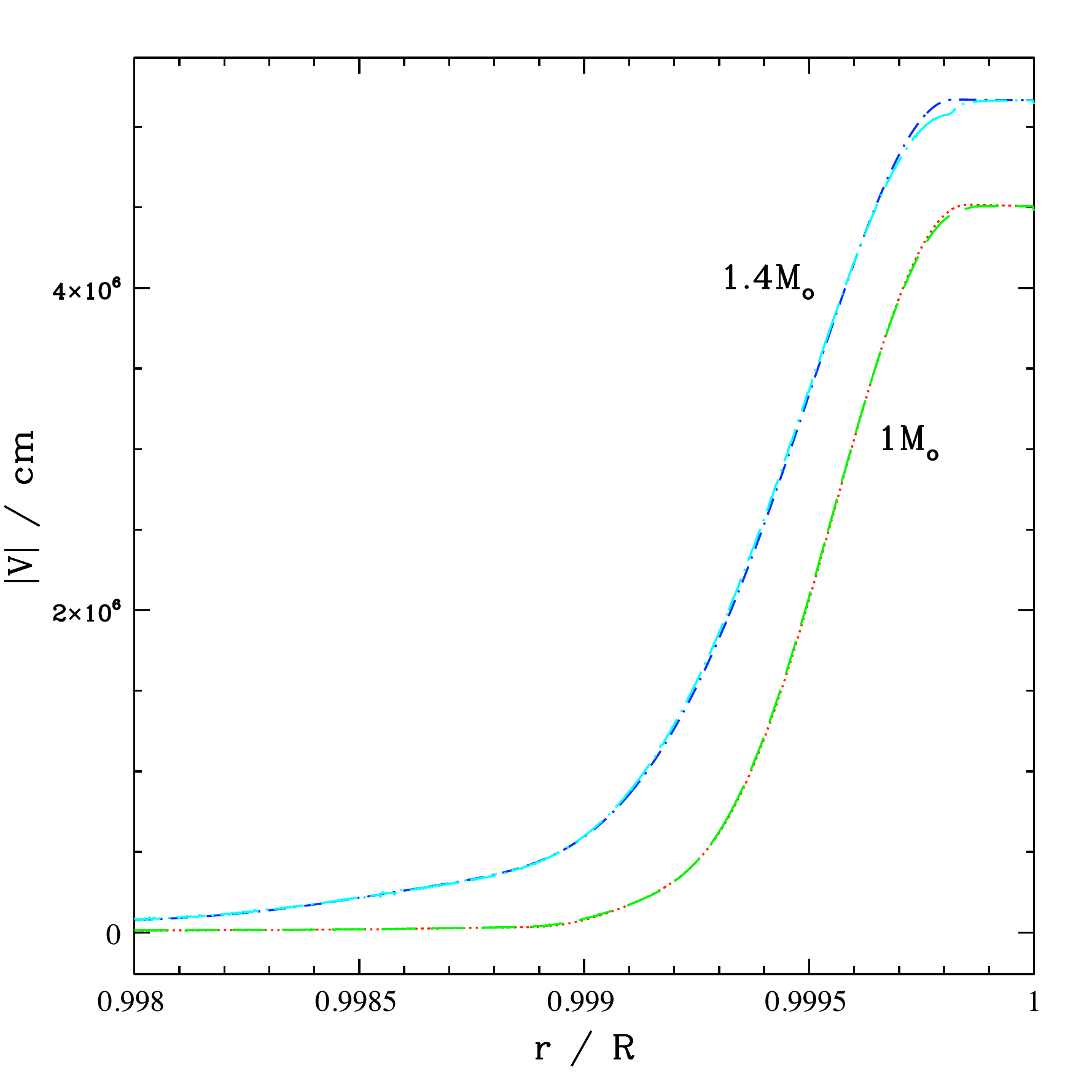}
        \caption{Comparison between the values of |V| given by equations~(\ref{eq:Methods:analytical:rearranged_cont_V}) and~(\ref{eq:Methods:analytical:rearranged_eom_V}) as a function of $r/R$ near the stellar surface for a Jupiter mass planet in a circular orbit with period 4.23 days, for the case of frozen convection.  We note $V_{\text{cont}}$ and $V_{\text{eom}}$ the values of $V$ given by the continuity equation~(\ref{eq:Methods:analytical:rearranged_cont_V}) and the equation of motion (\ref{eq:Methods:analytical:rearranged_eom_V}), respectively.  The red dotted line is $|V_{\text{eom}}|$ and the green dashed line is $|V_{\text{cont}}|$ for the $1\,\text{M}_{\odot}$ case; the dark blue short dash--dot line is $|V_{\text{eom}}|$ and the cyan long dash--dot line is$|V_{\text{cont}}|$ for the $1.4 \, \text{M}_{\odot}$ case.  In both cases, the two expressions agree well, with some small discrepancy at points with significant curvature, due to the numerical evaluation of derivatives in the continuity equation.
        }
\label{fig:V_comparison}
\end{figure*}

As a check of internal consistency, the values of $V$ calculated from equations~(\ref{eq:Methods:analytical:rearranged_cont_V}) and~(\ref{eq:Methods:analytical:rearranged_eom_V}) {using the numerical solution of equations (\ref{eq:cont_osc})--(\ref{eq:mom_osc}) for a Jupiter mass planet with an orbital period of 4.23 days to determine the right hand sides} are compared in Figure~\ref{fig:V_comparison} in the region close to the surface, for both a $1$~$\text{M}_{\odot}$ and $1.4$~$\text{M}_{\odot}$ stars, for the case of frozen convection. Note that $V$ is not directly specified on the grid by the numerical solution. In both cases the expressions agree very well, with the greatest discrepancy arising at the points where the magnitude of the second derivative is large, due to the numerical evaluation of the derivative in equation~(\ref{eq:Methods:analytical:rearranged_cont_V}).

\subsubsection{An illustrative analytic estimate of $V$}\label{illustrativeest}

An analytical estimate for $V$ can be derived through the second law of thermodynamics given by equation~(\ref{secondthermo}) (see appendix \ref{sec:app:derivation}) and the continuity equation~(\ref{eq:Methods:analytical:rearranged_cont_V}). The former can be adapted to read:


\begin{equation}
\label{eq:app:eq:N20_surface2}
\Delta p - \frac{\Gamma_1 p_0}{\rho_0} \Delta \rho
=
-\frac{{\rm i}(\Gamma_3-1)}{m\omega} \bm{\nabla} \cdot {\bf F}' .
\end{equation}

\noindent We shall  assume hydrostatic equilibrium for the perturbations but non--adiabatic behaviour.
This is enough to indicate significant departures from the standard equilibrium tide.
 Thus, making use of
 equations~(\ref{eq:app:eq:rho}) and~(\ref {eq:app:eq:p1})  in appendix~\ref{sec:app:eq}, we obtain:
 
\begin{equation}
\label{eq:app:eq:N20_surface3}
 \left( \xi_r - \xi_{r,\text{eq}} \right)\left (   \frac{d p_{0}}{d r}  -\frac{\Gamma_1 p_0}{\rho_0}  \frac{d \rho_{0}}{d r}  \right)
=
-\frac{{\rm i}(\Gamma_3-1)}{m\omega} \bm{\nabla} \cdot {\bf F}' ,
\end{equation}

\noindent where $\xi_{r,\text{eq}} = -\Phi_{\text{P}}/g$ is the radial component of the Lagrangian displacement for the standard equilibrium tide (see appendix \ref{sec:app:eq}).
This implies deviation from the equilibrium tide when the non--adiabatic contributions on the right hand
side of equation~(\ref{eq:app:eq:N20_surface3}) become important. Then we find that, for small $\omega,$
$\xi_r $ scales as  $|1/\omega |.$
The horizontal component can be found from the continuity equation~(\ref{eq:Methods:analytical:rearranged_cont_V}) which may be rewritten as:

\begin{equation} \label{eq:Methods:analytical:cont_V_deltarho}
V = \frac{r}{l (l+1)} \left(  \frac{\rho'}{\rho_{0}} + \frac{\xi_r}{\rho_0} \frac{ d\rho_0}{ dr}  + \frac{\partial \xi_{r}}{\partial r}  +2 \frac{\xi_{r}}{r}  \right).
\end{equation}

\noindent { In the limit that  $\xi_r $ scales as  $|1/\omega |,$     $V$ also has that scaling and $\rho'$ may be neglected, assuming a perturbed flux given by the equilibrium tide. If the perturbed flux is not given by the equilibrium approximation, the frequency dependence is likely to be more complex.}


\noindent As we are interested in the value of $V$ at the surface of the star, we fix $r=R$, the radius of the star.
Under the assumption that $\xi_r$ varies on a scale comparable to the density scale height or larger {(due to non-adiabatic behaviour causing deviation from the equilibrium tide)},  we obtain the expression:
\begin{equation} \label{eq:Methods:analytical:V/r_surface}
\frac{V}{\xi_{r}} \approx \frac{R}{l(l+1)}\left. \frac{1}{\rho_0} \frac{\partial \rho_0}{\partial r} \right|_{r=R}  .
\end{equation}

\noindent In order to make estimates we evaluate the scale height at $r_{\text{na}}$, where here $r_{\text{na}}$ stands for  either $r_{\text{na,f}}$ or $r_{\text{na,p}}$ depending on whether convection is frozen or perturbed. 
  The scale height actually decreases towards the surface, although by less than an order of magnitude.
  Therefore, adopting (\ref{eq:Methods:analytical:V/r_surface})
with $H_{\rho},$ being evaluatedj
at $r=r_{\text{na}}$ is expected to lead to a lower bound for $|V|$, while giving an estimate of the right order of magnitude. 
{  Assuming $\xi_r$ remains the same  order of magnitude as the equilibrium tide radial displacement $\xi_{r,\text{eq}},$ 
 this gives the final relation between $V$ and $\xi_{r,\text{eq}}$ 
to the same level of accuracy as:
\begin{equation} \label{eq:Methods:analytical:V/r_surface_estimate}
\frac{V}{\xi_{r,\text{eq}}} \gtrsim \frac{R}{l (l+1) H_{\rho}|_{r_{\text{na}}}}.
\end{equation}

\noindent This suggests a large ratio $\sim 500$ which is approached in our calculations with frozen convection.

We remark that as an alternative to the above considerations 
we might have $\bm{\nabla} \cdot {\bf F'}=0$ in equation~(\ref{eq:app:eq:N20_surface3}).
However, that in general would imply  forms for $p'$ and $\rho'$
such that the equilibrium tide would not apply.
For further discussion of these aspects see appendix~\ref{sec:app:nheq}. }

\section{Results}
\label{sec:Results}

The response of a star to the perturbation of a Jupiter--mass planet on a 4.23 day orbit  is presented here, primarily for a $1$~$\text{M}_{\odot}$   star, but some results are also shown for the case of a star of mass $1.4$~$\text{M}_{\odot}$.  The results are focussed on the behaviour near the surface of the star, where non--adiabatic effects become important.
Throughout this section the perturbed quantities refer to their radial parts, that is, $T'$ stands for $T'(r)$.  As pointed out in Section~\ref{sec:Methods:set-up},  for any solved--for perturbed quantity, $q'(r)$, which is in general complex, we can form a real quantity that can be written as $q'(r, \theta, \phi, t) = \Re \left[  q'(r) Y_{2}^{-2}(\theta, \phi) \text{e}^{2 \text{i} \omega t}  \right].$ 

This section has been split into two parts: section\,\ref{sec:Results:Frozen} shows the calculation for frozen convection, i.e. under the assumption that it does not change from
its background value and section\,\ref{sec:Results:Gradient} shows the results for perturbed convection, i.e. it is assumed that the time scale is short enough for it 
to relax to a quasi--steady  state obtained from MLT. 
 Similar figures have been produced in each case, but the ranges displayed have been adjusted to make the results as clear as possible.

\subsection{Frozen convection}
\label{sec:Results:Frozen}

\begin{figure*}
        \includegraphics[width=\columnwidth]{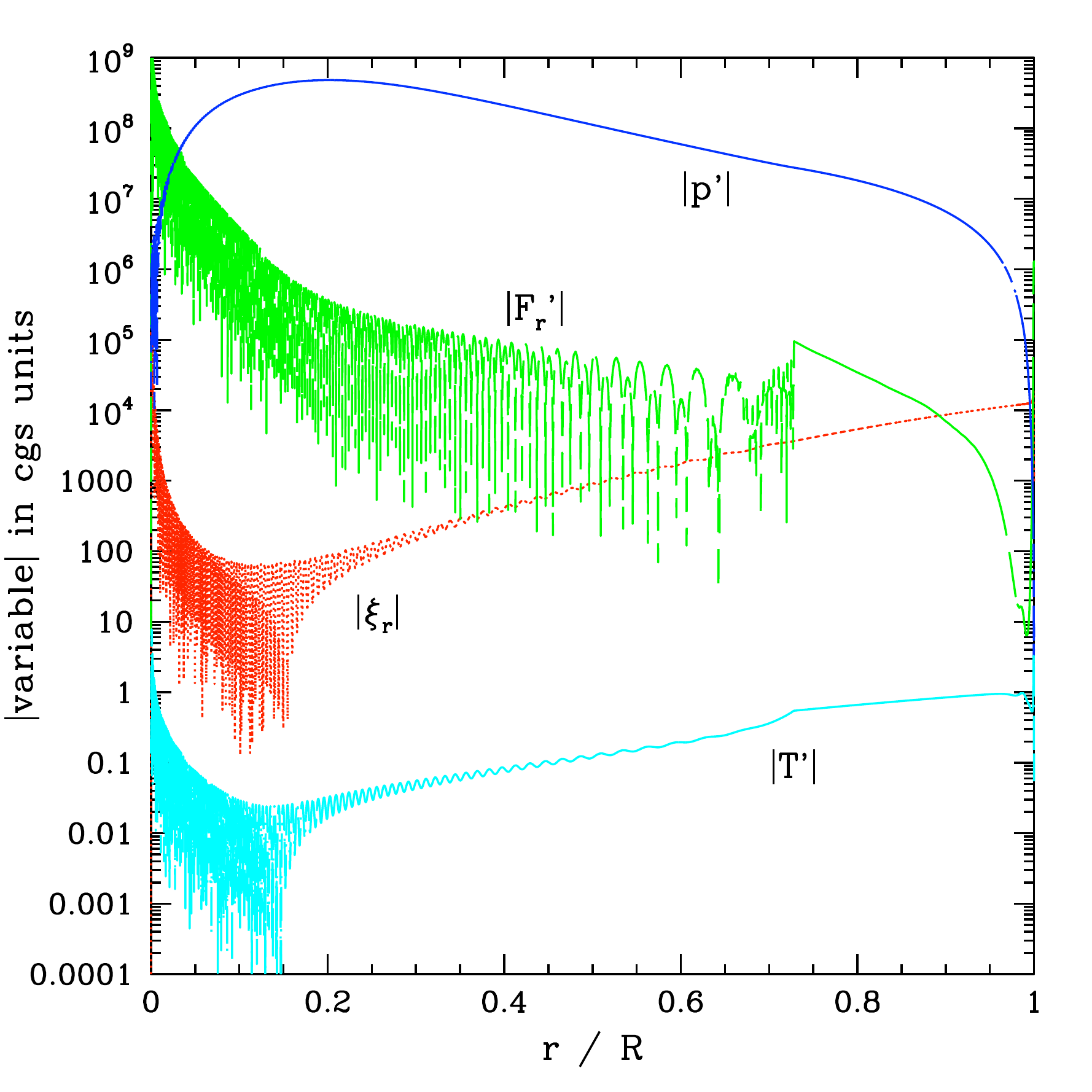}
    \caption{This shows, for the case of frozen convection, 
the magnitude of the four variables which are directly output from the code
 {\em versus} $r/R$, for the case $M = 1$~$\text{M}_{\odot}$: $\xi_{r}$,
 the radial displacement (red dotted line); $F_{r}'$, 
the perturbation to the radial radiative flux (green dashed line); $p'$,
 the perturbation to the pressure (dark blue short dash-dot line); and $T'$,
 the perturbation to the temperature (cyan long dash-dot line).
 The transition in behaviour at $r/R \sim 0.73$ is due to the onset of convection.
 Behaviour near the surface is examined more closely in Fig.~\ref{fig:log_mod_surface_variables}. Except for a thin surface layer, where non--adiabatic effects are important,
 the imaginary part is much smaller than
 the real part for all of these quantities. { Note that as the magnitude of quantities 
is shown on a logarithmic scale, unresolved minima may be smaller than indicated.}
}
    \label{fig:log_mod_variables}
\end{figure*}

\begin{figure*}
        \includegraphics[width=\columnwidth]{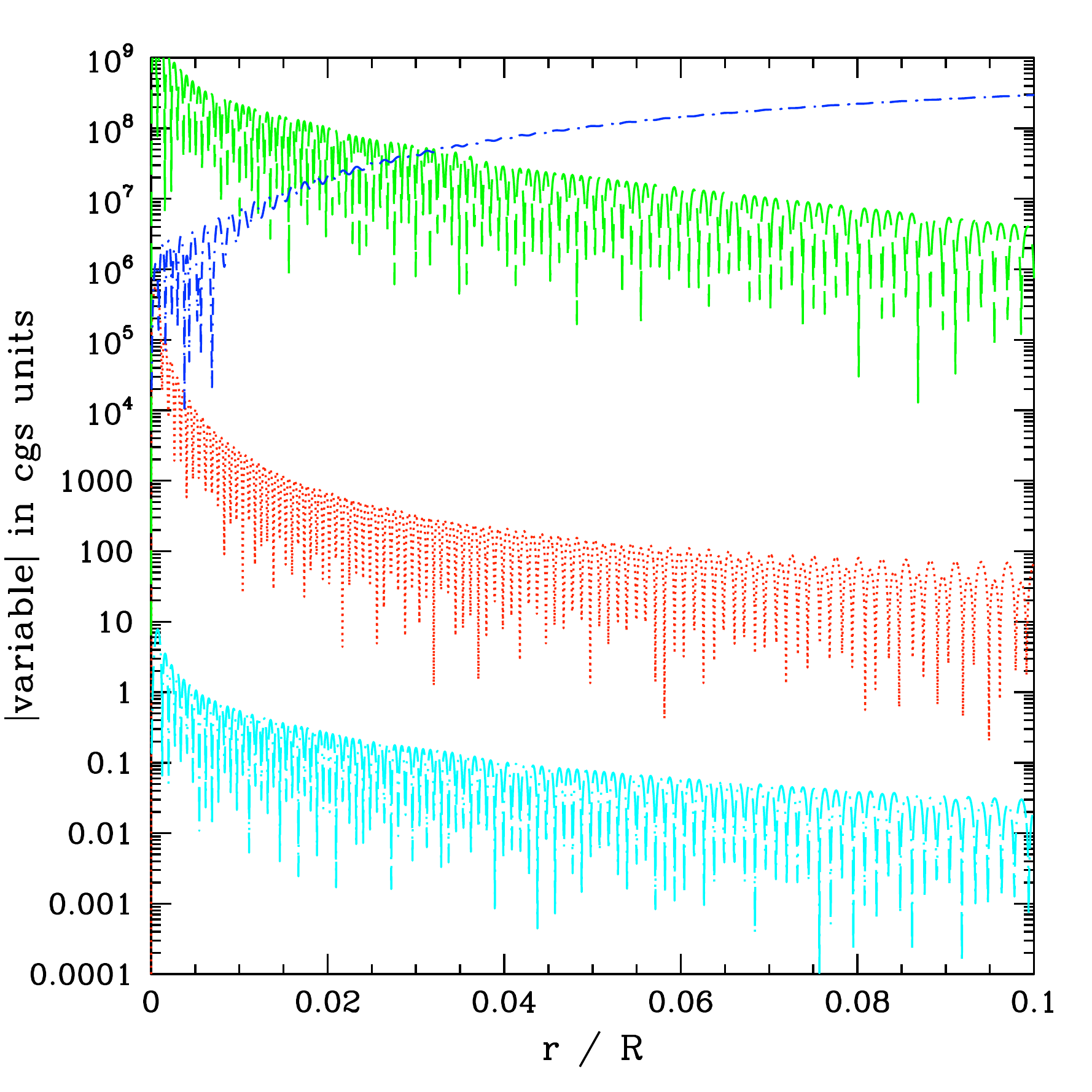}
    \caption{The quantities illustrated in Fig.~\ref{fig:log_mod_variables}
are plotted over a narrow radial range, $0 < r/R < 0.1$, for the 1~$\text{M}_{\odot}$ star, for the case of frozen convection,
in order to focus on the stellar core.
Shown are the magnitudes of $\xi_{r}$, the radial displacement
 (red dotted line); $F_{r}'$, the perturbation to the radial radiative flux (green dashed line);
$p'$, the perturbation to the pressure (dark blue short dash--dot line);
 and $T'$, the perturbation to the temperature (cyan long dash--dot line).
 The high--frequency spatial oscillations are resolved well even in the core.}
    \label{fig:log_mod_variables_core}
\end{figure*}

The magnitudes of the perturbed variables are shown for the full extent of the star in Figure~\ref{fig:log_mod_variables}. Within the body of the star, the response to the tidal potential agrees with previous work and oscillations are seen throughout the radiative core, becoming evanescent in the convection zone. Despite the high spatial frequency of the oscillations at the centre of the star, the peaks are still well resolved, as shown in Figure~\ref{fig:log_mod_variables_core}.

\begin{figure*}
        \includegraphics[width=\columnwidth]{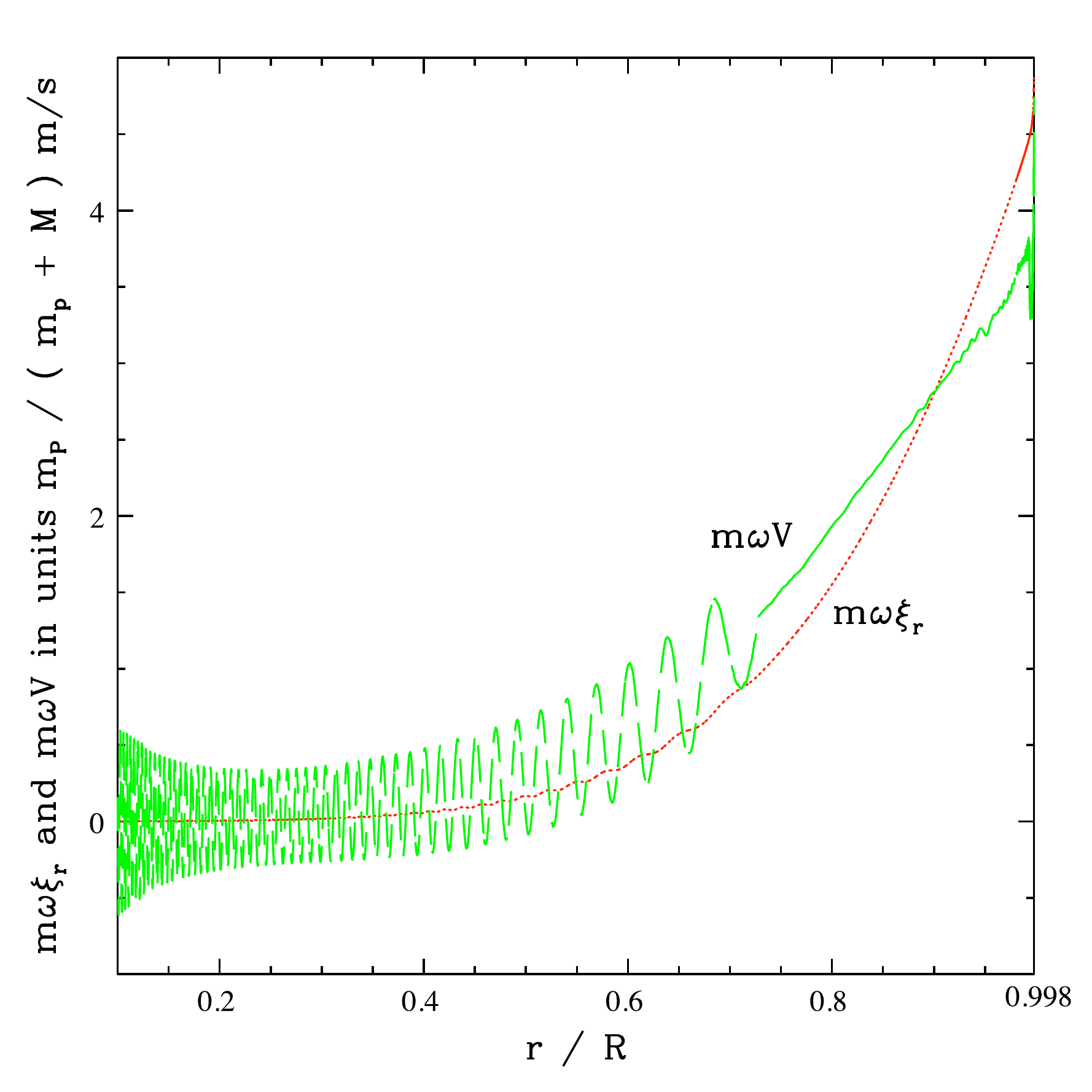}
    \caption{This figure shows, for the 1~$\text{M}_{\odot}$ case with frozen convection, the real parts of $m \omega \xi_{r}$ and $m \omega V$ in units of $[m_{p}/ (m_{p} + M)]$\,m\,s$^{-1}$ for $0.1 < r/R < 0.998$, thereby excluding the thin surface region where non--adiabatic effects are prominent. The red dotted line shows $m \omega \xi_{r}$, and the green dashed line shows $m \omega V$. The behaviour here is very similar to that shown in \citet{Terquem1998}, except for the fact that in our case $V$ continues to increase within the convection region, for $ r/R  > 0.73$.}
    \label{fig:Terquem_comparison}
\end{figure*}

Figure~\ref{fig:Terquem_comparison} shows the real parts of $m \omega \xi_{r}$ (radial perturbed velocity) and $m \omega V$, which were plotted by \citet{Terquem1998} and therefore allow a comparison to be made.  The radial and horizontal displacements are both of similar magnitude to the equilibrium radial displacement $\xi_{r, eq}$ away from the surface (which we expect, as $V_{\text{eq}} \approx \xi_{r, \text{eq}}$ outside the stellar core). The results shown in Figure~\ref{fig:Terquem_comparison} are in good agreement with  those of \citet{Terquem1998}, other than the fact that in our case $V$ continues to increase in the convection zone, more closely matching the equilibrium tide approximation (see appendix~\ref{sec:app:eq}).

\begin{figure*}
        \includegraphics[width=\columnwidth]{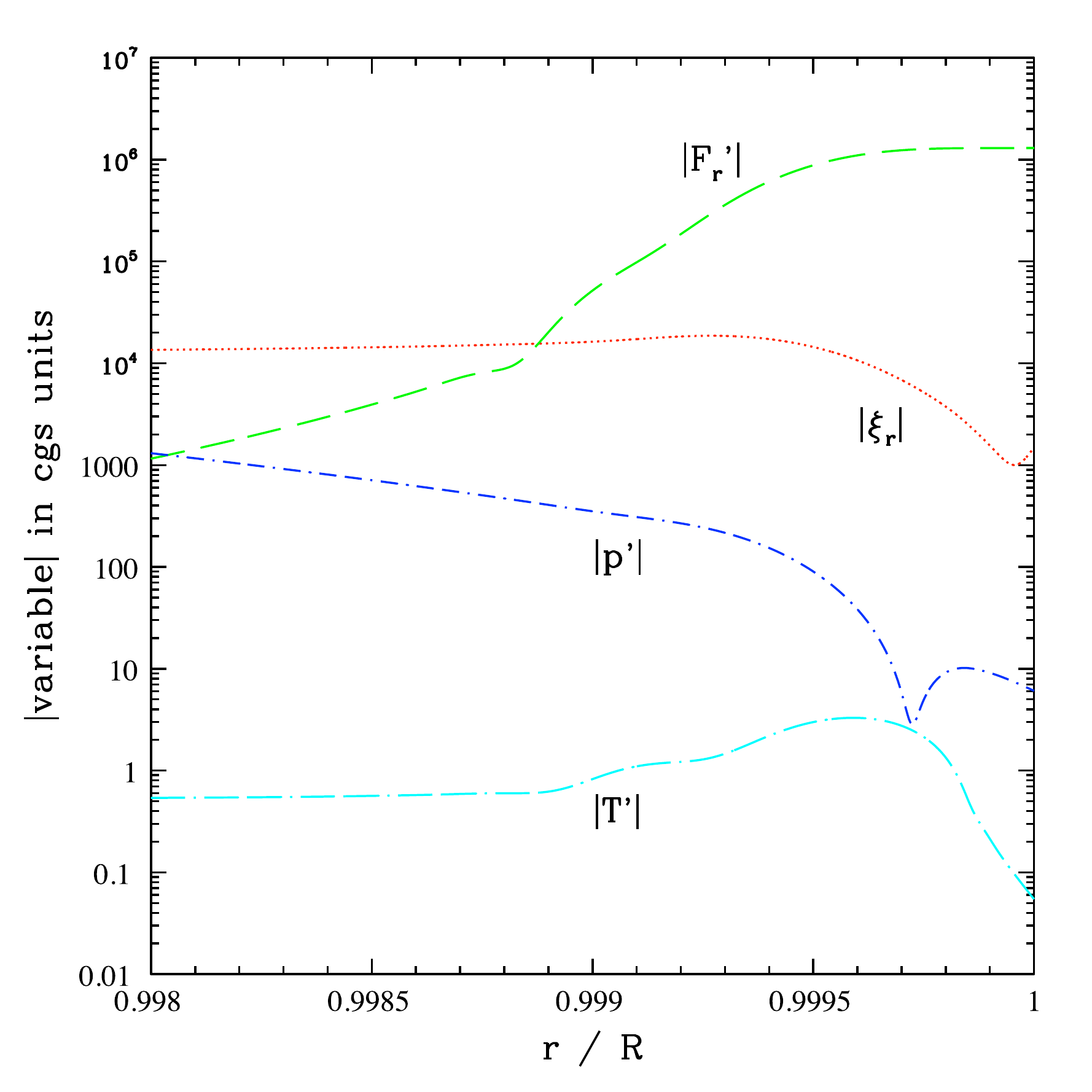}
    \caption{The quantities illustrated in Fig.~\ref{fig:log_mod_variables}
are plotted over a narrow radial range, $0.998 < r/R < 1$, for the 1~$\text{M}_{\odot}$ star, with frozen convection, in order to focus on the surface region.  Note the change in the range of the $y$--axis.
 Shown are the magnitudes of $\xi_{r}$,  the radial displacement
 (red dotted line); $F_{r}'$, the perturbation to the radial radiative flux (green dashed line);
$p'$, the perturbation to the pressure (dark blue short dash--dot line);
 and $T'$, the perturbation to the temperature (cyan long dash--dot line).
 The amplitude of the oscillations in the uppermost $0.2\%$ of the star changes rapidly on a small scale.
 In this region, the imaginary parts of the variables cannot be neglected.  This is consistent with the perturbation being non--adiabatic above $r_{\text{na,f}}=0.9995R$.}
    \label{fig:log_mod_surface_variables}
\end{figure*}

However, the surface behaviour is very different to both the equilibrium tide approximation and to the results of \citet{Terquem1998}, which did not fully incorporate non--adiabatic effects.  Within a thin region at the surface (of a similar width to the strongly superadiabatic convection region together with the overlying radiative zone near the surface of the star) 
the magnitude of the perturbations varies by orders of magnitude, as shown in Figure~\ref{fig:log_mod_surface_variables},
where the same quantities as illustrated in Figure~\ref{fig:log_mod_variables} are plotted over the radial range
$0.998 < r/R < 1$.   This is consistent with the estimated radius $r_{\text{na,f}}=0.9995R$ above which the perturbation is non--adiabatic.

\begin{figure*}
        \includegraphics[width=\columnwidth]{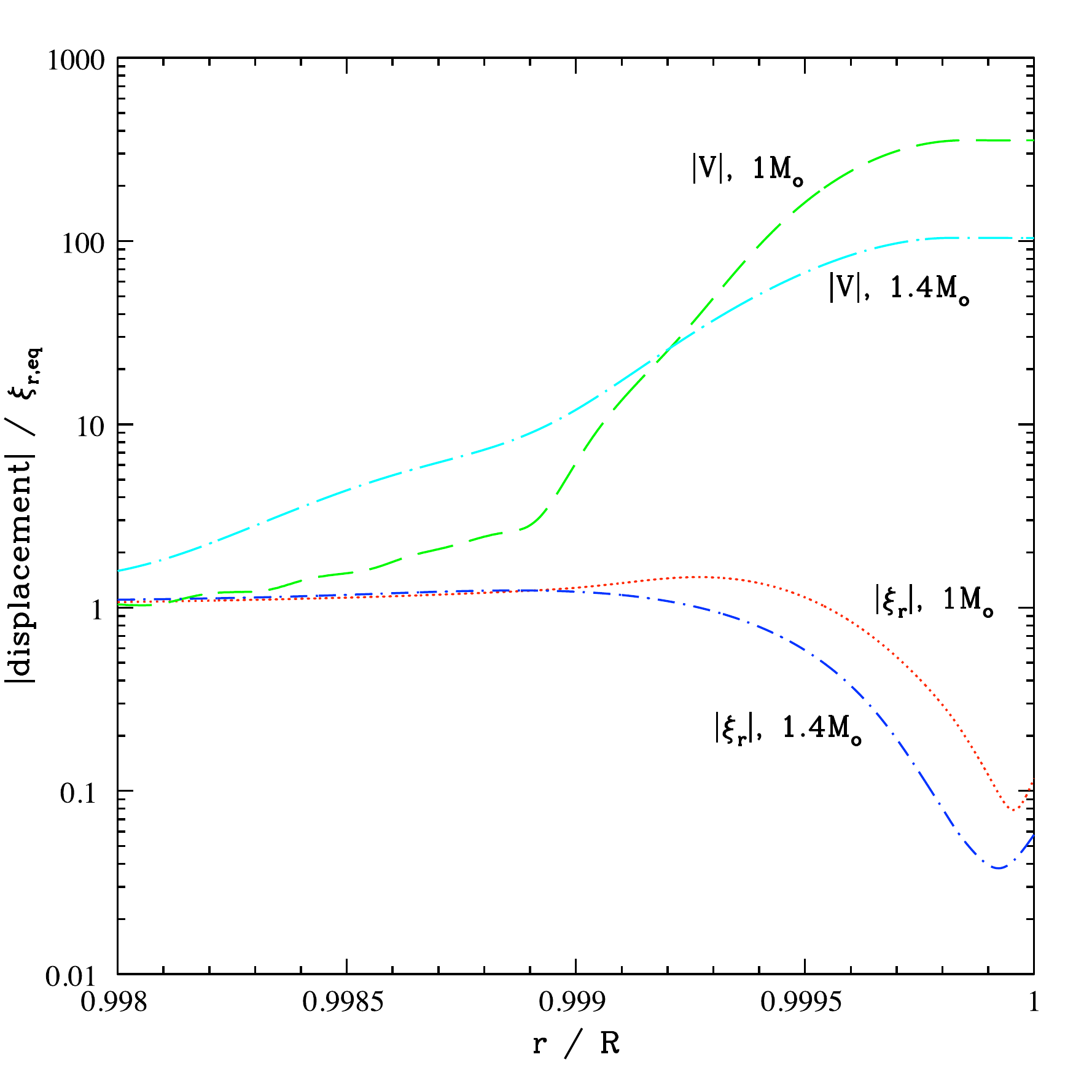}
    \caption{The radial and horizontal displacements in the case of frozen convection, scaled by the equilibrium tide radial displacement, $\xi_{r,\text{eq}},$
(given by  equation\,\ref{eq:app:eq:xi_r})
{\em versus} $r/R$ near the stellar surface.
 For the $1$~$\text{M}_{\odot}$ case, the red dotted line indicates $\xi_{r} /\xi_{r,\text{eq}}$ and the green dashed line  $V/\xi_{r,\text{eq}}$; for the $1.4$~$\text{M}_{\odot}$ case, the dark blue short dash--dot line is $\xi_{r}/\xi_{r,\text{eq}}$, and the cyan long dash--dot line is $V/\xi_{r,\text{eq}}$.
The radial displacement is suppressed compared to the equilibrium tide by a factor of $\sim 10$, whereas the horizontal displacement is greater than the
equilibrium tide by a factor of $\sim 100$. Whilst the exact amount of suppression or enhancement is different for the two stellar models, both exhibit the same qualitative behaviour.
This figure is consistent with departure from the equilibrium tide arising as a result of non--adiabaticity near the surface, although the plots indicate that non--adiabatic effects start to manifest themselves somewhat below $r_{\text{na,f}}$, which is $0.9995R$ for  the $1$~$\text{M}_{\odot}$ star and $0.9991R$ for  the $1.4$~$\text{M}_{\odot}$ star.
}
    \label{fig:surface_displacements}
\end{figure*}

In Figure~\ref{fig:surface_displacements}, we highlight the behaviour of the radial and horizontal displacements in the surface region in units of the radial displacement predicted by the equilibrium tide approximation, $\xi_{r,\text{eq}}$, for both the 1~$\text{M}_{\odot}$ and the 1.4~$\text{M}_{\odot}$ star (more details about the equilibrium tide can be found in appendix~\ref{sec:app:eq}).  At the surface, the radial displacement is suppressed by a factor of $\sim 10$ relative to the adiabatic case, and the horizontal displacement is amplified by a factor of $\sim 100,$ which is similar to the estimates provided in Section \ref{illustrativeest} and appendix \ref{sec:app:nheq}.  In order to investigate the effect of changing the surface boundary condition, we have rerun both the 1~$M_{\odot}$ case and the 1.4~$M_{\odot}$ case, replacing the condition $\Delta p = 0$ by the boundary condition given by \citet{Pfahl2008}. Results are very similar, apart from $|\xi_{r}|$ attaining even smaller values at the surface.  In the non--adiabatic zone, the horizontal displacement is therefore larger than the radial displacement by a factor of $\sim 1000$ or more.   The results displayed in Figure~\ref{fig:surface_displacements} are consistent with departure from the equilibrium tide arising as a result of non--adiabaticity above $\sim r_{\text{na,f}}$.

\begin{figure*}
        \includegraphics[width=\columnwidth]{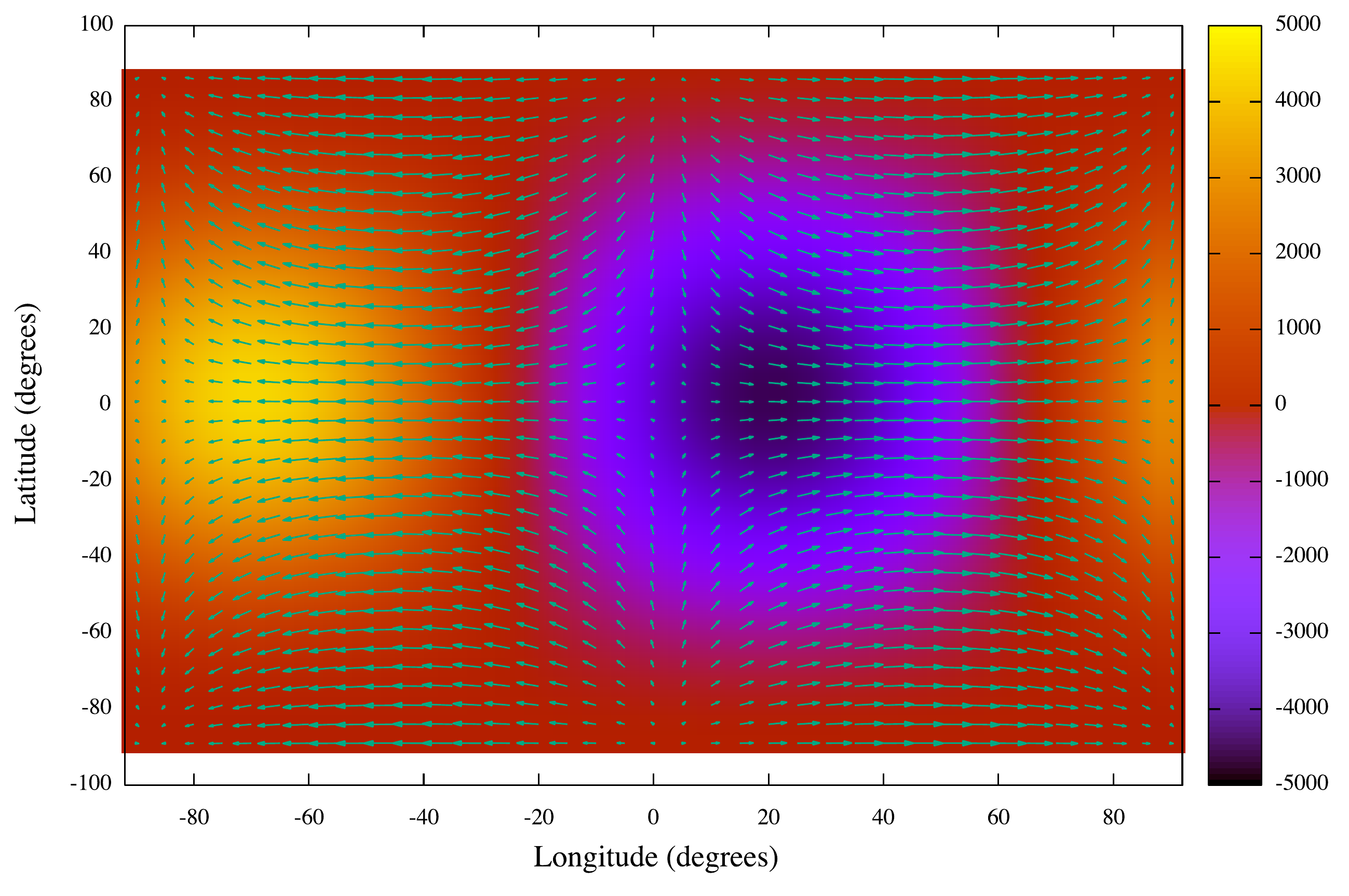}
    \caption{This plot illustrates the displacement vector at the surface of the star as a function of latitude and longitude for the $1 \text{M}_{\odot}$ star under the assumption of frozen convection. The vectors represent the horizontal components of the displacement, and the colour denotes the radial component of the displacement.
 The longitude range is $-90^{\circ}$ to $90^{\circ}$, and the sub--planetary
 point is at $(0,0)$, so the plot shows the visible disc of the star, as seen from the planet. Due to the non--adiabatic conditions at the surface, the tidal potential, radial displacement and horizontal displacement are all out of phase with each other. }
    \label{fig:colour_vector_180}
\end{figure*}

To illustrate the relationship between the radial and horizontal displacements, taking the angular dependence into account, Figure~\ref{fig:colour_vector_180} displays the horizontal displacements as vectors, and the radial displacement shown through the colour.  The plot displays the surface of one side of the star, covering latitudes from $-90^{\circ}$ to $90^{\circ}$, and longitudes from $-90^{\circ}$ to $90^{\circ}$ with $(0, 0)$ as the sub--planetary point.

\begin{figure*}
        \includegraphics[width=\columnwidth]{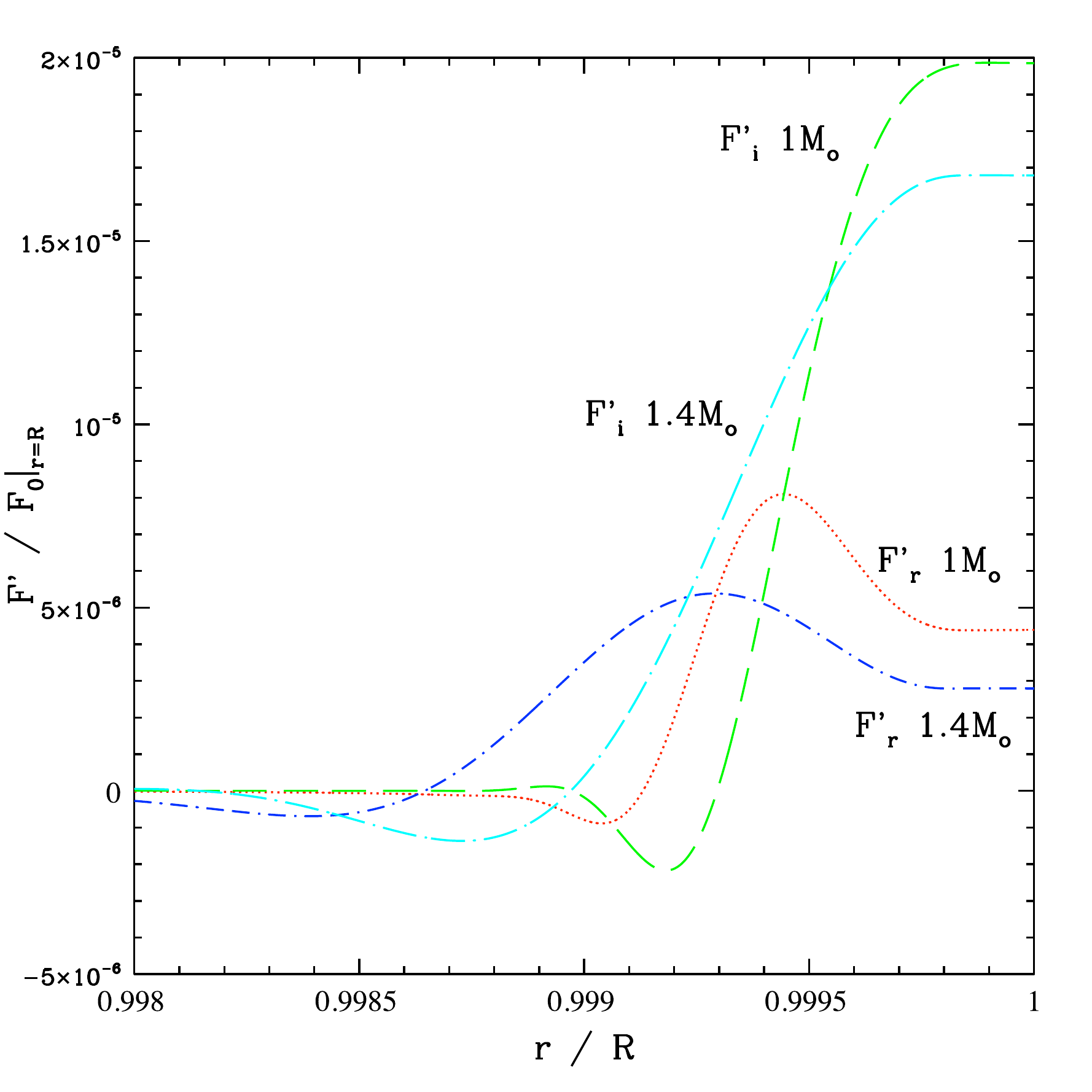}
    \caption{The perturbation to the radiative flux in the surface region,
in units of the equilibrium flux at the surface,
 showing both the real and imaginary parts for both the $1$~$\text{M}_{\odot}$ and the $1.4$~$\text{M}_{\odot}$ stars, with frozen convection.
For the $1$~$\text{M}_{\odot}$ case, the red dotted line shows the real part, and the green dashed line the imaginary part.
 For the $1.4$~$\text{M}_{\odot}$ case,   the dark blue short dash--dot line shows the real part, and the cyan long dash--dot line shows the imaginary part.  The imaginary parts grow rapidly in this surface layer, and dominate the perturbation to the flux at the surface, such that the peak flux will lag the planet by $\sim 40^{\circ}$.
 The details of this behaviour differ for the two stellar masses, but both show the {same qualitative behaviour and demonstrate the} importance of non-adiabatic effects.}
    \label{fig:surface_flux_perturbation}
\end{figure*}

The perturbation to the flux in the surface region is shown in Figure~\ref{fig:surface_flux_perturbation} 
for both the $1$~$\text{M}_{\odot}$ and the $1.4$~$\text{M}_{\odot}$ cases.
 Throughout this region, the perturbation to the flux grows rapidly, and the phase of 
the perturbation changes on a small scale,
 highlighting the non--adiabatic nature of this response. 
 At the surface, the imaginary component is dominant, but the real part remains non--negligible, resulting in a phase lag behind the planet of $\sim 40^{\circ}$.

\subsection{Perturbed convection}
\label{sec:Results:Gradient}

We present below the results corresponding to the case when the convective flux is perturbed, for both  approaches~A and~B described in Section~\ref{sec:Methods:convection}.  

\subsubsection{Perturbed convection following approach~A}

\begin{figure*}
        \includegraphics[width=\columnwidth]{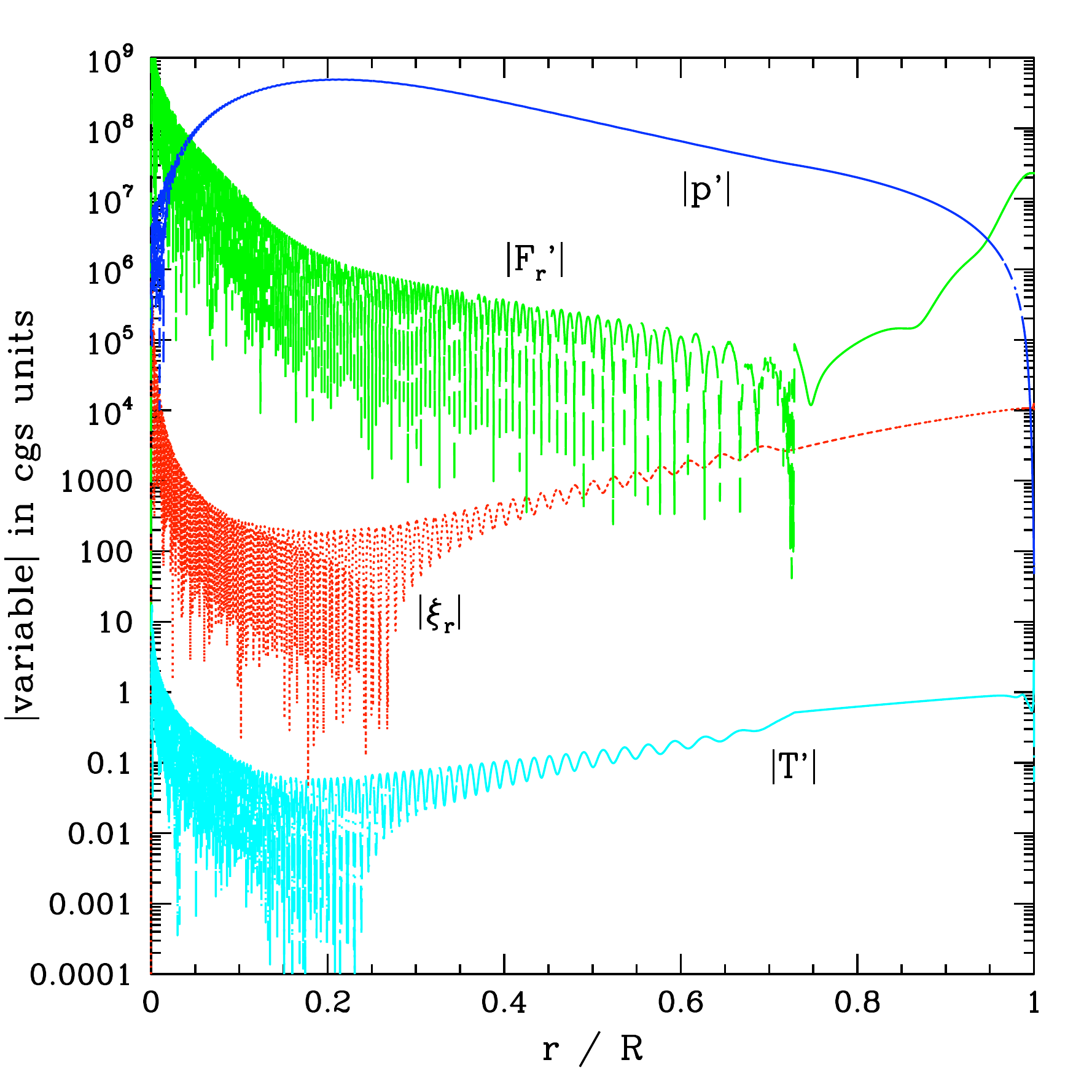}
        \caption{This shows, for the case of perturbed convection { following approach~A,} for a star with $M = 1$~$\text{M}_{\odot}$, the magnitude of the four variables which are directly output from the code {\em versus} $r/R$: $\xi_{r}$, the radial displacement (red dotted line); $F_{r}'$, the perturbation to the radial energy flux (green dashed line); $p'$, the perturbation to the pressure (dark blue short dash-dot line); and $T'$, the perturbation to the temperature (cyan long dash-dot line). The transition in behaviour at $r/R \sim 0.73$ is due to the onset of convection.  Behaviour near the surface is examined more closely in Fig.~\ref{fig:log_mod_variables_surface_BPT}.}
    \label{fig:log_mod_variables_BPT}
\end{figure*}

The response throughout the whole star for the $1$~$\text{M}_{\odot}$ star, in the case of perturbed convection following approach~A, is shown in Figure~\ref{fig:log_mod_variables_BPT}. Within the radiative core, the behaviour is oscillatory, transitioning to evanescent behaviour in the convection zone.  Despite the high spatial frequency towards the centre of the star,
 { just as in the case of frozen convection described above,
 the oscillations remain well resolved.}

\begin{figure*}
        \includegraphics[width=\columnwidth]{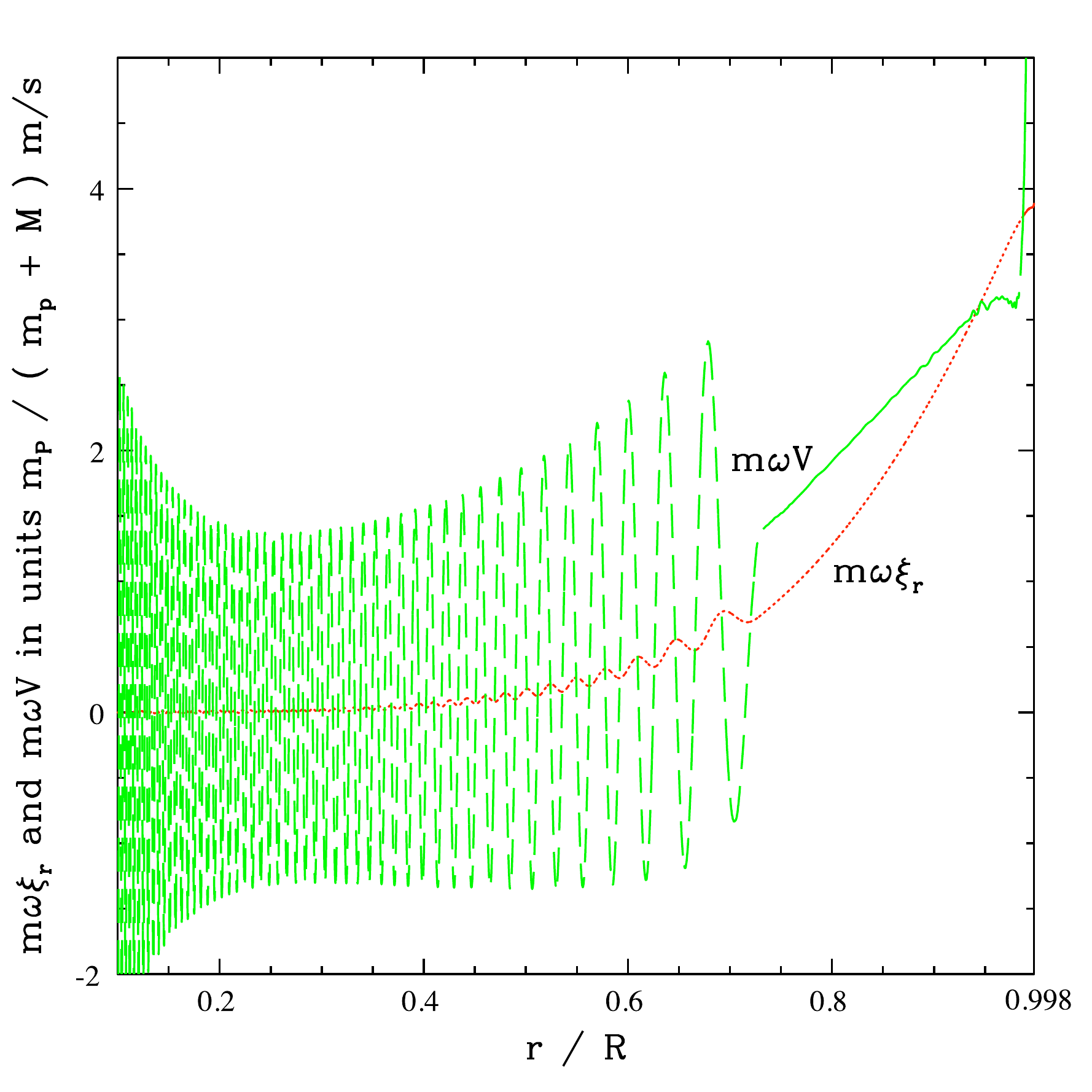}
        \caption{This figure shows, for the case of perturbed convection { following approach~A}, for a star with mass $M = 1$~$\text{M}_{\odot}$, the real parts of $m \omega \xi_{r}$ and $m \omega V$ in units of $[m_{p}/ (m_{p} + M)]$\,m\,s$^{-1}$ for $0.1 < r/R < 0.998$, thereby excluding the thin surface region where non--adiabatic effects are most prominent. The red dotted line shows $m \omega \xi_{r}$, and the green dashed line shows $m \omega V$. The behaviour here is similar to that shown in \citet{Terquem1998}, except for the fact that in our case $V$ continues to increase within the convection region, for $ r/R > 0.73$.}
    \label{fig:Terquem_comparison_BPT}
\end{figure*}

The radial and horizontal displacements are displayed in Figure~\ref{fig:Terquem_comparison_BPT},
  which can be compared  with the model shown in \citet{Terquem1998}.
The horizontal displacement has greater amplitude oscillations within the core, 
but both the radial and horizontal displacements are centred on the equilibrium tide, 
except at the very surface of the star.

\begin{figure*}
        \includegraphics[width=\columnwidth]{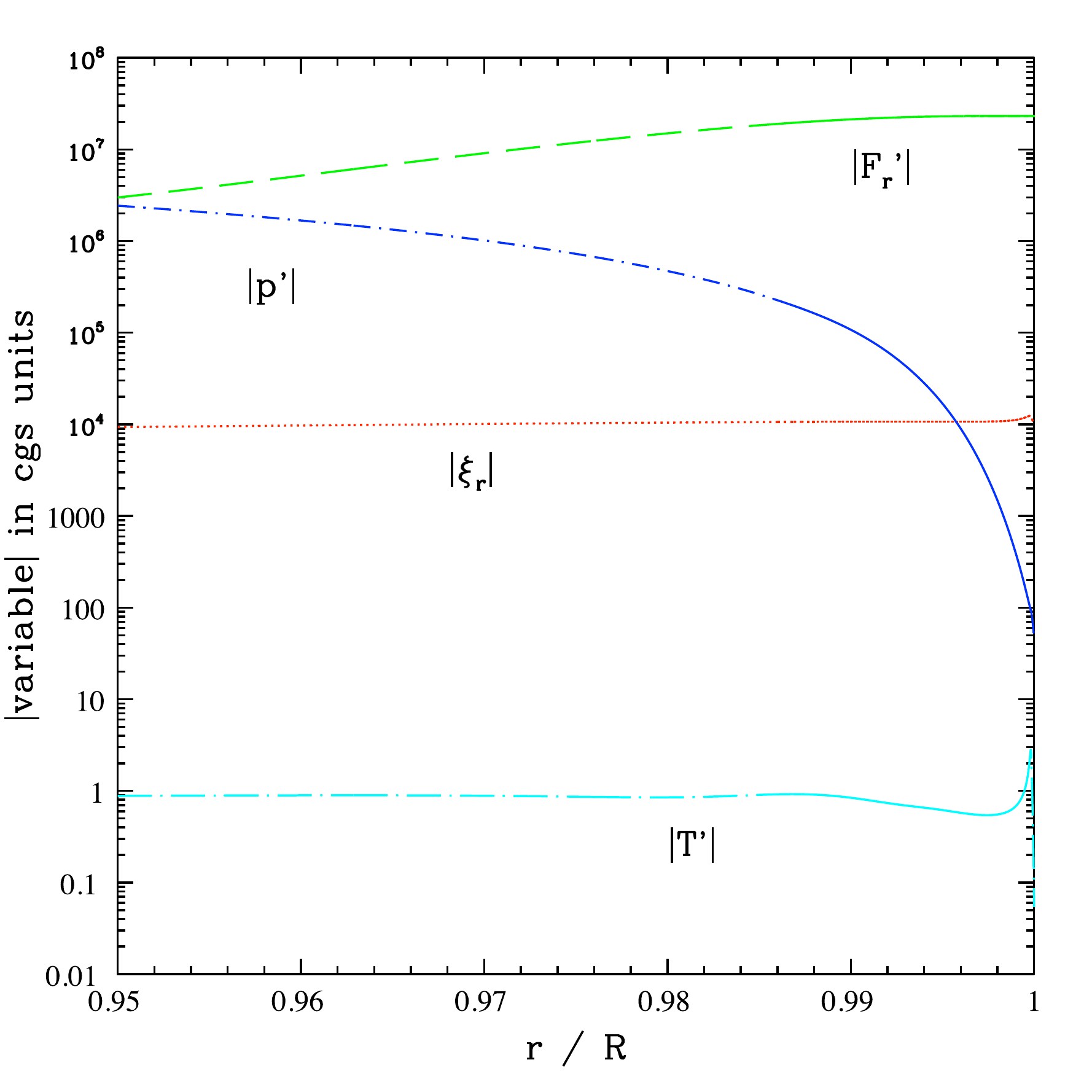}
    \caption{The quantities illustrated in Fig.~\ref{fig:log_mod_variables_BPT}
are plotted over a narrow radial range, $0.998 < r/R < 1$, 
for the 1~$\text{M}_{\odot}$ star with perturbed convection { following approach~A},
in order to focus on the surface region.  Note the change in the range of the $y$--axis.
 Shown are the magnitudes of $\xi_{r}$, the radial displacement
 (red dotted line); $F_{r}'$, the perturbation to the radial energy flux (green dashed line);
$p'$, the perturbation to the pressure (dark blue short dash--dot line);
 and $T'$, the perturbation to the temperature (cyan long dash--dot line).}
    \label{fig:log_mod_variables_surface_BPT}
  \end{figure*}

To highlight the surface behaviour, the variables which are directly solved for are shown over a narrow radial range ($0.95<r/R<1$) at the surface in Figure~\ref{fig:log_mod_variables_surface_BPT}. 
Over this range,  the response is fairly smooth, apart from
the small shifts in $\xi_{r}$ and $T'$ in the region of the radiative skin.
{ In particular, $F_r'$ is approximately constant, which is expected from the
  condition that $\bm{\nabla} \cdot {\bf F} \sim 0$ in the upper non--adiabatic layers.}

\begin{figure*}
        \includegraphics[width=\columnwidth]{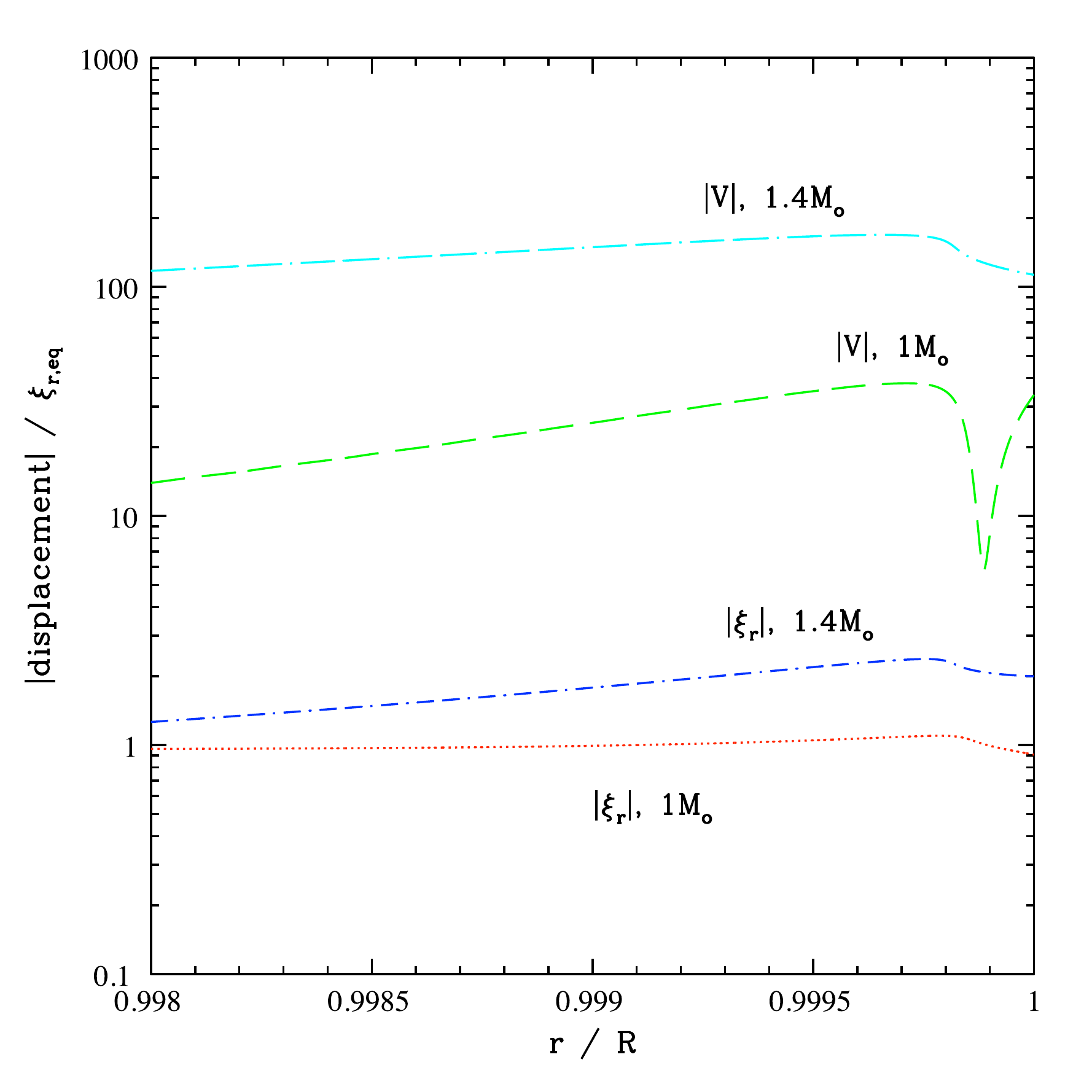}
    \caption{The radial and horizontal displacements, scaled by the equilibrium tide radial displacement $\xi_{r,\text{eq}},$
(given by  eq.~[\ref{eq:app:eq:xi_r}])
{\em versus} $r/R$ near the stellar surface, assuming perturbed convection { following approach~A}.
 For the case of $M = 1$~$\text{M}_{\odot}$, the red dotted line indicates $\xi_{r} /\xi_{r,\text{eq}}$ and the green dashed line  $V/\xi_{r,\text{eq}}$; for the $1.4$~$\text{M}_{\odot}$ case, the dark blue short dash--dot line is $\xi_{r}/\xi_{r,\text{eq}}$, and the cyan long dash--dot line is $V/\xi_{r,\text{eq}}$.
The radial displacements remain close to the equilibrium tide values, whereas the horizontal displacements are greater than the
equilibrium tide values by a factor of $\sim 30 - 100$. Whilst the exact behaviour is different for the two stellar models, both exhibit the same qualitative behaviour.}
    \label{fig:surface_displacements_BPT}
\end{figure*}

The radial and horizontal displacements in the surface region are 
shown in Figure~\ref{fig:surface_displacements_BPT} for a narrow radial range
 ($0.998< r/R <1$), for both the $1$ and $1.4$ solar mass stars.
 The radial displacements remain close to the equilibrium tide values, with small dips at the very surface. These features in $\xi_{r}$ at the surface correspond to much larger changes  in $V$.

Compared to the equilibrium tide, $\xi_{r}$ at the surface is changed by $\sim 10\%$ for the 1~$\text{M}_{\odot}$ star, and by a factor of 2 for the 1.4~$\text{M}_{\odot}$ star. The horizontal displacement at the surface is a factor of $\sim 30$ greater than the equilibrium tide value for the 1~$\text{M}_{\odot}$ star, and a factor of $\sim 100$ greater for the 1.4~$\text{M}_{\odot}$ star. { These values are smaller than in the case of frozen convection.}

\begin{figure*}
        \includegraphics[width=\columnwidth]{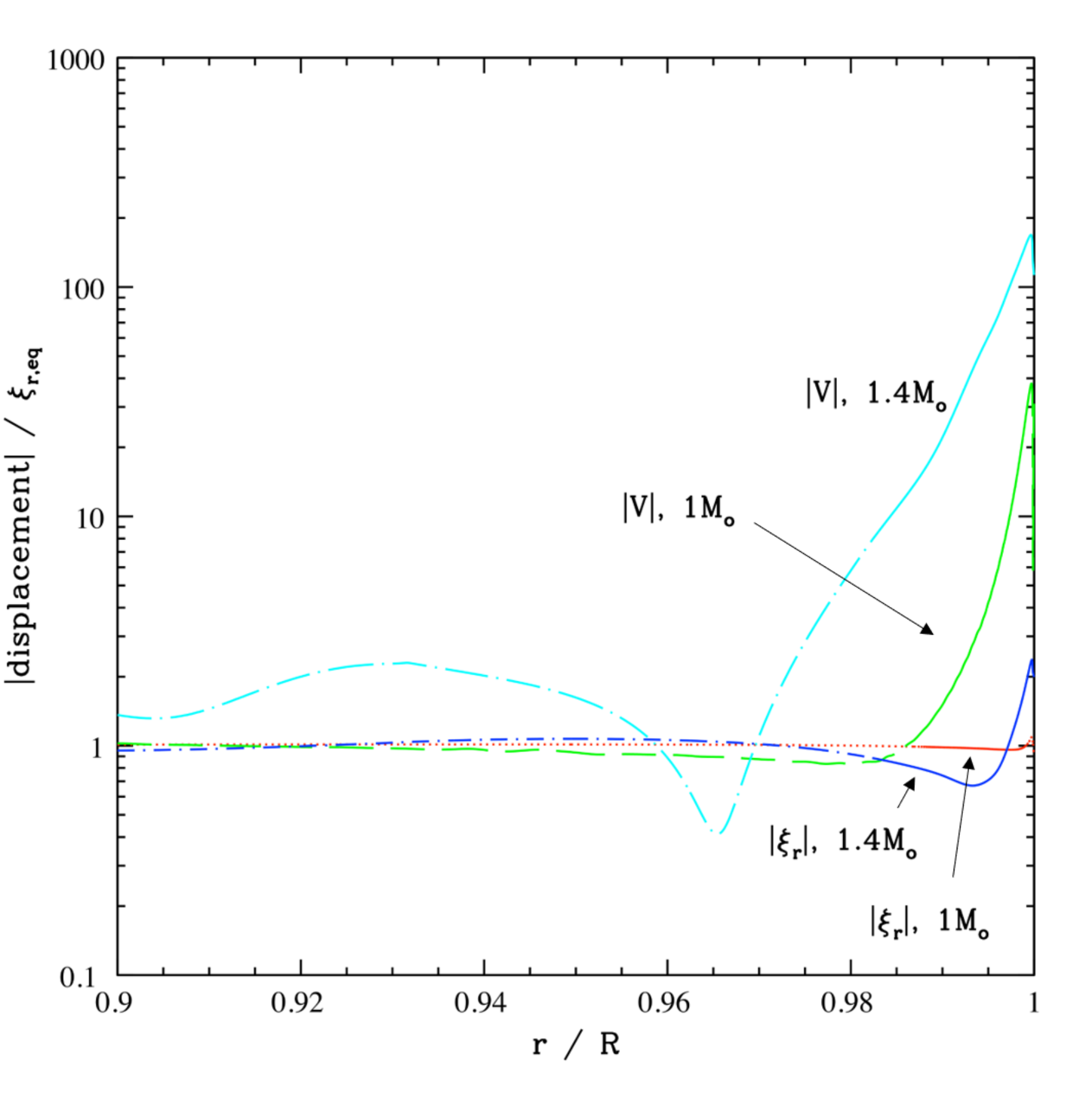}
    \caption{Same as Figure~\ref{fig:surface_displacements_BPT} but over a greater radial extent of the star, showing the departure from the equilibrium tide value. 
For both stellar masses, the horizontal displacement departs from the equilibrium tide value
 deeper than the radial displacement.
{Departure from equilibrium tide for $V$ is due to non--adiabaticity above $r\sim 0.99R$ for the $1~\text{M}_{\odot}$ star and $0.94R$ for the 1.4~$\text{M}_{\odot}$ star.   In the higher mass case, there may be an additional contribution from below $r_{\text{na,p}}$,  where the perturbation is adiabatic, due to $m^2\omega^2$ being,  although small, larger than $|N^2|$  (see appendix \ref{AppC1}).}
}
    \label{fig:surface_displacements_BPT_wide}
\end{figure*}

The location at which $V$ deviates from the equilibrium tide differs in the two mass cases, 
with the
$1.4~\text{M}_{\odot}$ star showing deviations deeper in the convection zone than the $1~\text{M}_{\odot}$ star, as seen in Figure~\ref{fig:surface_displacements_BPT_wide}.    For the $1~\text{M}_{\odot}$ star, these deviations originate {from around $ r= 0.99 R$ which is intermediate between  $r_{\text{na,p}}=0.97R$
and $r_{\text{na,p1}}=0.998R$}
 and therefore are most likely due to non--adiabaticity {( see discussion in Section \ref{Pertcon})} .  For the 1.4~$\text{M}_{\odot}$ star, $r_{\text{na,p}}=0.94R$, and the inner edge of the convection zone is at $0.93R$.  There is therefore only a narrow layer at the the bottom of the convection zone where the perturbation is adiabatic {if $r_{\text{na,p}}$
 defines the transition. However,  for $0.9R < r < 0.94R$ there is a region where $|N^2|$ is small and comparable to $m^2\omega^2$
 (as shown in Fig. \ref{fig:N2_surface}).
 In appendix \ref{AppC1}  it is argued that the standard equilibrium tide does not apply in this  limit  although it can be argued 
 from~\ref{eq:Methods:analytical:rearranged_eom1_V7} that when both these quantities are zero,
  fractional deviations of $\xi_{r}$ from the equilibrium tide are of order $H_{p} / r$, which are relatively small here. Note also that if $m^{2} \omega^{2} \ne 0$, but $N^{2} = 0$, there are also fractional corrections of order $m^{2} \omega^{2} r / g$, which is also small.}
{It may be that the departure from equilibrium tide observed for the 1.4~$\text{M}_{\odot}$ star is  due to a combination of these effects and non--adiabatic
effects.}


\begin{figure*}
        \includegraphics[width=\columnwidth]{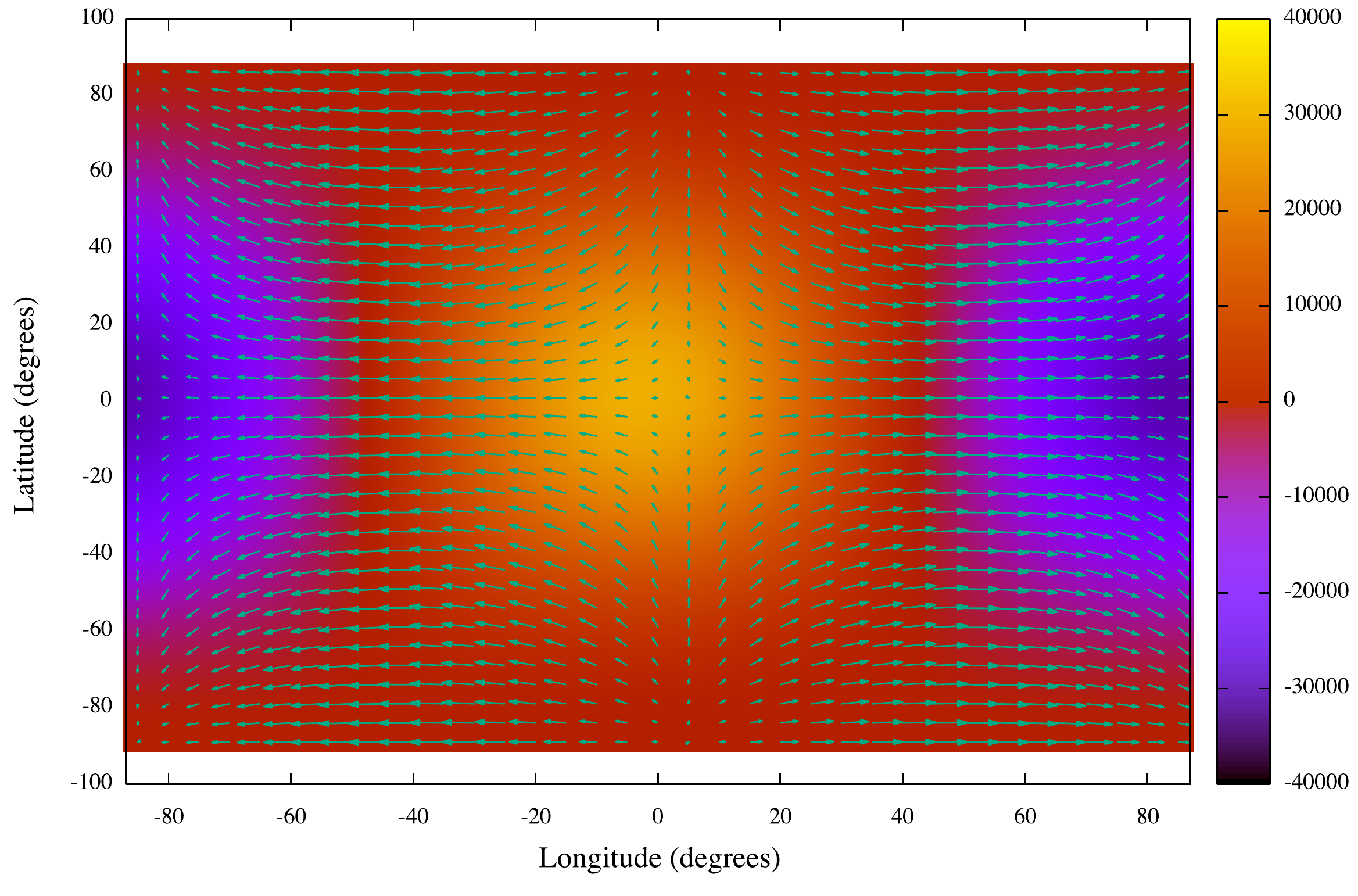}
    \caption{This plot illustrates the displacement vector at the surface of the star as a function of latitude and longitude for the $1 \text{M}_{\odot}$ star, for the case of perturbed convection { following  approach~A.}  The vectors represent the horizontal components of the displacement, and the colour denotes the radial component of the displacement.
 The longitude range is $-90^{\circ}$ to $90^{\circ}$, and the sub--planetary
 point is at $(0,0)$, so the plot shows the visible disc of the star, as seen from the planet. As the real parts of the displacements dominate at the surface, the radial and horizontal displacements are in phase with each other, and the planet.
}
    \label{fig:colour_vector_180_BPT}
\end{figure*}

The phase relations of the  displacement components are indicated in 
Figure~\ref{fig:colour_vector_180_BPT} for the $1~\text{M}_{\odot}$ case,
showing that the phase difference between  both the radial and horizontal components
with respect  to the planet's orbit is small.

\begin{figure*}
        \includegraphics[width=\columnwidth]{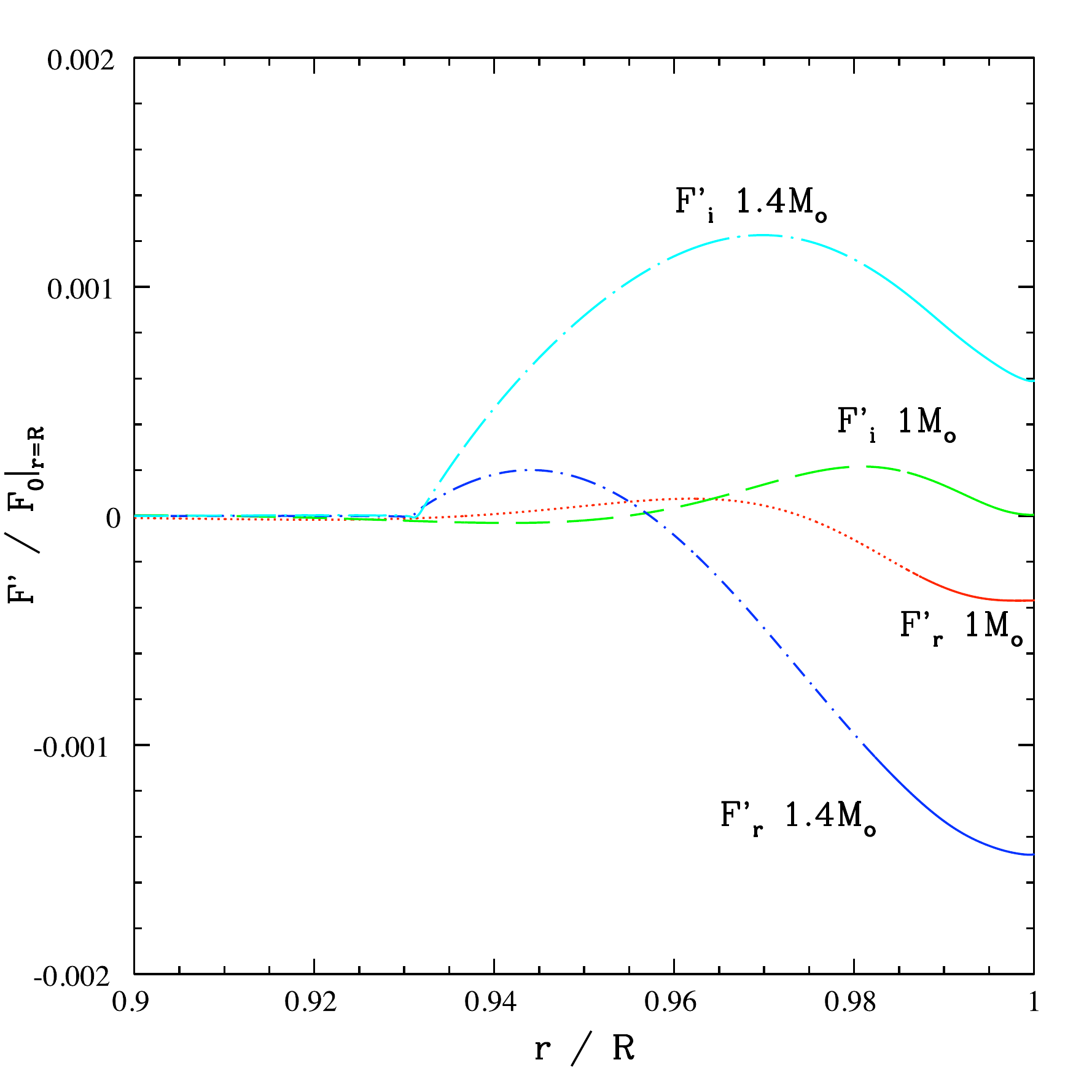}
    \caption{The perturbation to the radial component of the energy flux in the outer region of
the star, in units of the equilibrium flux at the surface,
showing both the real and imaginary parts for both the
$1$~$\text{M}_{\odot}$ and the $1.4$~$\text{M}_{\odot}$ stars,
with perturbed convection { following approach~A.}
For the $1$~$\text{M}_{\odot}$ case, the red dotted line shows the
real part, and the green dashed line the imaginary part:
significant perturbations occur well before the radiative skin of the star
reaching a roughly constant value around $r/R \sim 0.99$.
For the $1.4$~$\text{M}_{\odot}$ case,
the dark blue short dash--dot line shows the real part, and the cyan long dash--dot
line shows the imaginary part.
The details of this behaviour differ for the two stellar masses,
 but both demonstrate the importance of non--adiabatic effects {through the behaviour of the complex phase}.}
    \label{fig:surface_flux_perturbation_BPT}
\end{figure*}

The perturbation to the radial component of the energy flux is displayed in Figure~\ref{fig:surface_flux_perturbation_BPT}, 
showing the growth in $F'$ towards the surface  for both stellar masses. In both cases, the imaginary components are of similar size to the real components and cannot be neglected, which shows that non--adiabatic effects are important in this region  and result in the phase changing rapidly. 
{  In both cases, the plots are consistent with significant perturbations arising in the non--adiabatic regions above
  $r_{\text{na,p}}$}.

{ Whether convection is perturbed using approach~A or~B, it is found that the radial convective flux perturbation attains an almost constant  magnitude
  in the outer parts of the non--adiabatic zone where the perturbed radiative flux is still negligible.
  The magnitude of this constant  can be estimated by assuming that it can be determined from the perturbed convective flux for $r$ below but not far from $r_{\text{na,p}}$,  where it can be assumed that the equilibrium tide applies.  {To illustrate this we compare the perturbed flux in the convective zone to an estimate for the perturbation to the convective flux which would arise in the case that the behaviour is non-adiabatic, and that the radial displacement is given by the equilibrium tide. We evaluate the perturbation to the flux using approach~A, in the case that $\Delta s = 0$ and $\xi_{r} = \xi_{r,\text{eq}}$. Combining this with equation~\ref{eq:F_prime_conv_radial} gives $F'_{\text{c,eq}} = - A \frac{\partial}{\partial r} \left( - \xi_{r,\text{eq}} \frac{\partial s_{0}}{\partial r} \right)$.}
  
 {These are shown in Figure~\ref{fig:Feqcomparison}, where we plot} both the ratio of the magnitude of the radial component of the energy flux perturbation to the surface background value, $|F'/F_0 |_{r=R}|,$ evaluated using approach~A for the perturbed convective flux, and the ratio of the magnitude of the perturbed convective flux, evaluated assuming the equilibrium tide, to the background surface value of  the flux, $|F'_{\text{c,eq}} /F_{0}|_{r=R}|,$ for the 1~M$_{\sun}$ star. The left hand panel shows the interval $ 0.75 < r/R < 0.9998.$ A plot comparing these quantities in the non--adiabatic region very close to the surface is shown in the right hand panel.
  We expect the radiative flux to become significant at the radius  $r_{\text{na,f}}$ calculated assuming frozen convection, which is $0.9995R$ for the 1~M$_{\sun}$ star   (below $r_{\text{na,f}}$, convection is the dominant mode of transport of energy, whether convection is frozen or perturbed).  Therefore, the perturbed radiative flux is negligible throughout the range plotted (except very close to the outer edge), so that  $F' \simeq F'_{\text{c}}$. As can be seen from the left hand panel, the perturbed convective flux constructed from the equilibrium tide 
 does indeed track  the perturbed total flux throughout the range plotted.  
 We do not expect the match to be perfect, even below $r/R=0.97$ where the perturbations are adiabatic, because in this region $|N^2| < m^2 \omega^2$ and therefore there is departure from the  equilibrium tide (see appendix~\ref{AppC1}).
 Above $r/R \sim 0.98$ or so, we have  $|N^2| \gg m^2 \omega^2$ and therefore the convective timescale, over which flux perturbations are smoothed out, is much shorter than the period of the oscillations.   In this regime, the perturbed convective flux is approximately constant, as seen in the right hand panel.  Although this flux is not strictly constant all the way down to the base of the  non--adiabatic region 
 , we see from the the right hand panel that the variations are within a factor 2 only.   Just below $r_{\text{na,p}}$, the perturbations are adiabatic and therefore the equilibrium tide gives a reasonable estimate, within a factor 5 or so, as can be seen from the figure.   Therefore, making the crude approximation that the equilibrium tide approximation holds at $r_{\text{na,p}}$ and that the perturbed convective flux is constant above this value of $r$, we can get an estimate of the magnitude of this constant which is correct to within an order of magnitude.    
This indicates that the magnitude of the  perturbed  flux  may be understood in a simplified  way without reference to the  actual
 components of the Lagrangian displacement, and enables us to check that the numerical values of the convective flux we obtain are correct. 

\begin{figure*}
        \hspace{-2mm}
          \includegraphics[width=7cm,angle=0]{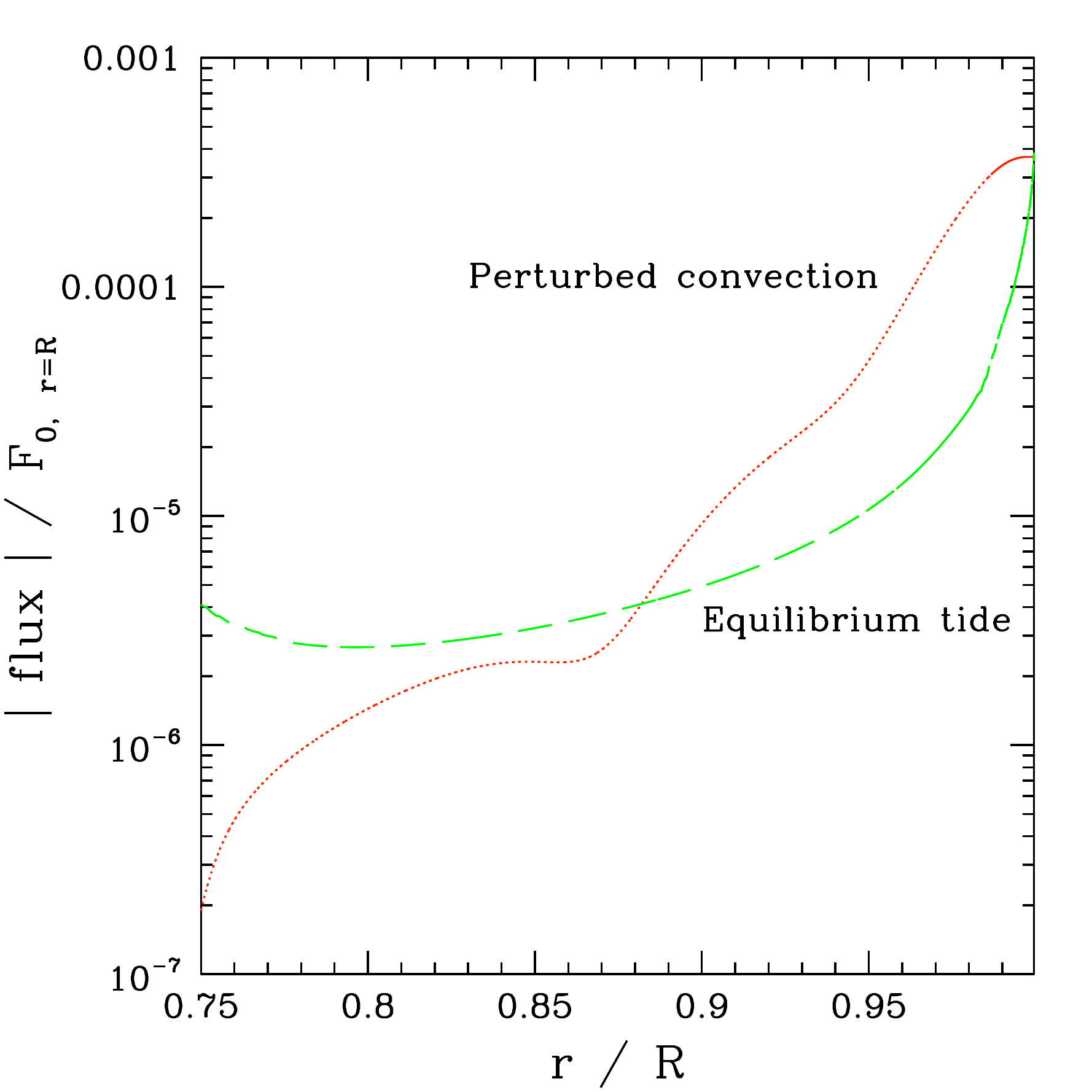}
         \includegraphics[width=7cm, angle=0]{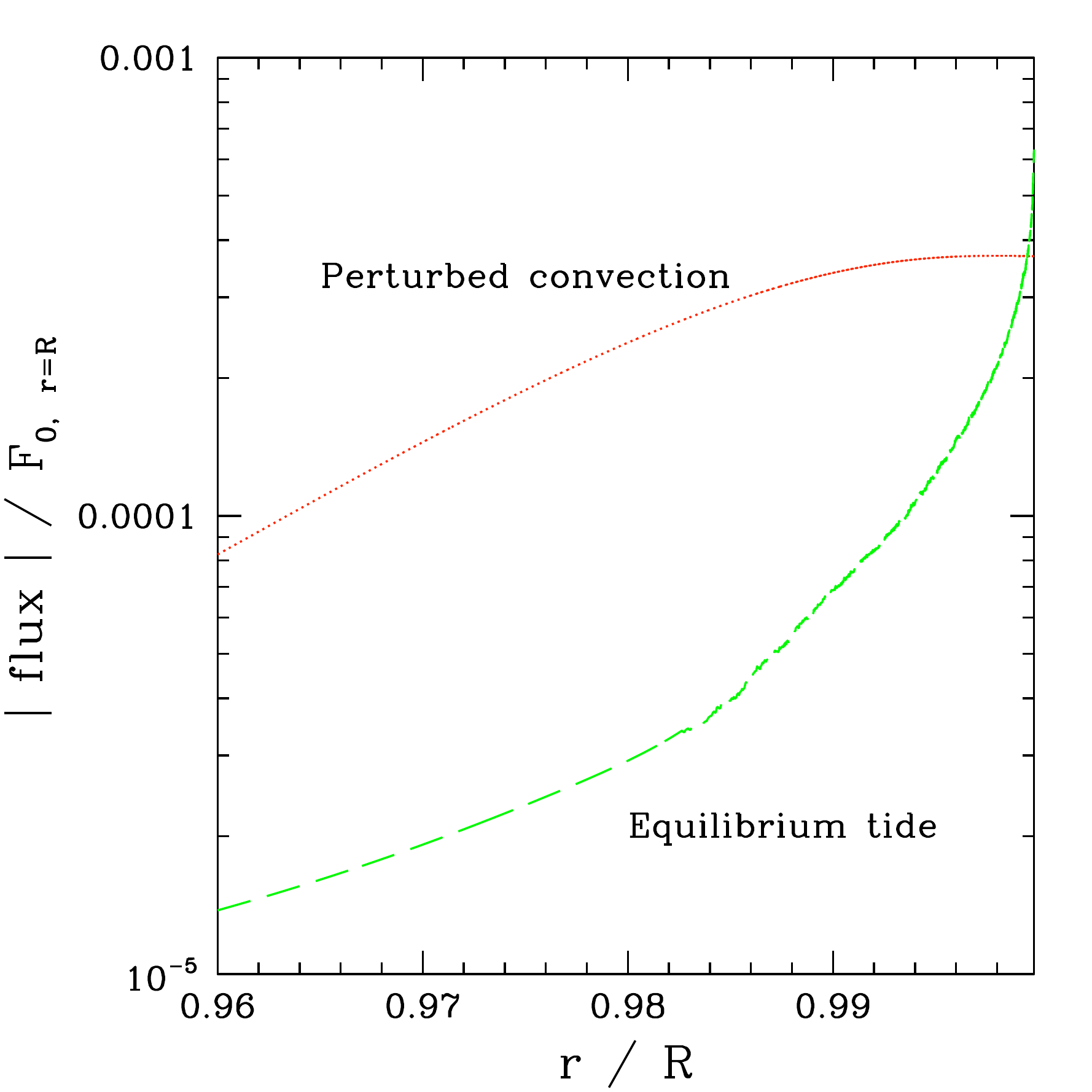}
        \caption{ The ratio of the magnitude of the radial component of the energy flux
 perturbation to the surface background value, $\left| F'/F_0|_{r=R} \right|,$ evaluated using approach~A for the perturbed convective flux (dotted red line) and the ratio of
 the magnitude of the radial component of the perturbed convective flux, evaluated assuming the equilibrium tide,
 to the  surface background  value of the flux, $\left| F'_{\text{c,eq}} /F_{,0}|_{r=R} \right| $ (dashed green line),    for the 1~M$_{\sun}$ star.
 The left hand panel shows these quantities for $ 0.75 < r/R < 0.9998,$ and the right hand panel
   shows them for $0.995 < r/R < 0.9998$
   where radiation becomes the dominant energy transport method in the background model.  }
        \vspace{2cm}
    \label{fig:Feqcomparison}
\end{figure*}

\subsubsection{Perturbed convection following approach~B}

Above we have described results for  perturbed convection following approach~A, for which $A$ in equation~(\ref{CONVFLUX}) is unperturbed.  We have also performed calculations following approach~B in which variations of $A$ are included.
These variations include perturbations to the state variables and the convective velocity but not the ad hoc parameters in the MLT
\citep{Salaris&Cassisi2008}. The results we obtain are similar to those found  following procedure~A, 
although details vary. To illustrate this,  we show the behaviour of  $|F'/F_0 |_{r=R}|$ 
for the two calculations in the left hand panel of Figure~\ref{fig:Fcomparison}
 in  the range $0.9  < r/R < 1$ for the 1~M$_{\sun}$ star.  In addition, we show $|V|/\xi_{r,eq}$    for the same  region in the right hand panel.
 The results are seen to be qualitatively similar with differences of about a factor of two at $r=R.$

\begin{figure*}
        \hspace{-2mm}
          \includegraphics[width=7cm,angle=0]{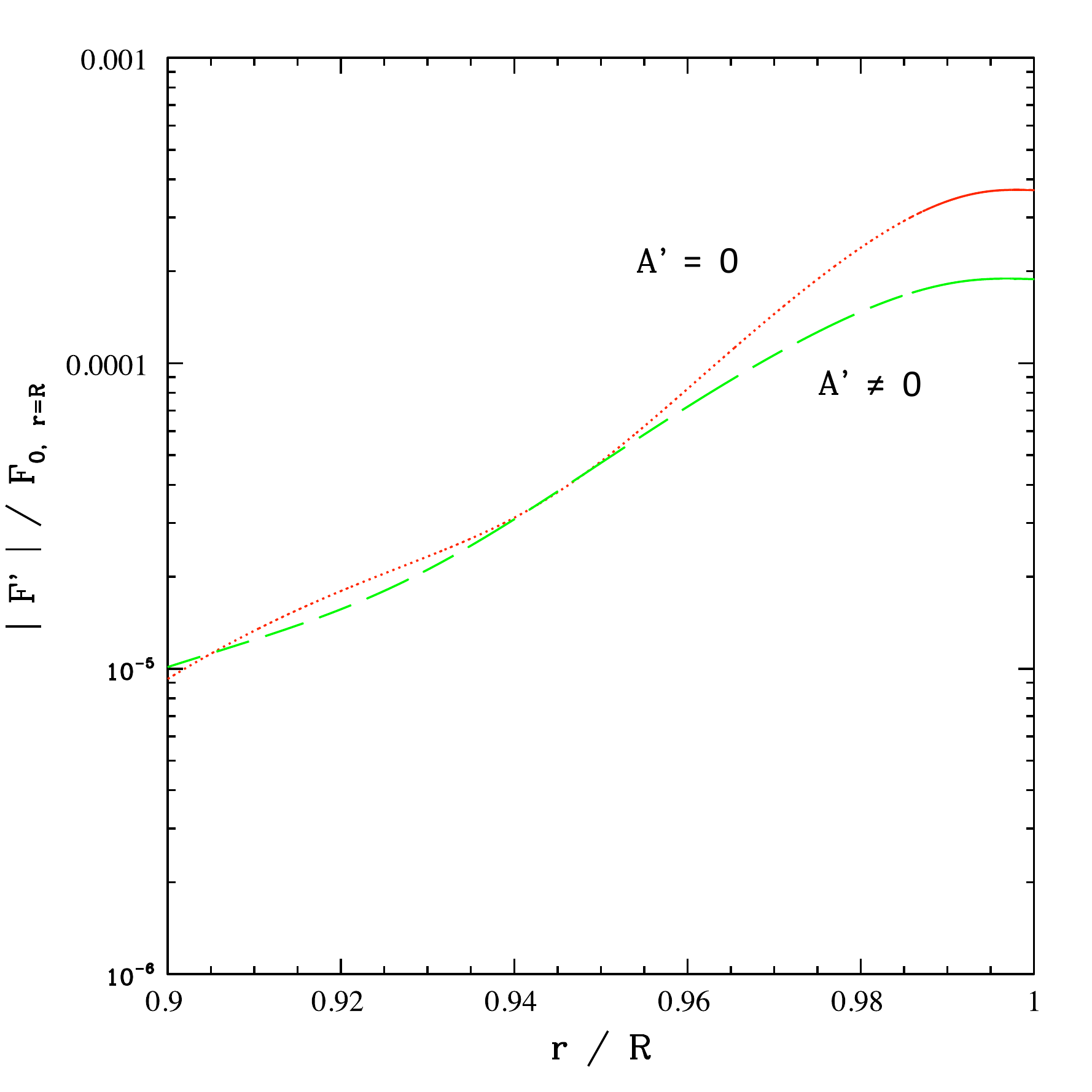}
         \includegraphics[width=7cm, angle=0]{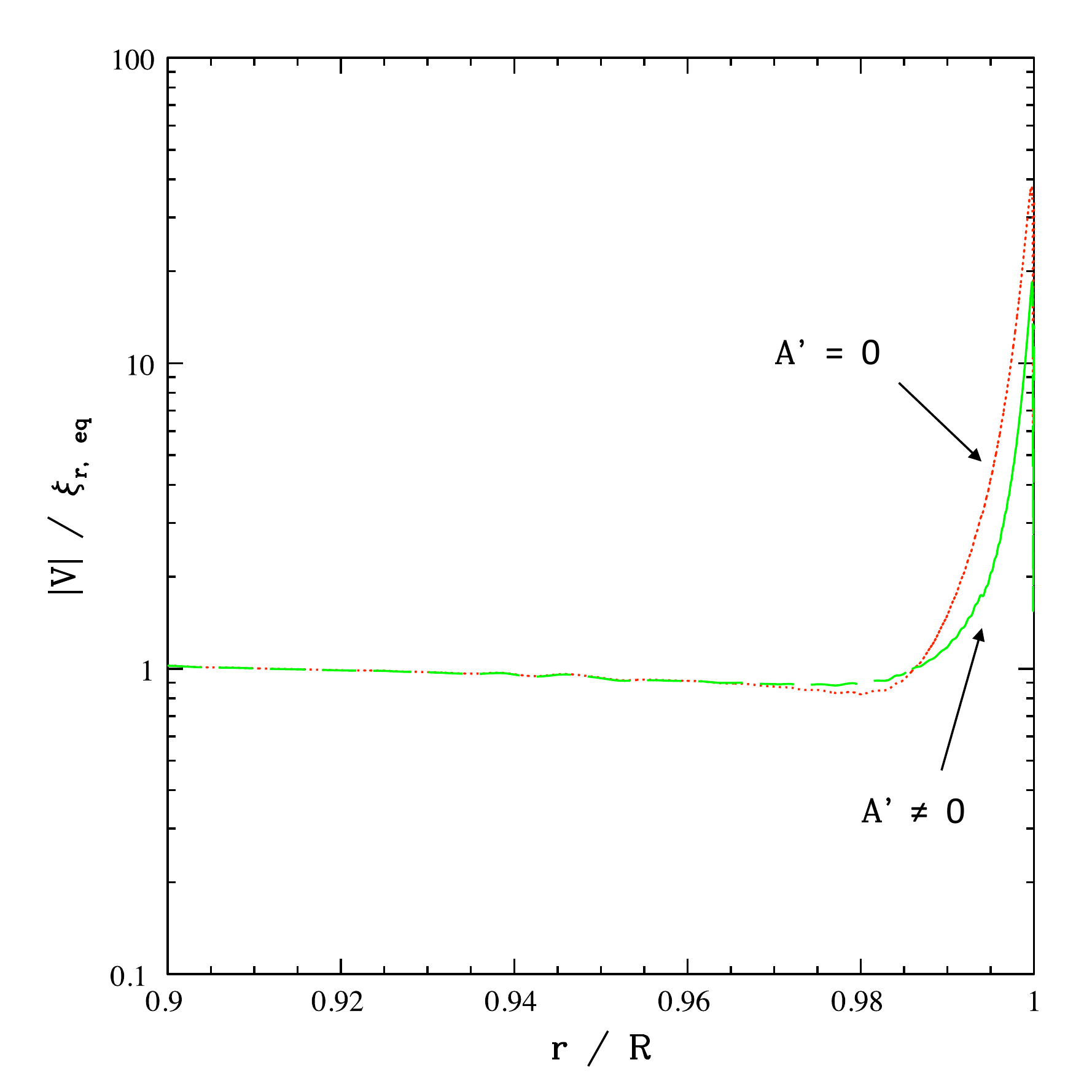}
        \caption{ 
          $\left| F'/F_0|_{r=R} \right|$  (left hand panel) and $|V|/\xi_{r,eq}$  (right hand panel) {\em versus} $r/R$ in the range
          $0.9  < r/R < 1$ 
          for the calculations with perturbed
          convection for both  approach~A (red dotted curves) and approach~B 
(green dashed  curves) for the 1~M$_{\sun}$ star.
      }
    \label{fig:Fcomparison}
\end{figure*}

\subsection{Resonance Survey}

As indicated in appendix~\ref{AppC1}, $g$--modes that are confined to the radiative core are expected to be excited and, being restricted to the region where
they are to a good approximation adiabatic and relatively free from dissipation, we expect resonances to be prominent.
In order to illustrate these, we perform a resonance survey by calculating the response of a 1~M$_{\sun}$ star over a fine grid
of orbital frequencies spanning the interval $[ 0.984\omega_0 , 1.016\omega_0 ]$  where $2\pi/\omega_0$ corresponds to the orbital period
of $4.23$~days  for which the calculations presented above were performed.   The calculations presented in this section are done for the perturbed convection case using approach~A.   The kinetic energy of the oscillation is plotted as a function of frequency in Figure~\ref {fig:Resonancesearch}.
Resonance spikes in which this quantity increases by up to five orders of magnitude, indicating a high quality factor, are clearly visible.
The frequency separation is uniform, as expected for high order $g$--modes, with an interval $\delta\omega_0/\omega_0 \sim  0.005,$ indicating an order $\sim 200.$

\begin{figure}
        \hspace{-2mm}
          \includegraphics[width=7cm,angle=0]{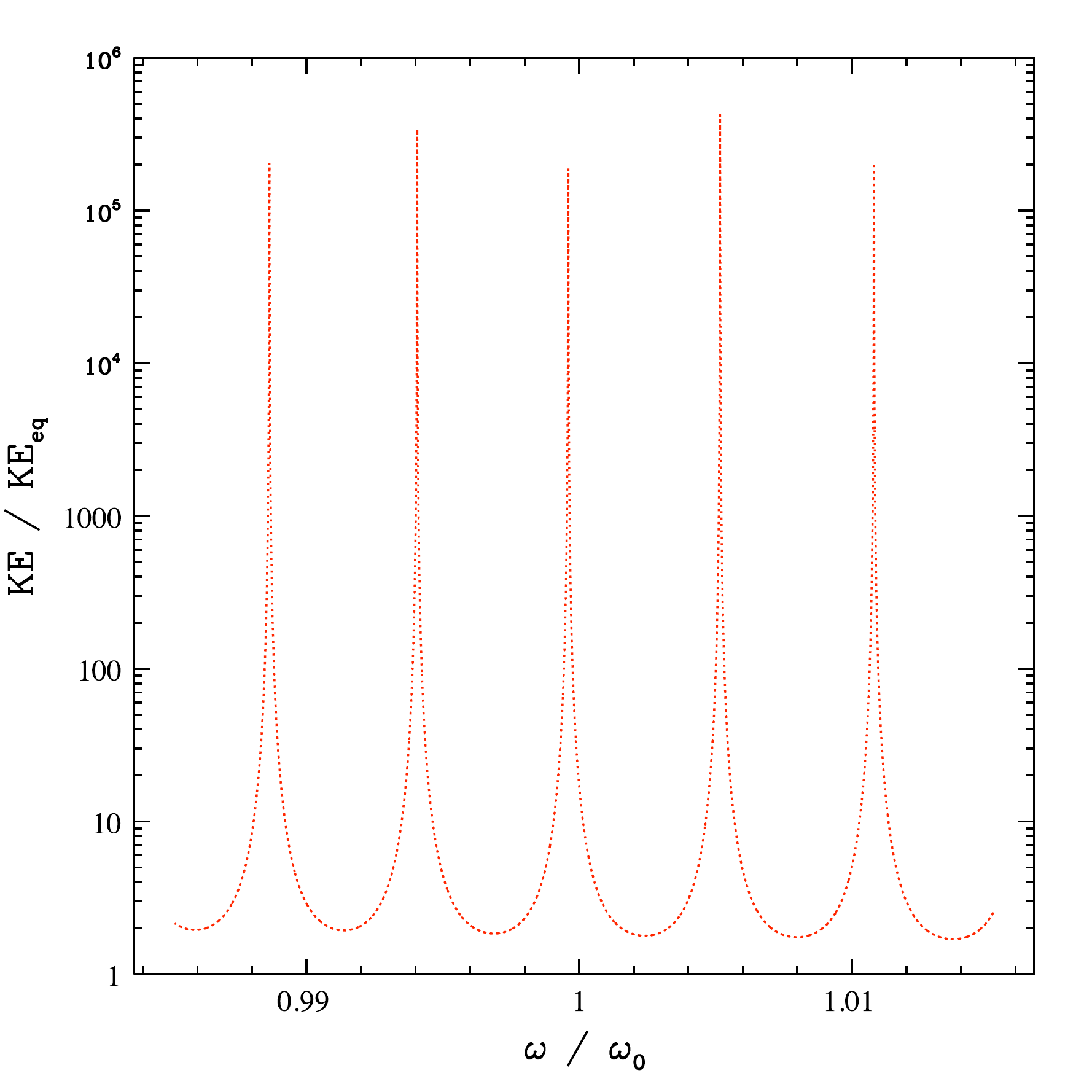}
        \caption{ This shows the kinetic energy of the oscillation divided by the value corresponding to the equilibrium tide as a function of orbital frequency
        for a narrow band of frequencies in the neighbourhood of $\omega_0,$ with $2\pi/\omega_0$
        corresponding to an orbital period of $4.23$~days.  The forcing frequency is twice the orbital frequency.  
        Prominent resonant spikes are clearly visible.  This is for a 1~M$_{\sun}$ star.}
        \vspace{0cm}
    \label{fig:Resonancesearch}
\end{figure}

Snapshots of the response obtained during the resonance search are presented in Figure~\ref{fig:Resonance}.
 The function $V$ is shown as a function of  $r,$  both in the region containing the radiative core and  in the non adiabatic surface layers,
  for the forcing frequency 
  $2\omega_0,$ which corresponds to the orbital period of $4.23$~days, and a close by  forcing frequency of $1.99965\omega_0,$ which corresponds to the centre of the 
  nearest resonance.   In the latter case, the resonantly excited $g$--mode that is confined to the radiative core can be clearly seen.
  However, the behaviour of $V$ close to the surface is  hardly  affected by this, indicating that for the most part  the response outside of the radiative core
 is robust and unaffected by internal resonances.}

\begin{figure*}
        \hspace{-2mm}
          \includegraphics[width=7cm,angle=0]{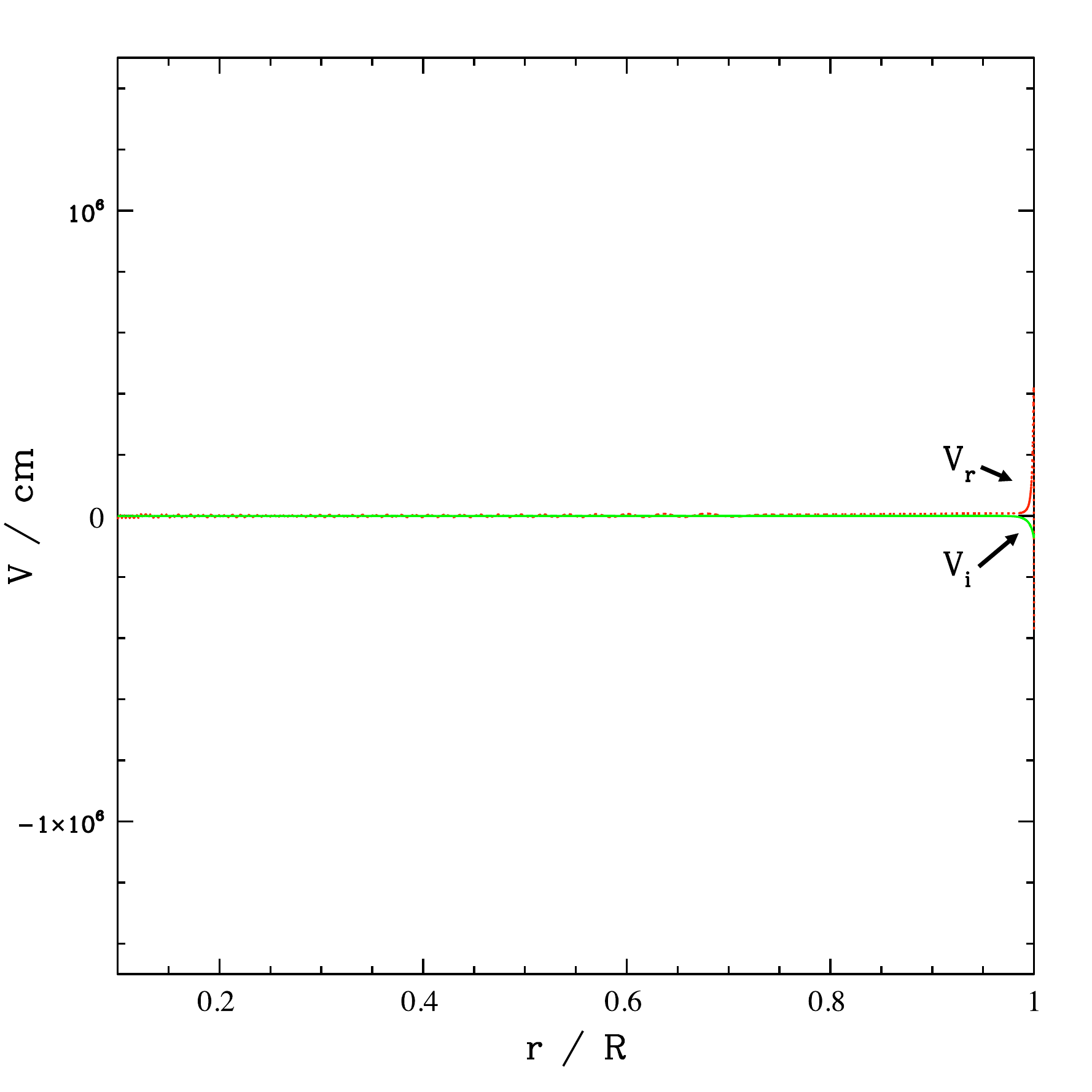}
         \includegraphics[width=7cm, angle=0]{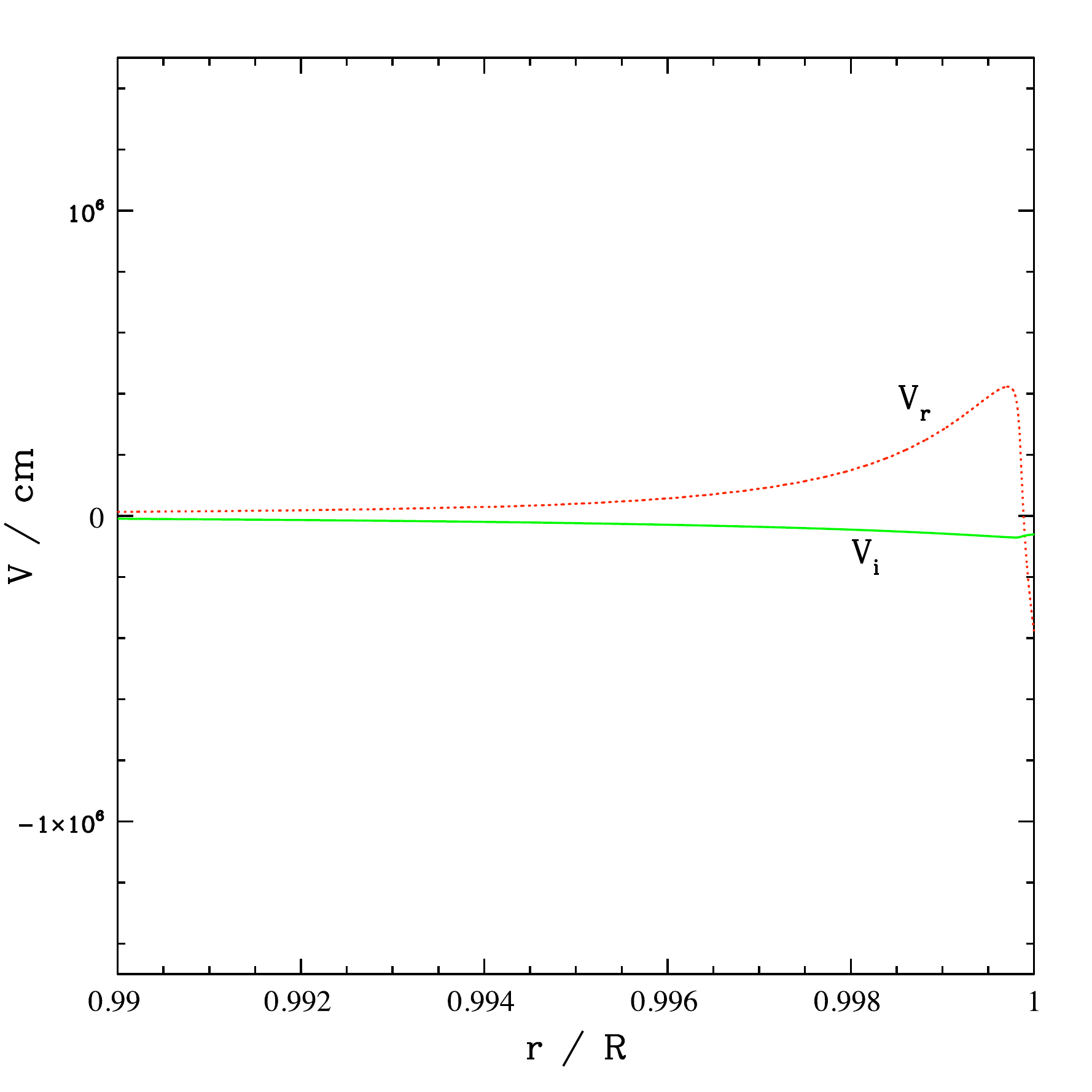}
                      \includegraphics[width=7cm]{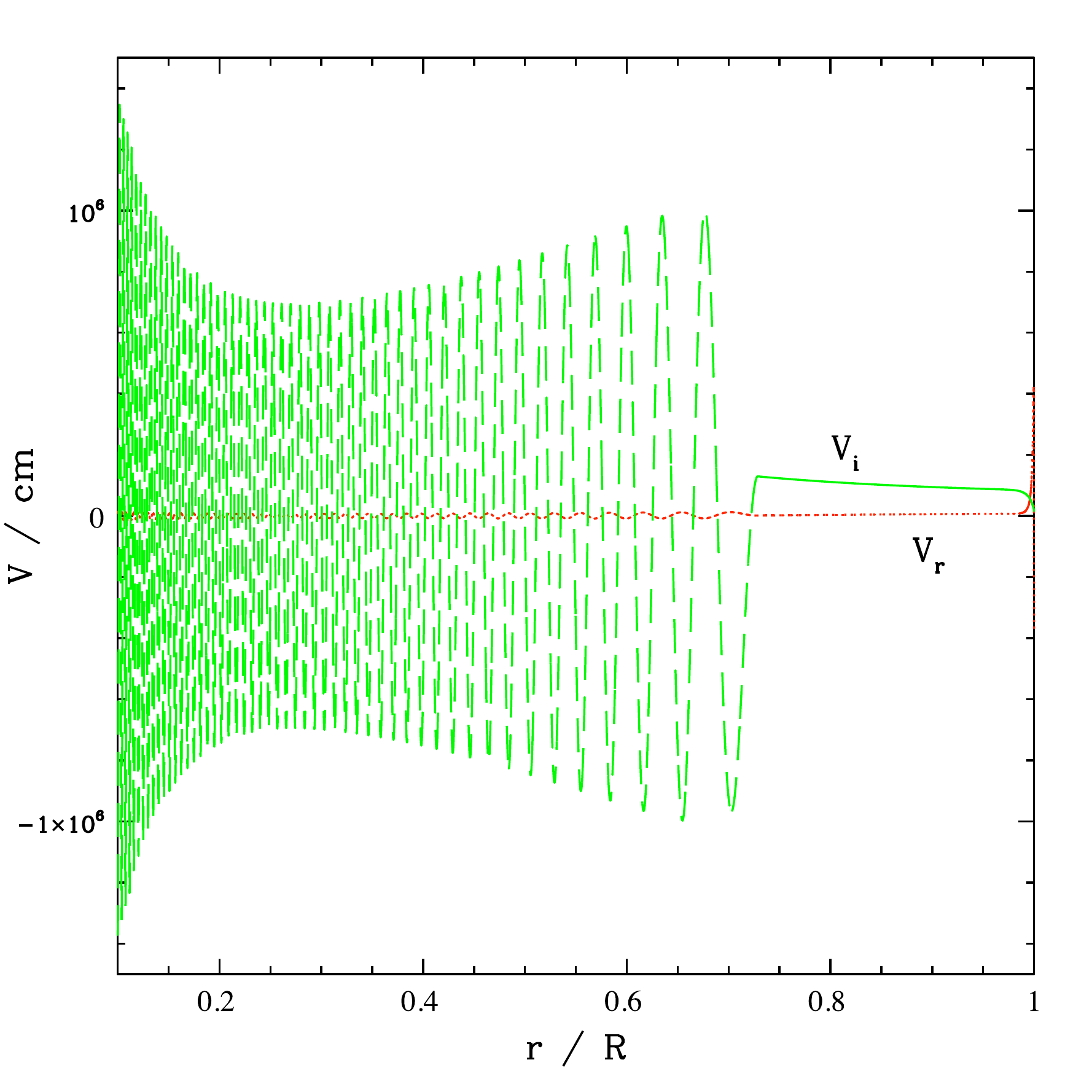}
        \includegraphics[width=7cm, angle=0]{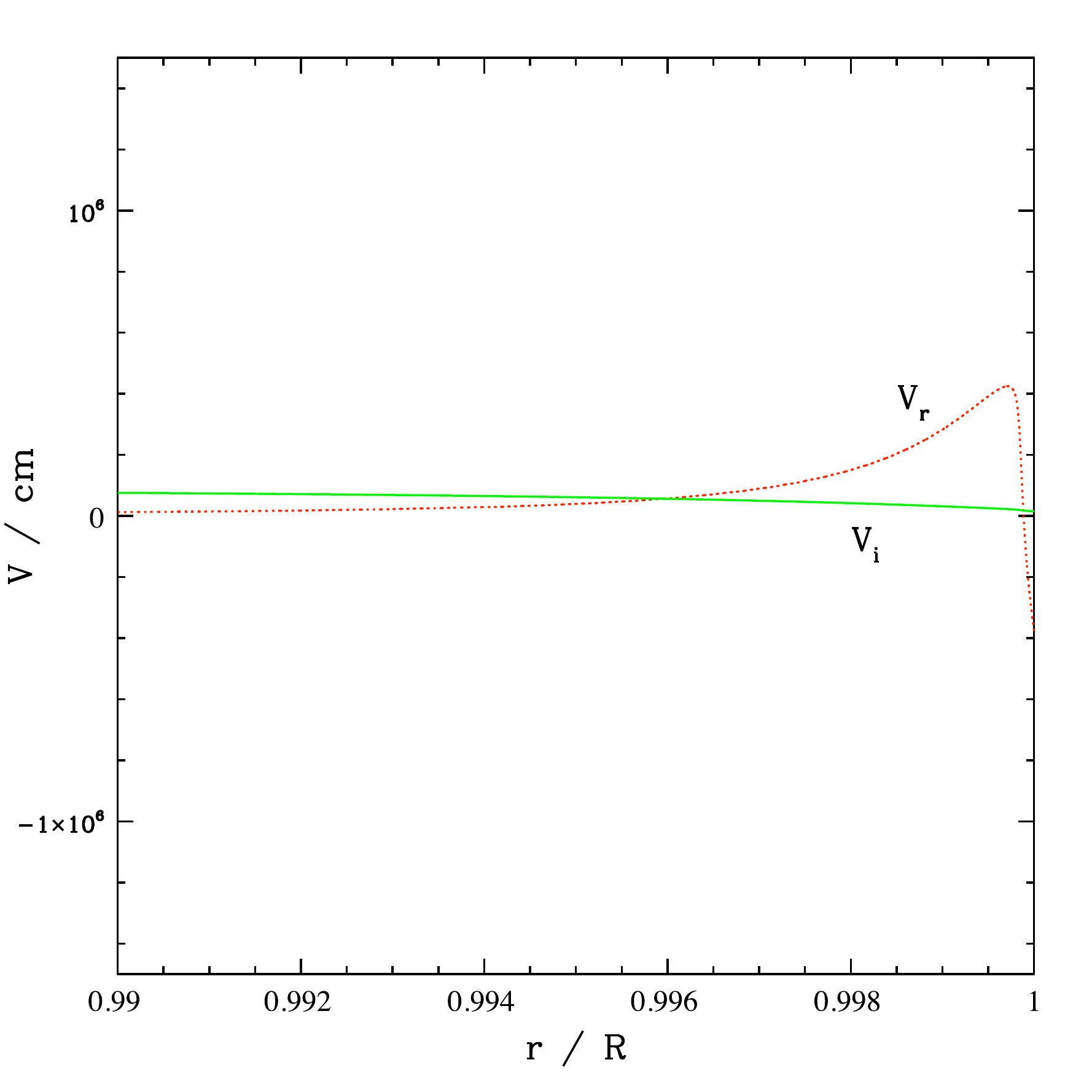}
        \caption{ This shows snapshots of the response obtained during the resonance search for a 1~M$_{\sun}$ star, with perturbed convection following approach~A.
        The real and imaginary components of the function $V$ are plotted as a function of $r/R$ for the range $0.1< r/R < 1$ 
in the left hand panels
and for the range $0.99< r/R < 1$ in right hand panels.
The upper panels correspond to the forcing frequency
$2\omega_0,$ and the lower panels to a forcing frequency of $1.99965\omega_0,$ 
which corresponds to the centre of the 
closest resonance, with the forcing frequency being twice the orbital frequency.    By comparing the upper and lower right hand panels It 
can be  seen that the behaviour of $V$ close to the surface is almost
unaffected by the resonant excitation below.}
    \label{fig:Resonance}
\end{figure*}

\section{Discussion}
\label{sec:Discussion}

The behaviour throughout the body of the star (shown in Figures~\ref{fig:log_mod_variables} and \ref{fig:log_mod_variables_BPT}) largely agrees with previous work, in that the response is oscillatory in the radiative core and evanescent in the convection envelope. { In the interior, away from a $g$--mode resonance,} the radial displacement follows the expectation from the equilibrium tide approximation well, oscillating around the equilibrium tide value. It may be reasonable to use the equilibrium tide approximation if the average response of the stellar interior is of interest.

{Bearing in mind the considerable uncertainties involved in modeling convection,  we considered  two extreme treatments
 Following a MLT approach, we considered  the case when the convection was unresponsive and frozen 
during perturbation and  the opposite limit when it was assumed to respond on a time scale short compared to the inverse forcing frequency and so attain a relaxed state.
In this last situation, we considered two approaches.  In the first, only the entropy gradient was perturbed while the state variables and convective velocity were held fixed.
In the second approach, these were allowed to vary.}
For all of the  treatments of convection, the behaviour in the surface region diverges from the expectation from the equilibrium tide approximation. 
The horizontal displacement greatly increases, being coupled with a change in the perturbed energy flux.  

This is the consequence of 
the importance of non--adiabatic effects, which are not consistent with the requirements for the standard equilibrium tide to be valid.  This is because hydrostatic equilibrium implies zero Lagrangian perturbations to the density and pressure,  which cannot be achieved when the perturbation is non--adiabatic.  As a result, there is a large amplification of the horizontal displacement.

Changing the exact treatment of convection used does produce quantitatively different results,
with a factor of $\sim10$ difference between the values of the horizontal displacement, 
and between the flux perturbations at the surface.
Another significant difference is in the extent of the region over which the perturbations
diverge from the equilibrium tide values.
In the case of frozen convection the perturbations do not have much scope to
grow in the convection zone as the flux cannot be perturbed there,
by assumption, and the perturbations are shown to change rapidly over
a length scale of a few times the depth of the surface radiative layer
once the flux can be perturbed once again. {This essentially restricts the
depth of the non adiabatic layer.}

Adopting a more plausible approximation of perturbed convection smooths out the changes
compared to those found using the assumption of frozen convection, as they occur over a greater
radial extent of the star instead of being confined to the radiative skin of the star.
This results from the fact that the convective flux can be perturbed,
and departures from the equilibrium tide  become evident once the  timescale $|1/N|$
exceeds the timescale of the oscillation (around
$r/R \sim 0.97$ for $M = 1$~$\text{M}_{\odot}$).
Nonetheless, in this case, the radial displacement remains close to the equilibrium value,
with a sudden, but much less pronounced, peak just below the surface,
and a phase shift of only around $1$~degree at the surface,
suggesting that non--adiabatic effects are only weakly evident in the radial displacement.
{Note that our resonance survey indicated resonant forcing frequencies
for which there were large spikes in  the kinetic energy of the oscillation,
associated with high order $g$--modes confined to the core. However, these were found to have little effect on
the solution near the surface.  Also,  in the absence of 
special conditions leading to resonance 
locking, non linear effects 
and increased dissipation 
are expected to lead to a low probability
of being found in resonance.}

The horizontal displacement still significantly deviates from the equilibrium approximation,
with a surface magnitude which is greater than the equilibrium approximation by a factor of $\sim 30$
{ when procedure~A was followed for convection.}
 { This could be reduced by a factor of three if procedure~B was followed instead.
 On the other hand, we remark that a test carried out where we returned to
procedure~A but simply increased the convective flux perturbation by a factor of $1.5$
 showed that surface horizontal displacement increased by a factor of three.} 
The phase shift at the surface is also more significant, of $\sim 10$~degrees.
These are both potentially testable predictions for observations, 
 as the horizontal displacement could be detected through spectroscopy.
 This could be done by utilising a disc--averaged effect, such as a periodic shift in the peak of the observed wavelength, such as in \citet{Dziembowski1977}. 
 Another potential method of detection could be to examine the Doppler broadening 
due to the different motions of different sections of the stellar surface. 
This would lead to a time--dependent broadening signal, and its magnitude and shape would depend upon the size of the oscillations, as well as the viewing angle.

The form of the horizontal displacement given 
in equation~\ref{eq:Methods:analytical:rearranged_eom_V} shows that
 $V \propto 1/\omega^{2}$. As the tidal perturbation goes as $D^{-3}$, the effect on $V$ from the change in $\omega$ will be counteracted by the change in the tidal perturbation, and we expect that the horizontal displacement will be constant for a given planetary mass,
independent of the orbital distance.
This has been found to be the case in our numerical solutions. The associated
radial velocity signal would therefore be proportional to $\omega$. 

The perturbation to the flux at the surface is also significantly different 
between the approximations of frozen and perturbed convection. 
Both show significantly non--adiabatic behaviour towards the surface, with large magnitudes
and non--negligible imaginary parts, although the detailed form does depend strongly upon the
 treatment of convection. { Nonetheless, in the case of perturbed convection, the magnitude of the
perturbed energy flux can be estimated in a simple way by evaluating the flux using the equilibrium tide near the
base of the non--adiabatic layer (see  Fig.\ref{fig:Feqcomparison}). 
}

 Whilst the magnitude of the radial displacement is not likely to result in a large change
 in the observed brightness of the star, the perturbation to the flux would result in an oscillating signal which could potentially be detectable.
 Note that this scales with the mass of the orbiting planet. 
 Such a detection would necessarily have to be disc--integrated, and therefore $V$ could
 not be utilised, as its effect cancels out when averaged over
 the whole disc \citep{Dziembowski1977}.
Non--transiting planets could possibly be detected photometrically using this method, as the signal scales as $\sin^{2} i$, where $i$ is the inclination of the orbital plane relative to the observer.
 Therefore, a planet which narrowly misses out on transiting its star would have a signal
 only a few percent smaller than a directly edge--on observation.

Such a detection of tidally induced oscillations would lead to a more
 precise determination of the parameters of known planetary systems. 
 It also  would make it possible to calculate the planetary mass:
 in the case of a transit detection this would be a direct inference,
 but in the case of a radial velocity detection this would be through breaking
 the degeneracy between mass and inclination. 
 The effects of the oscillations on observables will be the subject of a future paper. 

For a system in which a known hot Jupiter is already well constrained, this
modelling of the response of the star to the tidal potential of the planet would
be important if we were to look for additional planets in the system. 
By filtering out the photometric and spectroscopic variations associated with
the tidal oscillations, we can unambiguously associate additional
variations with other planets in the system.

{This approach could also be applied to stellar binaries, rather than to a planetary companion, which could give rise to larger signals. This could be used to investigate the theory itself, and possibly differentiate between the different approaches to perturbing the convective flux.}
				
Whilst these results may involve fewer assumptions than some previous work,
 there remain potentially significant areas for improvement. 
 The inclusion of rotation and its effects on the frequencies of the response,
 as well as on the form of the observational signal could be important,
 and therefore this paper is mostly aimed at slowly rotating stars.
  Also, as the perturbation to convection has been shown to have a significant effect on the calculation, further investigation into how best to model the perturbation to the convective flux,
 or the limitations of this approach, would be beneficial.
 In addition, the effect of the surface boundary conditions should be further investigated.

\section{Conclusions}
\label{sec:Conclusion}

In this paper, the non--adiabatic response of a star to a tidal perturbation due to a Jupiter--mass planet on a $4.23$\,day orbit 
was calculated.  We considered both a  $1$~M$_{\odot}$ and a $1.4$~M$_{\odot}$ star, under the assumptions of frozen convection,
and  convection that responded to the perturbation instantaneously.
 In all cases the behaviour in the interior of the star was found to
be similar to the adiabatic response and roughly followed the equilibrium tide approximation.

However, at the surface, non--adiabatic effects were found to be very important, 
causing the horizontal displacement to be amplified by a factor of $\sim 10-100$ compared to the equilibrium tide. 
This behaviour is due to non--adiabatic effects causing the perturbations to fail to be consistent with the
hydrostatic equilibrium assumption and/or the zero Lagrangian pressure and density perturbation assumption
 (which are the basis of the standard equilibrium tide theory).  When convection is frozen, these non--adiabatic effects are important  
in the thin superadiabatic convection region and the strongly stratified radiative region at the surface of the star.  {In the more realistic case that} convection is perturbed, non--adiabaticity extends further down{, and the departure from the equilibrium tide is significant, though  less than in the case of frozen convection}.
This departure from the equilibrium tide has major implications for the observation of tidally induced oscillations.

These oscillations may be observable through both photometric and spectroscopic techniques. Both the magnitude and the phase of the observed signal could be used to constrain and characterise the system. For known systems this could lead to a better constraint on  the mass of the planet, and it could also help to unambiguously detect additional planets where other methods have shortcomings (such as the requirement for a transit limiting the range of detectable inclinations).

Whilst this work has limitations, the fact that non--adiabatic effects have to be taken into account when calculating the response of the star at its surface is a robust result. The photometric and spectroscopic signals resulting from the oscillations of the stellar surface will be the subject of a forthcoming paper.

\section*{Acknowledgements}

AB is supported by a PhD studentship from the Science and Technology Facilities Council (STFC), grant ST/N504233/1.  JCBP thanks the Physics Department at Oxford University for the hospitality during the period when this project was done.  We thank the referee for comments that have significantly improved the paper.




\bibliographystyle{mnras}
\bibliography{library} 




\appendix

\section{Tidal potential}
\label{sec:app:Tidal}

The tidal potential considered here is the lowest order time--varying potential which is not constant across the star. To find this we evaluate the potential in the frame of the star,  accounting for the force acting at the stellar centre of mass (the indirect term) in that frame, as

\begin{equation}
\label{eq:app:tidal:start}
\Phi_{\text{P}} =  \Phi + \frac{G m_{\text{p}}}{D^3} \mathbfit{D} \bm{\cdot} \mathbfit{r},
\end{equation}

\noindent where $\Phi_{\text{P}}$ is the tidal potential, $\Phi$ is the gravitational potential due to the perturber, $G$ is the gravitational constant, $m_{\text{p}}$ is the mass of the perturbing body, $\mathbfit{D}$ is the vector from the centre of the star to the  perturber, approximated as a point mass, and $\mathbfit{r}$ is the position vector from the centre of the star to the point at which the potential is calculated.

This can be expanded in terms of $r/D$, with $D = |{\bf D}|$, which is a small quantity, to give

\begin{equation}
\label{eq:app:tidal:Taylor}
\Phi_{\text{P}} = - \frac{G m_{\text{p}}}{D} \left[ 1 + \left( \frac{r}{D} \right)^{2} \left( \frac{3}{2} (\hat{\mathbfit{D}} \cdot \hat{\mathbfit{r}})^{2} - \frac{1}{2} \right) \right] + \textit{O} \left( \frac{r^{3}}{D^{3}} \right)
\end{equation}

\noindent where the hats denote unit vectors.
Using the fact that $\hat{\mathbfit{D}} \bm{\cdot} \hat{\mathbfit{r}} = [ \cos(\omega t) \hat{\mathbfit{x}} + \sin(\omega t) \hat{\mathbfit{y}} ] \bm{\cdot} [ \sin(\theta) \cos(\phi) \hat{\mathbfit{x}} + \sin(\theta) \sin(\phi) \hat{\mathbfit{y}} + \cos(\theta) \hat{\mathbfit{z}} ]$, where we adopt Cartesian coordinates with origin at the stellar centre of mass and
$(x,y)$ plane coinciding with that of the orbit. The unit vectors in the coordinate directions are
 $\hat{\mathbfit{x}}$,   $\hat{\mathbfit{y}}$, $\hat{\mathbfit{z}}.$ 
Removing terms with non--zero time average (and therefore keeping only the oscillatory terms), we arrive at the expression

\begin{equation} \label{eq:app:tidal:second_order}
 \Phi_{\text{P}} \simeq - \frac{G m_{\text{p}}}{4 D} \left[ \left( \frac{r}{D} \right)^{2} 3 \sin^{2} (\theta) \cos \left[ 2 (\omega t - \phi) \right] \right].
\end{equation}

\noindent As $3 \sin^{2}(\theta) = P_{2}^{|2|}(\cos \theta)$ (the associated Legendre polynomial), $\Phi_{P}$ can be written as

\begin{equation}
\Phi_{\text{P}} \simeq \Re \left( - \frac{G M_{\text{p}}}{4 D^{3}} r^{2} Y^{-2}_{2} (\theta, \phi) \text{e}^{2 \text{i} \omega t} \right)
\end{equation} 

\noindent where $Y^{-2}_{2} (\theta, \phi) = P_{2}^{|2|}(\cos(\theta)) \text{e}^{- 2 \text{i} \phi}$ is a spherical harmonic.

\noindent Because this is the only source of time and angular dependence (as the equilibrium model is taken to be spherically symmetric and static) in the system of linear equations, any perturbed quantity in  equations  (\ref{eq:cont_osc}) to (\ref{eq:mom_osc}), $q'$, can be expressed in the form

\begin{equation}
\label{eq:app:tidal:sep_q'}
q' \left( r, \theta, \phi, t \right) = \Re \left(  q'(r) 3 \sin^{2}  \theta \text{e}^{2 \text{i} (\omega t - \phi)}  \right)
\end{equation}

\noindent where $q'(r)$ may itself be complex.

\section{Derivation of oscillation equations}
\label{sec:app:derivation}

Here we set out the skeleton for the derivation of the stellar oscillation equations used in this work, including stating the approximations used along the way.
We start from the continuity, momentum,  and energy { equations together with the expression  we used for the heat flux in the form}

\begin{equation}
\label{B1}
  \frac{\partial \, \rho}{\partial \, t} + \bm{\nabla} \cdot ( \rho \, \bm{u} ) = 0 ,
\end{equation}
\begin{equation}
\label{B2}
\rho \left( \frac{\partial}{\partial \, t} +  \bm{u} \cdot \bm{\nabla} \right)
\bm{u} = - \bm{\nabla} p - \rho \bm{\nabla} \Phi 
\end{equation}
\begin{equation}
\label{B3}
\rho T \left( \frac{\partial}{\partial t} + \bm{u} \cdot \bm{\nabla} \right) s =
- \bm{\nabla} \cdot \bm{F} 
\end{equation}
\begin{equation}
  \bm{F} = - K \bm{\nabla} T + \bm{F_{\text{c}}}
\end{equation}

\noindent where $\rho$ is the density, $\bm{u}$ is the vector velocity, $p$ is the pressure, $\Phi$ is the gravitational potential, $T$ is the temperature, $s$ is the specific entropy, $\bm{F}$ is the total flux, $\bm{F_{\text{c}}}$ is the convective flux and $\kappa$ is the opacity.

{ In this work, we assume that  the local convection time scale is much less than the inverse oscillation frequency
so that the convective flux can relax to an equilibrium value in regions where we need to take it into account.
Assuming a  standard form derived from mixing length theory,  the  expression for the
 convective flux we adopt is  given by:}
\begin{equation}
  \bm{F_{\text{c}}} = - \bm{\hat{n}} \frac{b \rho T v_{\text{c}} l}{1 + {c \sigma T^{3}} ( {\rho^{2} c_{p} l v_{\text{c}} \kappa})^{-1} } \bm{\hat{n} \cdot \vec{\nabla}} s
  \label{appB:Fconv}
 \end{equation}
where $\sigma$ is the Stefan-Boltzmann constant, $v_{\text{c}}$ is the convective velocity, { $\bm{\hat{n}}$ is the unit vector in the direction opposite to that of gravity, $c_{p} \equiv T \left( {\partial s}/{\partial T} \right)_{p}$ is the specific heat capacity at constant pressure, $l$ is the mixing length, and $b$ and $c$ are numerical factors which depend on the model of convection being used  \citep[see][for more details and discussion]{Salaris&Cassisi2008}.
Note that in the steady unperturbed state  $\bm{\hat{n}}  =  \bm{\hat{r}},$ with the latter pointing in the radial direction.}



Equations (\ref{B1}) - (\ref{appB:Fconv})   are linearised, and only first order terms are retained. As the background state is in equilibrium, we set $\bm{u} = \partial \bm{\xi}/{\partial t}$ where $\bm{\xi}$ is the Lagrangian displacement. This gives us the set of linearised equations:

\begin{equation}
\frac{\partial}{\partial t} \left( \rho' + \bm{\nabla} \bm{\cdot} \left( \rho_{0} \bm{\xi} \right) \right) = 0,
\end{equation}
\begin{equation}
\rho_{0} \frac{\partial^{2} \bm{\xi}}{\partial t^{2}}= - \bm{\nabla} p' - \rho_{0} \bm{\nabla} \Phi_{\text{P}}
- \rho' \bm{\nabla} \Phi_{0} ,
\end{equation}
\begin{equation}
\rho_{0} T_{0} \frac{\partial}{\partial t} \left( s' + \bm{\xi} \bm{\cdot} \bm{\nabla} s_{0} \right)
= - \bm{\nabla} \bm{\cdot} \bm{F}' ,
\label{secondthermo}
\end{equation}
\begin{equation}
\bm{F}' = - K' \bm{\nabla} T_{0} - K_{0} \bm{\nabla} T' + \bm{F'_{\text{c}}},
\end{equation}
\begin{equation}
\frac{K'}{K_{0}} = 3 \frac{T'}{T_{0}} - \frac{\kappa'}{\kappa_{0}} - \frac{\rho'}{\rho_{0}},
\end{equation}
\begin{equation}
\kappa' = \kappa_{\rho} \frac{\rho'}{\rho_{0}} + \kappa_{T} \frac{T'}{T_{0}},
\end{equation}
\begin{equation}
s'= c_{p} \left( \frac{T'}{T_{0}} - \nabla_{\text{ad}} \frac{p'}{p_{0}} \right),
\end{equation}
\begin{equation}
\frac{\rho'}{\rho_{0}} = \frac{1}{\chi_{\rho}} \left( \frac{p'}{p_{0}} - \chi_{T} \frac{T'}{T_{0}} \right),
\end{equation}

\noindent where $q'$ denotes the Eulerian perturbation to the variable $q$.
We have introduced 
$\nabla_{\text{ad}} \equiv \left( {\partial \ln T}/{\partial \ln p} \right)_{s}$, the adiabatic gradient; $\kappa_{\rho} \equiv \left( \partial \kappa/\partial \ln \rho \right)_{T}$; $\kappa_{T} \equiv \left( \partial \kappa/\partial \ln T \right)_{\rho}$; $\chi_{\rho} \equiv \left( \partial \ln p/{\partial \ln \rho} \right)_{T}$
and $\chi_{T} \equiv \left( {\partial \ln p}/{\partial \ln T} \right)_{\rho}$.
 
 To find the perturbation to the convective flux we take the unperturbed form to be
 \begin{equation}
  \bm{F_{\text{c}}} = - \bm{\hat{r}} A \bm{\hat{r} \cdot \vec{\nabla}} s,
 \end{equation}
 { where comparison with (\ref{appB:Fconv}) defines $A.$  We expect that} the contribution of  $\bm{F'_{\text{c}}}$ is small when the response is essentially adiabatic, and  only becomes important
 in a thin layer towards the surface.  Therefore we make the approximation that $\bm{F'_{\text{c}}}$ is dominated by the gradient term. This leads to a perturbed convective flux of the form
 \begin{equation}
  \bm{F'_{\text{c}}} = - \bm{\hat{r}} A \bm{\hat{r} \cdot \vec{\nabla}} s' - \bm{\Delta \hat{n}} A \bm{\hat{r} \cdot \vec{\nabla}} s
 \end{equation} 
 where $- \bm{\Delta \hat{n}}$ is the change in the direction of free-fall, which is perpendicular to the radial direction.  { The assumption that $A$ is taken to be unchanged by the perturbation rests on the assumed dominance of the gradient term which might  be expected for small scale perturbations in a thin layer. We remark that, 
 as discussed above, we have investigated cases where $A$ is allowed to vary,  also incorporating   the dependence of the convective velocity on the entropy gradient,
 and do not find qualitative changes of behaviour.}
  
\noindent  Following these procedures we obtain a set of 15 equations, with the associated 15 variables being: $p', T', \bm{\xi}, \bm{F}',
\rho', s', K', \kappa$ and $\bm{F'_{\text{c}}}$. We eliminate eleven of these with the aim of obtaining four equations for the four variables which we desire to remain, 
namely  $p', T', \xi_{r}$ and $F_{r}'$, where the subscripted $r$ denotes the radial component of the vector quantity.

\noindent The equations are all linear and have time--independent coefficients, so we can separate the time dependence from the spatial dependence. As the perturbing potential is proportional to $Y_{l}^{-m} (\theta, \phi) \text{e}^{\text{i} m \omega t}$, with $l = m = 2$, we look for solutions with the same angular and time dependence. Therefore, with $q'$ representing one of the variables we solve for, we have ${\partial q'}/{\partial t} = i m \omega q'$, ${\partial q'}/{\partial \phi} = - i m q'$ and $\nabla_{\perp}^{2} q' = -( {l ( l + 1) }/{r^2} )q'$, where $\bm{\nabla}_{\perp} = \bm{\nabla} - \hat{\bm{r}} {\partial}/{\partial r}$, so that $\nabla_{\perp}^{2}  = \nabla^{2} - ({1}/{r^{2}}) {\partial}/{\partial r} \left( r^{2} {\partial }/{\partial r} \right)$. Note that the equilibrium variables are purely radial, so $\bm{\nabla}_{\perp} q_{0} = 0$.

In order to make use of the properties of spherical harmonics, the equations must be formed in such a way as to ensure that $\bm{\nabla}_{\perp}$ only appears as $\nabla_{\perp}^{2}$.  To do this, the vector equations must be split into their radial and tangential components, and the divergence terms must also be split into the radial and tangential parts. By eliminating the unwanted unknowns, we are left with the oscillations equations, given in equations (\,\ref{eq:cont_osc}) to\,(\ref{eq:mom_osc}).

For solving these equations numerically, we convert to dimensionless variables $\tilde{\xi}_{r} = \xi_{r} / R$, $\tilde{F}_{r} = F_{r}' / F_{r_{\text{BC}}}$, $\tilde{p} = p' / p_{0}$ and $\tilde{T} = T' / T_{0}$, where $F_{r_{\text{BC}}} = F_{r_{0}}|_{r=R}$. The equations are also converted to a dimensionless form, whilst avoiding any potential singularities, giving

\begin{multline} \label{eq:cont_osc_dim}
\frac{1}{\rho_{0} R} \frac{\partial}{\partial r} ( r^{2} \rho_{0} \tilde{\xi}_{r} )  
+ \left( \frac{r^{2}}{\chi_{\rho} R^{2}} - \frac{l (l+1) p_{0}}{m^{2} \omega^{2} R^{2} \rho_{0} } \right) \tilde{p} \\
- \frac{\chi_{T}}{\chi_{\rho}} \frac{r^{2}}{R^{2}} \tilde{T}
=
\frac{l (l+1)}{m^{2} \omega^{2} R^{2}} \Phi_{P}
\end{multline}
\begin{multline} \label{eq:ent_osc_dim}
i \frac{1}{c_{p}} \frac{r^{2}}{R} \frac{d s_{0}}{dr} \tilde{\xi}_{r}
+ \frac{ F_{r_{\text{BC}}} }{ m \omega \rho_{0} T_{0} c_{p} R^{2} }  \frac{\partial}{\partial r} \left( r^{2} \tilde{F}_{r} \right) \\
- i \nabla_{\text{ad}} \frac{r^{2}}{R^{2}} \tilde{p}
+ \left( i \frac{r^{2}}{R^{2}} + \frac{l (l+1) K_{0}}{\rho_{0} m \omega c_{p} R^{2}} \right) \tilde{T}
=
0
\end{multline}
\begin{multline} \label{eq:flux_osc_dim}
- \frac{dr}{dT_{0}} \frac{F_{r_{\text{BC}}}}{K_{0}} \tilde{F}_{r}
- \frac{dr}{dT_{0}} T_{0} \frac{\partial \tilde{T}}{\partial r}  
+  \left[ - 4 + \frac{\kappa_{T}}{\kappa_{0}}  - \frac{\chi_{T}}{\chi_{\rho}}  \left(  1 +  \frac{\kappa_{\rho}}{\kappa_{0}}  \right)      \right] \tilde{T}  \\
  +   \frac{dr}{dT_{0}}  \frac{dr}{ds_{0}} \frac{F_{\text{c}, r,0}}{K_{0}}  \frac{d}{dr}(c_{p} \tilde{T})
  + \frac{1}{\chi_{\rho}} \left( 1 + \frac{\kappa_{\rho}}{\kappa_{0}} \right) \tilde{p} \\
-  \frac{dr}{dT_{0}}  \frac{dr}{ds_{0}} \frac{F_{\text{c}, r,0}}{K_{0}}  \frac{d}{dr}(c_{p} \nabla_{\text{ad}} \tilde{p})
=
0
\end{multline}
\begin{multline} \label{eq:mom_osc_dim}
- \tilde{\xi}_{r} 
+ \frac{1}{m^{2} \omega^{2} R} \left(  \frac{1}{\rho_{0}} \frac{\partial \left( p_{0} \tilde{p} \right)}{\partial r} +  \frac{d \Phi_{0}}{d r} \frac{\tilde{p}}{\chi_{\rho}}  \right) \\
-  \frac{d \Phi_{0}}{d r} \frac{1}{m^{2} \omega^{2} R} \frac{\chi_{T}}{\chi_{\rho}} \tilde{T}
=
- \frac{1}{m^{2} \omega^{2} R} \frac{\partial \Phi_{P}}{\partial r}
\end{multline}
\\
\noindent where equations~\ref{eq:cont_osc_dim} to~\ref{eq:mom_osc_dim} correspond to equations~(\ref{eq:cont_osc}) to~(\ref{eq:mom_osc}), multiplied by $r^{2} / (\rho_{0} R^{2})$, $r^{2} / ( m \omega \rho_{0} T_{0} c_{p} R^{2} )$, $- dr / dT_{0}$, and $1 / ( m^{2} \omega^{2} \rho_{0} R )$ respectively.

\section{The equilibrium tide and the low frequency limit in the adiabatic region}
\label{sec:app:eq}

This section briefly covers the equilibrium tide and its prediction for the radial and horizontal displacements, so that the equilibrium tide and the non--adiabatic dynamical tide can be compared.

\subsection{The situation when $N^2$ is non--zero}

The equilibrium tide is calculated in the low--frequency limit, such that $\omega \approx 0$, in which the equation of motion reduces to the condition for hydrostatic equilibrium, { which, when
linearised, takes}  the form:
\begin{equation}
\label{eq:app:eq:eom}
0 = - \bm{\nabla} p' - \rho_{0} \bm{\nabla} \Phi_{\text{P}} - \rho' \bm{\nabla} \Phi_{0},
\end{equation}

\noindent where $\Phi_{\text{P}} (r) = - G m_{\text{P}} r^2/ (4 D^{3})$ is the radial  part of the tidal potential  obtained after  factoring out the angular dependent factor $Y^{-2}_{2} (\theta, \phi) \text{e}^{2 \text{i} \omega t}$ (see Appendix \ref{sec:app:Tidal}). The same factor is also removed from the perturbation response.
 
  Here $\Phi_{0}$ is the equilibrium gravitational potential, $p'$ is the perturbation to the pressure, $\rho_{0}$ is the equilibrium density, 
 and $\rho'$ is the perturbation to the density.

\noindent We assume that the background star is at hydrostatic equilibrium, giving:

\begin{equation}
\label{eq:app:eq:hydrostatic}
\frac{d p_{0}}{d r} = - g \rho_{0}
\end{equation}

\noindent where $g \equiv d \Phi_{0} /dr $ is the magnitude of the gravitational acceleration.

\noindent The horizontal component of equation~(\ref{eq:app:eq:eom}) yields:

\begin{equation}
 \label{eq:app:eq:p}
 p' = -\rho_0 \Phi_{\text{P}} ,
\end{equation}  

\noindent which can be { substituted into}  the radial component of equation~(\ref{eq:app:eq:eom}) { giving an expression for} $\rho'$ as:

\begin{equation}
\label{eq:app:eq:rho}
\rho' = \frac{d \rho_{0}}{d r} \frac{\Phi_{\text{P}}}{g} .
\end{equation}
{ Similarly (\ref{eq:app:eq:p}) can be written as:
\begin{equation}
 \label{eq:app:eq:p1}
 p' =   \frac{d p_{0}}{d r} \frac{\Phi_{\text{P}}}{g} .
\end{equation}  
Then from (\ref{eq:app:eq:rho} ) and (\ref {eq:app:eq:p1}) we obtain:
\begin{equation}
\label{eq:app:eq:ad}
 \Delta p - \frac{\Gamma_1 p_0}{\rho_0} \Delta \rho =    \left( \xi_r +\frac{\Phi_{\text{P}}}{g} \right)\left (   \frac{d p_{0}}{d r}  -\frac{\Gamma_1 p_0}{\rho_0}  \frac{d \rho_{0}}{d r}  \right).
\end{equation}  
From this we can conclude that for adiabatic perturbations at a location for which  $N^2\ne 0,$ we have}:
\begin{equation}
\label{eq:app:eq:delta}
\xi_{r} = - \frac{\Phi_{\text{P}}}{g},
\end{equation}

\noindent where $\xi_{r}$ is the radial displacement from the equilibrium position.
{ This is the standard form of the radial displacement for the equilibrium tide.}
 We remark  that taken together with equations~(\ref{eq:app:eq:rho})
and~(\ref{eq:app:eq:p1}),  this implies that the Lagrangian perturbations to the pressure and density, $\Delta P$ and $\Delta \rho$, are both zero.
This is consistent with these quantities being related by a condition for adiabatic change. But note that other relations between these quantities, such as may occur in a non adiabatic zone, may lead to an inconsistency, with the consequence that hydrostatic equilibrium may not be assumed and the standard equilibrium tide discussed { above} will not be applicable.

\noindent Expressing the horizontal displacement as $\bm{\xi}_{\perp} = r \bm{\nabla}_{\perp} V$, where $V$ has the same dependences upon $\theta$ and $\phi$ as the other variables under consideration (see section~\ref{sec:Methods:analytical}), gives the continuity equation as:
\begin{equation}
\label{eq:app:eq:cont}
\rho'  +  \frac{1}{r^{2}} \frac{\partial}{\partial r} \left(  r^{2} \rho_{0} \xi_{r}  \right)  -  \rho_{0} \frac{l \left( l + 1 \right) V }{r} = 0,
\end{equation}

\noindent where we have used the fact that $V \propto Y^{-m}_{l} (\theta, \phi)$ such that $\nabla^{2}_{\perp} V = -  l \left( l + 1 \right) V/{r^{2}}$.

\noindent Using $\Delta \rho =0$, this can be rearranged and simplified to give
\begin{equation}
\label{eq:app:eq:V_general}
V = \frac{r}{l \left( l + 1 \right)} \left(  2 \frac{\xi_{r}}{r} + \frac{\partial \xi_{r}}{\partial r}  \right).
\end{equation}

\noindent To examine the behaviour towards the surface, we use the approximation that $g = G M / r^{2}$, where $G$ is the gravitational constant, and $M$ is the stellar mass. This approximation is increasingly valid as we approach the surface due to the low density of surface material. Combining this with the expression for $\Phi_{\text{P}}$ gives the radial displacement as:
\begin{equation}
  \label{eq:app:eq:xi_r}
  \xi_{r} =  \frac{m_{\text{P}} r^{4}}{4 M D^{3}}.
\end{equation}

\noindent This then gives the horizontal displacement function as:
\begin{equation}
\label{eq:app:eq:V_surface}
V \approx  \frac{3 m_{\text{P}} r^{4}}{2 l \left( l + 1 \right) M D^{3}},
\end{equation}

\noindent which, combined with the fact that $l=2$, gives the result that $V = \xi_{r}$ towards the surface. Importantly, the scale of variation of these quantities
is of the order of the radius.

\noindent As mentioned already, this { was derived } under the assumptions of hydrostatic equilibrium, applying in the limit $m\omega \rightarrow 0$, and that 
{ either the perturbations are adiabatic or the Lagrangian variation of the density and pressure
 are zero with $N^2 \ne 0$. However, a different situation arises when both $m^2\omega^2$ and $N^2$ are small and tend to zero simultaneously.
 This is potentially significant in convection zones in which the convective heat transport is efficient as in the inner solar convective envelope.
 This we now explore.

 \subsection{The situation when $m^2\omega^2$ is small 
   and $N^2$ is either smaller or  of the same orders }
 \label{AppC1}

We begin with equation (\ref{eq:Methods:analytical:rearranged_eom_V}) which reads:
\begin{equation} 
\label{eq:Methods:analytical:rearranged_eom1_V1}
  V = \frac{1}{r m^{2} \omega^{2}} \left( \frac{p'}{\rho_{0}} + \Phi_{\text{P}} \right),
\end{equation}
and  rewrite it in the form:
\begin{equation} 
\label{eq:Methods:analytical:rearranged_eom1_V2}
  V = \frac{1}{r\rho_0 m^{2} \omega^{2}}  {\cal W} {\cal F} ,
\end{equation}
which defines ${\cal W}$, and where ${\cal F} $ is defined through the relation:
\begin{equation} 
\label{eq:Methods:analytical:rearranged_eom1_V3}
 \frac{ d{{\cal F}/dr} }{{\cal F}}= \frac{ d p_0/dr} {\Gamma_1 p_0}.
\end{equation}
Using the adiabatic relation $\Delta p = (\Gamma_1 p_0/\rho_0)\Delta \rho $ together with the radial component of the linearised  equation of motion
gives:
\begin{equation} 
\label{eq:Methods:analytical:rearranged_eom1_V4}
  \left( { m^{2} \omega^{2}} - N^2 \right ) \xi_r   =   \frac{ {\cal F}}{\rho_0} \frac{d{\cal W}}{dr} +\frac{  \Phi_{\text{P}}N^2}{g}
\end{equation}
From equations (\ref{B1}) and (\ref{B3}) under the adiabatic assumption
(i.e. $\nabla\cdot {\bf F}$ set to zero), we find that:
\begin{equation} 
\label{eq:Methods:analytical:rearranged_eom1_V5}
  p' =   -\frac{\Gamma_1 p_0}{ {\cal F}}
  \nabla\cdot({\bmath{\xi}} {\cal F} ) .
\end{equation}
Making use of  equations~ (\ref{eq:Methods:analytical:rearranged_eom1_V1}), (\ref{eq:Methods:analytical:rearranged_eom1_V2}),
(\ref{eq:Methods:analytical:rearranged_eom1_V4}), and (\ref{eq:Methods:analytical:rearranged_eom1_V5}),   
we obtain to within corrections of order $m^4\omega^4$ on the right hand side:
\begin{align} 
\label{eq:Methods:analytical:rearranged_eom1_V6}
 & \left( { m^{2} \omega^{2}} - N^2 \right ) \xi_r - \frac{ {\cal F}m^2\omega^2}{\rho_0} \frac{d }{dr} \left(\frac{\rho_0} {l(l+1){\cal F}^2}\frac{d(r^2{\cal F}\xi_r)}{dr}\right)= 
 \nonumber\\
& \frac{  \Phi_{\text{P}}N^2}{g}    -\frac{ {\cal F}}{\rho_0} \frac{d }{dr} \left(\frac{\Phi_{\text{P}}\rho_0^2r^2m^2\omega^2}{{\cal F}l(l+1)\Gamma_1 p_0}\right)
\end{align}
From the above, we see that if $m^2\omega^2 \rightarrow 0$ with $N^2$ remaining finite, we recover the
equilibrium tide solution $\xi_r = -\Phi_{\text{P}}/g,$ as expected.

However, if they remain of the same order, equation (\ref{eq:Methods:analytical:rearranged_eom1_V6}) must be considered in its
entirety and the equilibrium tide cannot be assumed.
In the limit $N^2\rightarrow 0,$ with $\omega^2$ remaining finite or tending to zero more slowly, equation (\ref{eq:Methods:analytical:rearranged_eom1_V6}) becomes:
\begin{align} 
\label{eq:Methods:analytical:rearranged_eom1_V7}
 &  \frac{ {\cal F}}{\rho_0} \frac{d }{dr} \left(\frac{\rho_0} {l(l+1){\cal F}^2}\frac{d(r^2{\cal F}\xi_r)}{dr}\right) -\xi_r=   
  \frac{ {\cal F}}{\rho_0} \frac{d }{dr} \left(\frac{\Phi_{\text{P}}\rho_0^2r^2}{{\cal F}l(l+1)\Gamma_1 p_0}\right) . \nonumber\\
&  
\end{align}
 This corresponds to the isentropic limit.
 It was  discussed in section\,2.2 of \citet{Terquem1998} who consider an equivalent situation  when a barotropic equation of state applies.
 In our case we expect it to apply in the deeper layers of the convective envelope which are extensive in the  1~$M_{\odot}$ case
 but much less so in the 1.4~$M_{\odot}$  case (see Fig. \ref{fig:N2_surface}).
 
 More generally, equation (\ref{eq:Methods:analytical:rearranged_eom1_V6}) takes the form of  a forced oscillator with gravity waves that can be excited and
  propagate in the region  where $N^2 \ge m^2\omega^2.$ These can be confined in an inner radiative cavity decaying exponentially
 into the convection zone forming standing waves that can be resonantly  excited  as seen in the calculations. We should not
 expect the equilibrium tide to hold then. 
 
 

}

\section{Low frequency limit when the standard equilibrium tide does not apply and perturbations are not adiabatic}
\label{sec:app:nheq}
{ We  now investigate situations where the standard equilibrium tide does not apply in the low frequency limit in the non adiabatic layer close to the surface.
The reason for this is that  the density and temperature perturbations 
do not even approximately satisfy the adiabatic condition $\Delta p =(\Gamma_1p_0)/\rho_0\Delta \rho$, which is required for
hydrostatic equilibrium and  the standard equilibrium tide to hold
(see equations (\ref{eq:app:eq:ad}) and (\ref{eq:app:eq:delta}) and the discussion immediately below).

To begin, we note that} equation~(\ref{eq:Methods:analytical:rearranged_eom_V}), derived from the horizontal components of the linearised equation of motion, can be written in the form:
\begin{equation}
\label{eq:app:eq:V1_surface}
m^2\omega^2\rho_0V r = p'     +h\frac{dp_0}{dr},
\end{equation}
where the quantity $ h = -\Phi_{\text{P}}/g$ is  equal to the radial displacement component of the standard equilibrium tide.
The radial component of the equation of motion (\ref{eq:mom_osc}) can, with the help of
equation~(\ref{eq:app:eq:V1_surface}), be written in the form:
\begin{equation}
\label{eq:app:eq:r1_surface}
m^2\omega^2\left(  \rho_0\xi_ r  - \frac{\partial( \rho_0V r)}{\partial r} \right)= \left(\rho'     +h\frac{d\rho_0}{dr}\right)g,
\end{equation}
For adiabatic perturbations  expected in the deep interior we  may write: 
\begin{equation}
p' -\frac{\Gamma_1 p_0}{\rho_0}\rho'
+\xi_r\frac{\Gamma_1 N^2 p_0}{g}
=0 
\label{adiabcon}
\end{equation}

In order to investigate more general relations between the density and pressure perturbations
that may not be consistent with hydrostatic equilibrium as may occur in the non adiabatic zone near the surface
in a simplified manner, we modify (\ref{adiabcon}) by introducing  parameters $\beta_1$ and $\beta_2$  so that it reads:
\begin{equation}
\label{eq:app:eq:N20_surface}
p' -\frac{ \Gamma_1 p_0}{\rho_0}\rho'+ \xi_r\frac{\Gamma_1 N^2 p_0}{g}
+\beta_1 \left(\frac{\beta_2(\Gamma_1-1) p_0}{\rho_0}\rho'- \xi_r\frac{\Gamma_1 N^2 p_0}{g}\right)
=
0.
\end{equation}
{ From the second law of thermodynamics we have:
\begin{equation}
\label{eq:app:eq:N20_surface1}
\beta_1 \left(\frac{\beta_2(\Gamma_1-1) p_0}{\rho_0}\rho'- \xi_r\frac{\Gamma_1 N^2 p_0}{g}\right)
=
-\frac{{\rm i}(\Gamma_3-1)\nabla\cdot{\bf F}'}{m\omega}
\end{equation}
which implies that  $\beta_1$ and/or $\beta_2$ may be complex.
In addition we note that,
when $\beta_1=0,$ we have the usual adiabatic condition (\ref{adiabcon}), 
and for $\beta_1=1,$ with $\beta_2=1,$ we have, assuming an ideal gas, the isothermal condition $p' = p_0\rho'/\rho_0.$
In addition when  $\beta_1=1,$ with $\beta_2=  0 $ we have the condition for zero entropy
perturbation that $p'/p_0 = \Gamma_1\rho'/\rho.$

Although the introduction of the parameters $\beta_1$ and $\beta_2$ is used to
simplify  matters by allowing one to ignore  the details of radiation transport,
this approach is able to indicate how the significant departures 
from predictions made from consideration of the adiabatic equilibrium tide,  that
 we see in the surface layers,  can come about.}
Using equations~(\ref{eq:app:eq:V1_surface}) and (\ref{eq:app:eq:r1_surface})
to eliminate $p'$ and $\rho'$ 
in equation~(\ref{eq:app:eq:N20_surface}), we obtain a consistency condition connecting the standard 
equilibrium tide radial displacement and the actual displacement components in the form:
\begin{align}
\label{eq:app:eq:N21_surface}
 &(1 - \beta_1) (\xi_r-h)\frac{N^2 \Gamma_1 p_0}{g}      -\beta_1 h\left(   \frac{dp_0}{dr}    - \frac{\beta_3p_0}{ \rho_0} \frac{d\rho_0}{dr}\right )  =\nonumber\\
&-m^2\omega^2\rho_0V r+\frac{ \beta_4 p_0}{\rho_0}\frac{m^2\omega^2}{g}\left(  \rho_0\xi_ r  - \frac{\partial( \rho_0V r)}{\partial r} \right),
\end{align}
where
$\beta_3 = (\beta_2+\Gamma_1(1-\beta_2))$ and

\noindent $\beta_4 =(\beta_2(\beta_1 +(1-\beta _1)\Gamma_1)+\Gamma_1(1-\beta_2 )).$
{ Let us suppose now that we can take the limit $\omega \rightarrow 0$ and recover the equilibrium tide
prediction for the radial displacement that $\xi_r =h.$
Equation (\ref{eq:app:eq:N21_surface}) then implies that, if $\beta_1\ne 0$  and  $V$ is non--singular, we must have
$ dp_0/{dr} = (\beta_3 p_0/\rho_0 ){d\rho_0}/{dr},$ 
corresponding to a specific relation between $p_0$ and $\rho_0.$
Thus the standard equilibrium tide description cannot be recovered in general.}

Writing equation~(\ref{eq:app:eq:N21_surface}) in terms of the pressure scale height $H_p \ll r,$ we obtain:
\begin{align}
\label{eq:app:eq:N211_surface}
  &(1 - \beta_1) (\xi_r-h)\frac{g }{r} \left( \frac{\Gamma_1p_0}{\rho_0} \frac{d\rho_0}{d p_0} - 1\right)   +\beta_1 h\frac{g}{ r }\left( 1  -\frac{\beta_3p_0}{\rho_0} \frac{d\rho_0}{d p_0} \right)  = \nonumber \\
 & m^2\omega^2  \left(    \beta_4\left[ V \frac{p_0}{\rho_0} \frac{d\rho_0}{d p_0}  +\frac{H_p}{r} \left( \xi_r  - \frac{\partial( V r)}{\partial r} \right)\right]-V\right).
\end{align}
 We remark that equation~(\ref{eq:app:eq:N211_surface}) implies significant departures from the standard equilibrium tide 
 when the conditions for it to apply are not satisfied (see appendix \ref{sec:app:eq} above).
 
   When  $\xi_r$ does not exceed $V$ in characteristic magnitude and we 
assume that $V$ changes on a scale significantly exceeding $H_p,$ as occurs for the equilibrium tide,  then the term proportional
to $H_p$ is negligible. { If $\xi_r =h,$ the standard equilibrium tide value, and there are significant departures from 
adiabatic behaviour, with $|\beta_1|$ of order unity,
 we would conclude that
$|V/ h| =  g/( m^2\omega^2r) \gg 1.$  This large increase in $|V|$ in comparison to $h$,
a factor $\sim 400$ for our solar mass model,
would have to occur in the thin non--adiabatic surface layer.  Accordingly, the characteristic scale
of increase would be the scale height there or less.
Rearranging the expression $|V/ h| =  g/( m^2\omega^2r)$ at the surface, and using $G M = D^{3} \omega^{2}$, from the properties of the orbit, we get $|V/ R| =  m_{p}/( 4 m^2 M)$. This is independent of the orbital distance, and depends only on the mass ratio between the planet and the star.

An alternative possibility from equation~(\ref{eq:app:eq:N211_surface})  that would avoid large values of  $|V|$ occurs if the left hand side is zero.
However, this implies significant departures of $\xi_r$ from the equilibrium tide occurring in the thin non adiabatic zone.}


\bsp	
\label{lastpage}
\end{document}